\documentclass[aps,prb,twocolumn,superscriptaddress]{revtex4-2}

\usepackage{amsmath}    
\usepackage{longtable}
\usepackage{makecell}
\usepackage{amsfonts}
\usepackage{amssymb}
\usepackage[colorlinks=true,citecolor=blue,linkcolor=red]{hyperref}
\usepackage{graphicx}
\usepackage{bbold}					
\usepackage[dvipsnames]{xcolor}

\usepackage{bm} 
\usepackage{dcolumn}
\usepackage{mathtools}

\usepackage{array,booktabs,ragged2e}	
\usepackage{soul}						
\usepackage{cancel}						
\usepackage{tabularx}
\usepackage{multirow}
\usepackage{hhline}
\usepackage{adjustbox}
\usepackage{sidecap}
\usepackage{threeparttable}
\usepackage{colortbl} 
\usepackage{tabularx}
\usepackage[utf8]{inputenc}
\usepackage[T1]{fontenc}
\usepackage{siunitx}
\usepackage[normalem]{ulem}
\usepackage{xcolor}
\definecolor{darkgreen}{rgb}{0.0, 0.8, 0.0} 

\def\degree {^{\circ}}

\newcolumntype{P}[1]{>{\raggedleft\arraybackslash}p{#1}}
\newcolumntype{R}[1]{>{\centering\arraybackslash}p{#1}}

\usepackage{bm,color,amsmath,amssymb,mathrsfs,latexsym,graphicx,psfrag,mathtools}
\usepackage{color}
\usepackage[dvipsnames]{xcolor}

\usepackage{empheq}

\newcommand{\bsl}[1]{\boldsymbol{#1}}

\newcommand{\ii}{\mathrm{i}}

\newcommand{\dsZ}{\mathbb{Z}}
\newcommand{\dsN}{\mathbb{N}}
\newcommand{\dsR}{\mathbb{R}}

\newcommand{\Tr}{\mathop{\mathrm{Tr}}}

\newcommand{\U}{\mathrm{U}}

\newcommand{\eqnref}[1]{Eq.\,\eqref{#1}}

\newcommand{\refcite}[1]{Ref.\,\cite{#1}}

\newcommand{\mat}[1]{\left(\begin{matrix}#1\end{matrix}\right)}

\newcommand{\eq}[1]{\begin{equation} #1 \end{equation}}
\newcommand{\eql}[1]{\begin{widetext}\begin{align}\begin{split} #1 \end{split}\end{align}\end{widetext}}
\newcommand{\eqa}[1]{\begin{align}\begin{split} #1 \end{split}\end{align}}

\let\oldAA\AA
\renewcommand{\AA}{\text{\normalfont\oldAA}}

\newcommand{\ie}{{\emph{i.e.}}}

\newcommand{\A}{\mathcal{A}}

\newcommand{\K}{\text{K}}

\newcommand{\tmt}{\text{$t$MoTe$_2$}}
\newcommand{\tws}{\text{$t$WSe$_2$}}

\newcommand{\cellnum}[2]{\makecell{$#1$ \\ $#2\mathrm{i}$}}
\usepackage{comment}

\usepackage{cleveref}
\crefname{appendix}{App.\,}{Apps.\,}
\crefname{equation}{Eq.}{Eqs.\,}
\crefname{figure}{Fig.\,}{Figs.\,}
\crefname{table}{Tab.\,}{Tabs.\,}
\crefname{section}{Sec.\,}{Secs.\,}
\creflabelformat{appendix}{[#2#1#3]}

\begin{document}

\title{Universal Moir\'e-Model-Building Method without Fitting: Application to Twisted MoTe$_2$ and WSe$_2$}

\author{Yan Zhang}
\thanks{These authors contributed equally.}
\affiliation{Beijing National Laboratory for Condensed Matter Physics and Institute of Physics,
Chinese Academy of Sciences, Beijing 100190, China}
\affiliation{University of Chinese Academy of Sciences, Beijing 100049, China}

\author{Hanqi Pi}
\thanks{These authors contributed equally.}
\affiliation{Beijing National Laboratory for Condensed Matter Physics and Institute of Physics,
Chinese Academy of Sciences, Beijing 100190, China}
\affiliation{University of Chinese Academy of Sciences, Beijing 100049, China}

\author{Jiaxuan Liu}
\thanks{These authors contributed equally.}
\affiliation{Beijing National Laboratory for Condensed Matter Physics and Institute of Physics,
Chinese Academy of Sciences, Beijing 100190, China}
\affiliation{University of Chinese Academy of Sciences, Beijing 100049, China}

\author{Wangqian Miao}
\affiliation{Department of Physics, The Hong Kong University of 
Science and Technology, Clear Water Bay, Hong Kong, China}

\author{Ziyue Qi}
\affiliation{Beijing National Laboratory for Condensed Matter Physics and Institute of Physics,
Chinese Academy of Sciences, Beijing 100190, China}
\affiliation{University of Chinese Academy of Sciences, Beijing 100049, China}

\author{Nicolas Regnault}
\affiliation{Center for Computational Quantum Physics, Flatiron Institute, 162 5th Avenue, New York, NY 10010, USA}
\affiliation{Department of Physics, Princeton University, Princeton, New Jersey 08544, USA}
\affiliation{Laboratoire de Physique de l’Ecole normale sup\'erieure,
ENS, Universit\'e PSL, CNRS, Sorbonne Universit\'e,
Universit\'e Paris-Diderot, Sorbonne Paris Cit\'e, 75005 Paris, France}

\author{Hongming Weng}
\affiliation{Beijing National Laboratory for Condensed Matter Physics and Institute of Physics,
Chinese Academy of Sciences, Beijing 100190, China}
\affiliation{University of Chinese Academy of Sciences, Beijing 100049, China}
\affiliation{Songshan Lake Materials Laboratory, Dongguan, Guangdong 523808, China}
\date{\today}

\author{Xi Dai}
\affiliation{Department of Physics, The Hong Kong University of 
Science and Technology, Clear Water Bay, Hong Kong, China}

\author{B. Andrei Bernevig}
\affiliation{Department of Physics, Princeton University, Princeton, New Jersey 08544, USA}
\affiliation{Donostia International Physics Center, P. Manuel de Lardizabal 4, 20018 Donostia-San Sebastian, Spain}
\affiliation{IKERBASQUE, Basque Foundation for Science, Bilbao, Spain}

\author{Quansheng Wu}
\email{quansheng.wu@iphy.ac.cn}
\affiliation{Beijing National Laboratory for Condensed Matter Physics and Institute of Physics,
Chinese Academy of Sciences, Beijing 100190, China}
\affiliation{University of Chinese Academy of Sciences, Beijing 100049, China}

\author{Jiabin Yu}
\email{yujiabin@ufl.edu}
\affiliation{Department of Physics, University of Florida, Gainesville, FL, USA}
\affiliation{Department of Physics, Princeton University, Princeton, New Jersey 08544, USA}
\date{\today}

\begin{abstract}
We develop a comprehensive method to construct analytical continuum models for moir\'e systems directly from first-principle calculations \emph{without any parameter fitting}. 
The core idea of this method is to interpret the terms in the continuum model as a basis, allowing us to determine model parameters as coefficients of this basis through Gram-Schmidt orthogonalization.
We apply our method to twisted MoTe$_2$ and WSe$_2$ with twist angles ranging from 2.13$^\circ$ to 3.89$^\circ$, producing continuum models that exhibit excellent agreement with both energy bands and wavefunctions obtained from first-principles calculations.
We further propose a strategy to integrate out the higher-energy degrees of freedom to reduce the number of the parameters in the model without sacrificing the accuracy for low-energy bands. 
Our findings reveal that decreasing twist angles typically need an increasing number of harmonics in the moir\'e potentials to accurately replicate first-principles results.
We provide parameter values for all derived continuum models, facilitating further robust many-body calculations.
Our approach is general and applicable to any commensurate moir\'e materials accessible by first-principles calculations.

\end{abstract}

\maketitle

\section{Introduction}
The groundbreaking discovery of superconductivity in twisted bilayer graphene~\cite{cao_Unconventional_2018, yankowitz_Tuning_2019, codecido_Correlated_2019, lu_Superconductors_2019, arora_Superconductivity_2020, stepanov_Untying_2020, saito_Independent_2020, oh_Evidence_2021} has ignited a surge of research into moir\'e materials as promising platforms for hosting a plethora of exotic correlated and topological phenomena~\cite{balents_Superconductivity_2020, andrei_Marvels_2021, regnault_Catalogue_2022, checkelsky_Flat_2024}.
Subsequently, experiments have observed (or showed strong signatures of) fractional Chern insulators in moir\'e materials~\cite{cai_Signatures_2023, zeng_Thermodynamic_2023, park_Observation_2023, xu_Observation_2023, kang_Evidence_2024, ji_Local_2024, redekop_Direct_2024, xu_Interplay_2024, lu_Fractional_2024, xie_Even_2024, park_Ferromagnetism_2024, choi_Electric_2024, lu_Extended_2024}, following earlier observations of fractional-quantum-Hall-like states in lattice systems under magnetic fields~\cite{spanton_Observation_2018, xie_Fractional_2021}.

A defining characteristic of moir\'e materials is their ability to host nearly flat topological bands, which significantly enhance electron-electron interactions and facilitate the emergence of correlated phases. This unique electronic structure underscores the necessity for robust single-particle models that can accurately capture the low-energy physics intrinsic to these systems. The continuum model, pioneered by \refcite{lopesdossantos_Graphene_2007,bistritzer_Moire_2011,lopesdossantos_Continuum_2012} has become a fundamental tool in this context. Analogous to the $k \cdot p$ models~\cite{luttinger_Motion_1955, kane_Chapter_1966, liu_Model_2010}, the continuum model effectively encapsulates the physics arising from the large moir\'e lattice constant relative to the underlying atomic lattices.

Despite the generality afforded by symmetry considerations in formulating continuum models, the determination of precise model parameters remains a critical challenge. Traditionally, these parameters are extracted by fitting the continuum model's band structure to that obtained from Density Functional Theory (DFT) calculations~\cite{bistritzer_Moire_2011, jung_Initio_2014, moon_Electronic_2014, chittari_GateTunable_2019, song_All_2019, carr_Exact_2019, herzog-arbeitman_Moire_2024, wu_topological_2019, koshino_Effective_2020, angeli_Valley_2021, reddy_fractional_2023, wang_fractional_2024, xu_maximally_2024, jia_moire_2024, xu_Multiple_2024, devakul_magic_2021}. In graphene-based systems, parameters can be derived through Fourier analysis (or perturbation analysis) of hopping functions~\cite{bistritzer_Moire_2011, nam_Lattice_2017, koshino_Maximally_2018, balents_deform_2019, carr_Exact_2019, guinea_Continuum_2019, chittari_GateTunable_2019, koshino_Effective_2020, bernevig_Twisted_2021, vafek_Continuum_2023, kang_Pseudomagnetic_2023, xie_tbg_relax_2024}. However, the determination of the hopping parameters in the hopping functions still require fitting to DFT results~\cite{moon_Energy_2012, fang_Electronic_2016}, a process that can introduce significant uncertainties and inconsistencies.

Parameter fitting introduces several critical limitations: firstly, a model optimized solely for band structure alignment may fail to accurately reproduce the corresponding wavefunctions; secondly, when the number of moir\'e bands involved is large (for example, due to large fillings or strong interaction), achieving an exact match between model and DFT wavefunctions through fitting becomes very challenging due to the large number of needed parameters.
Disparate parameter sets can lead to different flat band topology~\cite{bistritzer_Moire_2011, jung_Initio_2014, wu_topological_2019, park_Topological_2023, wang_topology_2023, goldman_ZeroField_2023, reddy_fractional_2023, wang_fractional_2024, abouelkomsan_band_2024, xu_maximally_2024, jia_moire_2024, wang_Higher_2024, xu_Multiple_2024, zhang_polarization-driven_2024, crepel_Chiral_2024} due to the different wavefunctions.
Different wavefunctions can also lead to divergent predictions of the nature of correlated phases induced by interactions~\cite{qiu_interaction-driven_2023, dong_composite_2023, reddy_fractional_2023, wang_topology_2023, goldman_ZeroField_2023, reddy_toward_2023, abouelkomsan_band_2024, wang_fractional_2024, yu_fractional_2024, xu_maximally_2024, wang_Higher_2024, xu_Multiple_2024, kwan_When_2024, reddy_NonAbelian_2024, ahn_NonAbelian_2024, wang_Diverse_2024, pan_band_2020, wang_Correlated_2020, devakul_magic_2021, crepel_Anomalous_2023, guo_Superconductivity_2024, xia_Superconductivity_2024, kim_Theory_2024, foutty_Mapping_2024b, schrade_Nematic_2024, tuo_Theory_2024, belanger_Superconductivity_2022, klebl_Competition_2023, wietek_Tunable_2022, wu_PairDensityWave_2023, zegrodnik_Mixed_2023, akbar_Topological_2024, christos_Approximate_2024, zhu_Theory_2024, myerson-jain_SuperconductorInsulator_2024, xie_Superconductivity_2024, guerci_Topological_2024}, since the wavefunctions directly determine the form of the projected interactions, which dominate the correlation physics in the moir\'e flat-band systems.

These challenges underscore the urgent need for a novel approach to constructing faithful continuum models that obviates the reliance on parameter fitting. Such an approach would not only enhance the accuracy and reliability of the models but also streamline the process of exploring and predicting new correlated and topological phenomena in moir\'e materials.
Recently, directly Wannierization of the moir\'e DFT Hamiltonian has been done in \refcite{wang_Higher_2024}, which can provide a precise numerical low-energy Hamiltonian.
Nevertheless, this method cannot provide a continuum model, as the continuum model is built on plane wave basis instead of the moir\'e Wannier functions.

In this work, we introduce a universal framework for constructing \textit{ab initio} continuum models for moir\'e materials, eliminating the need for parameter fitting. 
Our method leverages the fact that different terms in the continuum model serve as the basis of the Hamiltonian, and the model parameters are essentially the coefficients of the basis, which can be determined from DFT calculations via projections.
The method removes the ambiguities and inaccuracies typically associated with traditional fitting procedures.
Our method is complementary to \refcite{wang_Higher_2024}.

To demonstrate the efficacy of our approach, we apply it to twisted bilayer MoTe$_2$ ({\tmt}) and twisted bilayer WSe$_2$ ({\tws}) with twist angles ranging from 2.13$^\circ$ to 3.89$^\circ$. Our continuum models exhibit exceptional agreement with DFT results---for example, the maximal energy deviation is less than 0.5 meV for 2.13$^\circ$ {\tmt} and 1.3 meV for 2.13$^\circ$ {\tws} for the top four energy bands, and the wavefunction overlap probability exceeds 97\% across all these bands, excluding the regions where bands touch. 
These results validate the accuracy and reliability of our continuum models, providing a reliable foundation for numerical studies of the correlated phases. 
Moreover, we propose a strategy to integrate out the higher-energy degrees of freedom to reduce the number of the terms (and thus parameters) in the model without sacrificing essential physical characteristics. 
This simplification produces a reduced continuum model that has smaller parameters; for example, the reduced continuum model for {\tmt} has 23 parameters compared to 148 parameters of the full continuum model (148 parameters represent the minimum required to keep energy deviations below 0.5 meV and wavefunction overlaps above 97\% for the top four energy bands). Notably, the reduced model still well capture the top three DFT bands. 
The reduced continuum model can facilitate future analytical studies.

All the parameter values of the full and reduced continuum models for  {\tmt} and {\tws} with twist angles ranging from 2.13$^\circ$ to 3.89$^\circ$ are provided in \cref{table:mote2_52orb_full_diag,table:mote2_52orb_full_inter,table:mote2_52orb_full_intra,table:mote2_52orb_reduced_diag,table:mote2_52orb_reduced_inter,table:mote2_52orb_reduced_intra,table:mote2_71orb_full_diag,table:mote2_71orb_full_intra,table:mote2_71orb_full_inter,table:mote2_71orb_reduced_diag,table:mote2_71orb_reduced_inter,table:mote2_71orb_reduced_intra,table:wse2_47orb_full_diag,table:wse2_47orb_full_intra,table:wse2_47orb_full_inter,table:wse2_47orb_reduced_diag,table:wse2_47orb_reduced_inter,table:wse2_47orb_reduced_intra} and on GitHub~\cite{zhang_continuum_2024}.
The parameter values indicate that as the twist angle decreases, we typically need an increasing number of harmonics in the moir\'e potentials to capture the DFT results.
For example, the number of intralayer (interlayer) harmonics for {\tmt} increases from 4 (4) for $3.89^\circ$ to 5 (6) for 2.13$^\circ$ for the full model, and increases from 1 (2) for $3.89^\circ$ to 5 (4) for 2.13$^\circ$ for the reduced model.

\begin{figure*}[!th]
    \centering
    \includegraphics[width=\linewidth]{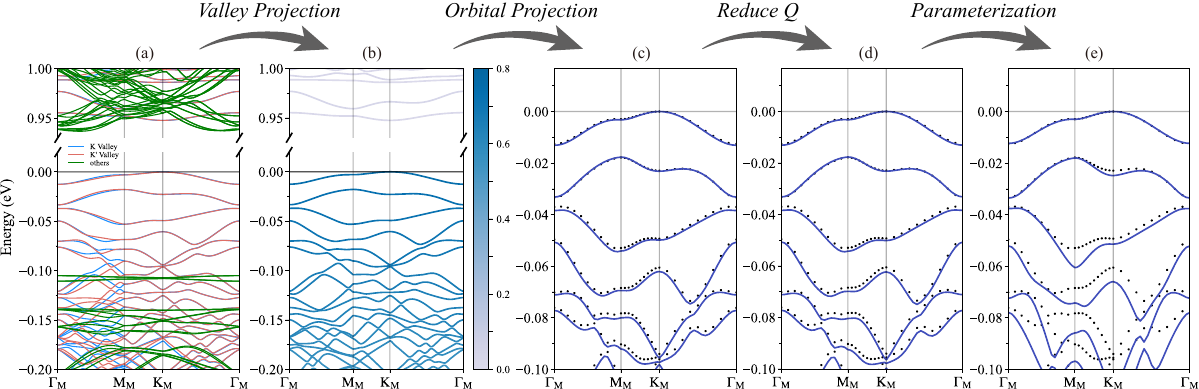}
    \caption{\textbf{DFT to Reduced Model.} 
    (a) Band structure of the full Hamiltonian for 3.89$\degree$ {\tmt} computed using OpenMX. Blue lines denote the K valley, red lines represent the K$^{\prime}$ valley bands, and green lines correspond to the bands of other valleys, including the $\Gamma$ valley. 
    (b) Valley-projected band structure obtained via the Truncated Atomic Plane Wave (TAPW) method, exclusively displaying the K valley. The color intensity indicates the contribution of Mo's $d_{xy}$ and $d_{x^2+y^2}$ orbitals to each band. 
    (c) Orbital projection focusing solely on Mo's $d_{xy}$ and $d_{x^2+y^2}$ orbitals.
    (d) Band structure derived from the reduced DFT model.
    (e) Band structure obtained from the reduced continuum model. 
    In panels (c) to (e), the black dotted lines correspond to the DFT results for comparison.
    }
    \label{fig:model_contruction}
\end{figure*}

\section{DFT-Based Structural and Electronic Calculations}\label{sec:DFT}

We first present a method for efficiently obtaining reliable electronic structures of moir\'e systems from first-principles calculations. We start by obtaining a relaxed structure of a moir\'e system.
Traditional methods face convergence challenges when dealing with moir\'e systems due to the large number of atoms in the unit cell (e.g., 4326 atoms for a 2.13$^\circ$ {\tmt}). 
To address this, we employed a Machine Learning Force Field (MLFF) trained on ab-initio data~\cite{jia_moire_2024,xu_Multiple_2024,mao_Transfer_2024,zhang_polarization-driven_2024}. After using VASP's MLFF module~\cite{kresse_Initio_1993,kresse_Normconserving_1994a,blochl_Projector_1994,kresse_Efficient_1996,kresse_Ultrasoft_1999,jinnouchi_Fly_2019} to generate training data from molecular dynamics simulations, we then utilized Allegro~\cite{musaelian_Learning_2023}, an E(3)-equivariant neural network, to train an accurate MLFF. Allegro's architecture ensures that both the input and output of each neural network layer are equivariant under rotations, reflections, and translations in three-dimensional space. This capability allows our MLFF to accurately predict forces and energies while effectively handling the rotational and translational symmetries inherent in the relaxed structures of moir\'e systems. 

After relaxing the crystal structures, we performed DFT calculations using OpenMX software ~\cite{ozaki_Variationally_2003, ozaki_Numerical_2004,ozaki_Efficient_2005} with norm-conserving pseudopotentials and localized basis functions, which is particularly efficient for large moir\'e systems due to its optimized handling of localized orbitals and its capability to handle large-scale computations. 
Consequently, we obtained an \textit{ab initio} tight binding Hamiltonian based on non-orthogonal pseudo-atomic orbitals (PAOs), resulting in a generalized eigenvalue problem. 
Alternatively, other software packages using numerical atomic orbitals such as SIESTA~\cite{soler_SIESTA_2002} and ABACUS~\cite{li_Largescale_2016,chen_Systematically_2010}  can also be used to perform these DFT calculations.
A detailed description of the relaxation and DFT calculation methodologies can be found in \cref{appendix:DFT_details} 

Solving a generalized eigenvalue problem in moir\'e systems presents substantial computational challenges due to the Hamiltonian's high dimensionality. 
To overcome this, we adopt the truncated atomic plane wave (TAPW) method, initially proposed to study twisted bilayer graphene systems in \refcite{miao_truncated_2023, chen_epc_2023,shi_tbg_epc_2024}.
In this work, we extend this method to non-orthogonal basis.

Using {\tmt} as a case study, where the low-energy physics is dominated by two atomic valleys (labeled by $\pm$K), we project the \textit{ab initio} tight binding Hamiltonian onto a group of atomic plane waves localized around the K valley to significantly reduce computational costs (the $-$K valley Hamiltonian can be obtained using time-reversal symmetry), as shown in \cref{fig:model_contruction}(a) to (b). 
Specifically, \cref{fig:model_contruction}(a) displays the full Hamiltonian's band structure of 3.89$^\circ$ {\tmt} computed using OpenMX, while  \cref{fig:model_contruction}(b) presents the K valley-projected band structure. This approach effectively captures the essential electronic properties of the moir\'e system with reduced computational complexity.

The atomic plane wave basis for the $l$-th layer at the K valley is defined as:
\eq{
\label{eq:FT_DFT}
|\psi_{l,\widetilde{\bsl{G}}_l,\alpha}(\bsl{k})\rangle = \frac{1}{\sqrt{N_\text{m}N_\text{a}}}\sum_{\text{I}, i} \mathrm{e}^{\mathrm{i}(\bsl{k}+\widetilde{\bsl{G}}_{l})(\mathbf{R}_{\text{I}}+\tau_{li\alpha})}\left|\phi_{\alpha,\mathbf{R}_{\text{I}}+\tau_{li \alpha}}\right\rangle\ ,
}
where $\widetilde{\bsl{G}}_{l}$ denotes a moir\'e reciprocal lattice vector for the $l$-th layer in the DFT convention (which has a large shift compared to the model convention as discussed in \cref{sec:continuum_model}), and $\bsl{k}$ represents the momentum in the first moir\'e Brillouin zone. 
$\mathbf{R}_{\text{I}}$ is the moir\'e lattice vector, $\tau_{li\alpha}$ indicates the displacement of the $\alpha$-th orbital in the $i$-th non-twisted unit cell of the $l$-th layer, $\left|\phi_{\alpha,\mathbf{R}_{\text{I}}+\tau_{li \alpha}}\right\rangle$ is the $\alpha$-th PAO centered at $\mathbf{R}_{\text{I}}+\tau_{li\alpha}$. $N_{\text{m}}$ represents the number of moir\'e unit cells and $N_{\text{a}}$ refers to the number of non-twisted unit cells within a single moir\'e unit cell.

The matrix element of the Hamiltonian and overlap integral can be obtained under the TAPW basis defined in Eq.~(\ref{eq:FT_DFT}),
\eq{
\begin{aligned}
    \label{eq:TAPW}
    \text{H}_{ln\alpha,l'm\beta}^{\text{TAPW}}(\bsl{k}) =\sum_{ij}\text{X}^\dagger_{li\alpha,ln\alpha}\text{H}_{li\alpha,l'j\beta}(\bsl{k})\text{X}_{l'j\beta,lm\beta} \ , \\
    \text{S}_{ln\alpha,l'm\beta}^{\text{TAPW}}(\bsl{k}) =\sum_{ij}\text{X}^\dagger_{li\alpha,ln\alpha}\text{S}_{li\alpha,l'j\beta}(\bsl{k})\text{X}_{l'j\beta,lm\beta} \ ,
\end{aligned}
}
where $\text{H}(\bsl{k})$ and $\text{S}(\bsl{k})$ represent the full non-orthogonal \textit{ab initio} tight binding matrix and overlap matrix, respectively, and $\text{X}$ denotes the plane wave projector under continuum approximation \cite{miao_truncated_2023}. \cref{eq:TAPW} represents for an unitary transformation when all atomic plane waves are preserved. However, in moir\'e systems, we can truncate the atomic plane waves near specific atomic valley we concern to simplify the problem.
The explicit expressions for $\text{H}(\bsl{k})$, $\text{S}(\bsl{k})$, and $\text{X}$, along with further details on the TAPW method, are provided in \cref{appendix:DFT_details}. By defining the normalized Hamiltonian as:
\eqa{
H_{\text{DFT}}(\bsl{k}) = (\text{S}^\mathrm{TAPW}(\bsl{k}))^{-1/2} \text{H}^\mathrm{TAPW}(\bsl{k}) (\text{S}^\mathrm{TAPW}(\bsl{k}))^{-1/2},
}
we can reduce the generalized eigenvalue problem to a standard eigenvalue problem.

\section{Constructing Continuum Model}
\label{sec:continuum_model}

\begin{figure*}[!t]
    \centering
    \includegraphics[width=\linewidth]{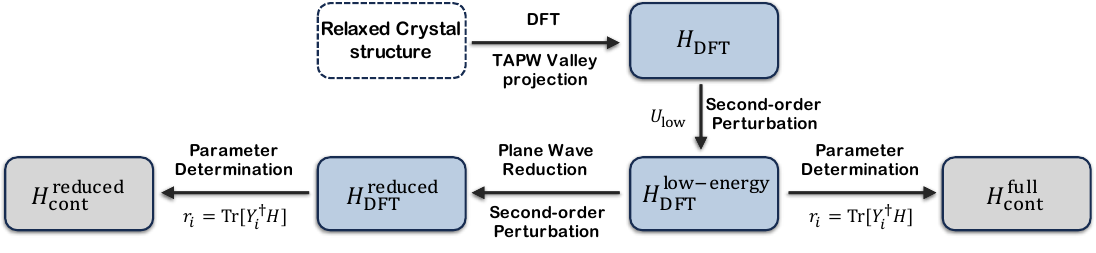}
    \caption{\textbf{Workflow overview}. A workflow to obtain the continuum model directly from DFT calculations without parameters fitting.}
    \label{fig:Methodology}
\end{figure*}

Based on DFT calculations, we propose a general method of constructing continuum model without parameter fitting for all twisted materials.
We first outline the workflow of the general method in \cref{fig:Methodology}.
After performing DFT calculations on the relaxed crystal structure and TAPW method, we get a $H_{\text{DFT}}$ that focus on the specific valley. Then we use second-order perturbation theory to obtain a low-energy DFT Hamiltonian $H_{\text{DFT}}^{\text{low-energy}}$ from the TAPW DFT Hamiltonian $H_{\text{DFT}}$.
Then, we use the orthogonalization and projection method to determine the parameter values of a full continuum model $H^{\text{full}}_{\text{cont}}$ from $H_{\text{DFT}}^{\text{low-energy}}$.
Given that fact that the full continuum model may require a large number of parameters, we use the second-order perturbation theory to integrate out the high-energy degrees of freedom, which can reduce the number of plane waves in $H_{\text{DFT}}^{\text{low-energy}}$ and give a reduced DFT Hamiltonian $H_{\text{DFT}}^{\text{reduced}}$.
(The procedure was also called Löwdin's partitioning method~\cite{lowdin_Note_1951}.)
We can again use the orthogonalization and projection method to determine the parameter values of a reduced continuum model $H_{\text{cont}}^{\text{reduced}}$ from $H_{\text{DFT}}^{\text{reduced}}$.
We provide more details for the genral methods in \cref{general method}).

In the rest of this section, we will illustrate the method by considering two specific cases, \textit{i.e.}, AA-stacked {\tmt} and {\tws}. 
In AA-stacked {\tmt} and {\tws}, the low-energy moir\'e bands originate from the highest valence bands near the $\K$ and $-\K$ valleys in each layer. %
Strong spin-orbit coupling constrains the spin-up states to the $\K$ valley and the spin-down states to the $-\K$ valley. %
This section focuses on the continuum model at the $\K$ valley, as the $-\K$ valley model can be derived through time-reversal symmetry~\cite{cao_valley-selective_2012,xiao_coupled_2012}.

\subsection{Low-energy DFT Hamiltonian}

We note that the DFT Hamiltonian around the $\K$ valley can be expressed as: 
\eql{
\label{main_eq:DFT_Hamiltonian}
H_{\text{DFT}} &  = \sum_{\bsl{k}}\sum_{l,l'}\sum_{\widetilde{\mathbf{G}}_l \widetilde{\mathbf{G}}_{l'}} \sum_{\alpha_1 \alpha_2}c^\dagger_{\bsl{k}+\widetilde{\mathbf{G}}_l,\alpha_1} c_{\bsl{k}+\widetilde{\mathbf{G}}_{l'},\alpha_2}  \left[ H_{\text{DFT}}(\bsl{k}) \right]_{l\widetilde{\mathbf{G}}_l\alpha_1,l' \widetilde{\mathbf{G}}_{l'} \alpha_2}\\
&  = \sum_{\bsl{k}}\sum_{\bsl{Q}_1 \bsl{Q}_2} \sum_{\alpha_1 \alpha_2}c^\dagger_{\bsl{k}+\mathbf{K}_{l_{\bsl{Q}_1}} - \bsl{Q}_1,\alpha_1} c_{\bsl{k}+\mathbf{K}_{l_{\bsl{Q}_2}'} - \bsl{Q}_2,\alpha_2}  \left[ H_{\text{DFT}}(\bsl{k}) \right]_{\bsl{Q}_1\alpha_1,\bsl{Q}_2 \alpha_2} \ ,
}
where $c^\dagger_{\bsl{k}+\widetilde{\mathbf{G}}_l,\alpha}$ denotes the DFT basis, $\widetilde{\mathbf{G}}_l$ is defined in \cref{eq:FT_DFT}, and $\mathbf{K}_l$ represents the monolayer K-point of the $l$-th layer.
$\bsl{Q}$ is a combination of the layer index and $\widetilde{\mathbf{G}}_l$, and $l_{\bsl{Q}}$ indicates the layer that $\bsl{Q}$ corresponds to.
Specifically, we have $\bsl{Q} = -\widetilde{\mathbf{G}}_{l_{\bsl{Q}}} + \mathbf{K}_{l_{\bsl{Q}}}$.
The DFT basis include $N_{\alpha}$ spin-$\uparrow$ orbitals per monolayer unit cell. %
For instance, in {\tmt} calculations, we use the PAO basis sets Mo-$s3p2d2$ and Te-$s3p2d2f1$, yielding a total of $N_{\alpha} = 71$ functions. %
The basis selection adheres to OpenMX guidelines, with alternatives detailed in \cref{appendix:DFT_details}. %

\begin{figure}[!t]
    \centering
    \includegraphics[width=\linewidth]{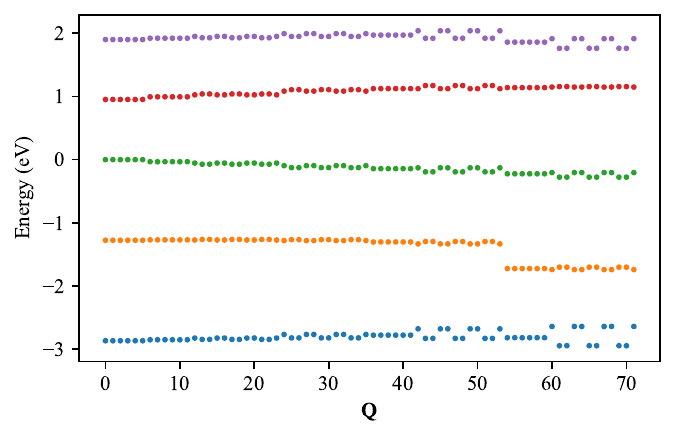}
    \caption{\textbf{Eigenvalues of each diagonal block of $H_{\text{DFT}}(\bsl{k_0})$ of 2.13$^\circ$ \tmt}. The horizontal axis represents the $\bsl{Q}$-label index, and the vertical axis shows the energy spectrum obtained by diagonalizing each $\bsl{Q}$-block ($D_{\bsl{Q}}(\bsl{k}_0)$) from $H_{\text{DFT}}(\bsl{k}_0)$. The green states near the Fermi level, which is set to zero, correspond to the energy states of interest. Here, $\bsl{k_0}$ is chosen at the $\Gamma_M$ point. The $\bsl{Q}$-label index corresponds to specific $\bsl{Q}$-points, which are detailed in \cref{fig:combined_figures} of Appendix. 
    }
    \label{fig:block eig}
\end{figure}

The DFT Hamiltonian in \cref{main_eq:DFT_Hamiltonian} contains a lot of high-energy degrees of freedom that are irrelevant to the low-energy physics of interest.
To simplify \cref{main_eq:DFT_Hamiltonian}, we recall that the low-energy physics happens among the low-energy states of monolayer Hamiltonians---we always choose the low-energy states of monolayer Hamiltonians as the basis of the continuum model.
In other words, the most dominant energy scale in the continuum model is the kinetic energy directly inherited from the monolayer Hamiltonians, which do not couple different $\bsl{Q}$'s for a fixed $\bsl{k}$.
Therefore, to isolate the low-energy states, we can simply diagonalize each $N_{\alpha} \times N_{\alpha}$ block of $H_{\text{DFT}}(\bsl{k})$ for a fixed $\bsl{Q}$, denoted as $D_{\bsl{Q}}(\bsl{k})$ with elements $\left[D_{\bsl{Q}}(\bsl{k}) \right]_{\alpha_1, \alpha_2} =  \left[ H_{\text{DFT}}(\bsl{k}) \right]_{\bsl{Q}\alpha_1, \bsl{Q} \alpha_2}$. 
This yields eigenvalues $\lambda_{n}^{\bsl{k}, \bsl{Q}}$ and eigenvectors $U_{n}^{\bsl{k}, \bsl{Q}}$. %
For each $\bsl{k}$ and $\bsl{Q}$, we identify a single eigenvalue close to the Fermi level and distinctly separated from other states. These states are illustrated with green dots in \cref{fig:block eig}, which shows the eigenvalues of each $\bsl{Q}$-block of $H_{\text{DFT}}(\Gamma_M)$. The corresponding eigenvector, $U_{\text{low}}^{\bsl{k}, \bsl{Q}}$, is then used to construct the low-energy basis:
\eqa{
\label{main_eq:continuum_basis_DFT}
\psi^\dagger_{\bsl{k},\bsl{Q}} = \sum_{\alpha} c^\dagger_{\bsl{k}+\K_{l} - \bsl{Q},\alpha} \left[ U_{\text{low}}^{\bsl{k},\bsl{Q}} \right]_{\alpha} \ . 
}
$U_{\text{low}}^{\bsl{k},\bsl{Q}}$ is dominated by the orbital $d_{xy}+\ii d_{x^2 + y^2}$, consistent with the orbital structure of the top valence band around $\K$ in monolayer MoTe$_2$ ~\cite{xiao_coupled_2012,wu_topological_2019,pan_band_2020,zhang_electronic_2021,devakul_magic_2021,wang_topological_2023,reddy_fractional_2023,dong_composite_2023,qiu_interaction-driven_2023,wang_topology_2023,reddy_toward_2023,wang_fractional_2024,yu_fractional_2024,xu_maximally_2024,abouelkomsan_band_2024,jia_moire_2024,zhang_polarization-driven_2024}. 
For convenience, we fix the $d_{xy}$ component of the eigenvector $U_{\text{low}}^{\bsl{k},\bsl{Q}}$ to be positive definite by fixing the random phase of $U_{\text{low}}^{\bsl{k},\bsl{Q}}$.
Consequently, $\psi^\dagger_{\bsl{k},\bsl{Q}}$ has exactly the same symmetry representations as the continuum basis used in the construction of the moir\'e continuum model~\cite{xiao_coupled_2012,wu_topological_2019,pan_band_2020,zhang_electronic_2021,devakul_magic_2021,wang_topological_2023,reddy_fractional_2023,dong_composite_2023,qiu_interaction-driven_2023,wang_topology_2023,reddy_toward_2023,wang_fractional_2024,yu_fractional_2024,xu_maximally_2024,abouelkomsan_band_2024,jia_moire_2024,zhang_polarization-driven_2024}.
Thus, $\psi^\dagger_{\bsl{k},\bsl{Q}}$ in \cref{main_eq:continuum_basis_DFT} serves as the basis for the continuum model employed in these studies \cite{xiao_coupled_2012,wu_topological_2019,pan_band_2020,zhang_electronic_2021,devakul_magic_2021,wang_topological_2023,reddy_fractional_2023,dong_composite_2023,qiu_interaction-driven_2023,wang_topology_2023,reddy_toward_2023,wang_fractional_2024,yu_fractional_2024,xu_maximally_2024,abouelkomsan_band_2024,jia_moire_2024,zhang_polarization-driven_2024}.
Using the continuum basis, we can further employ second-order perturbation to derive the low-energy Hamiltonian $h_{\text{DFT}}(\bsl{k})$, resulting in the following low-energy model in the continuum basis directly from DFT calculations:
\eq{
\label{main_eq:DFT_lowenergy}
H_{\text{DFT}}^{\text{low-energy}} = \sum_{\bsl{k}}\sum_{\bsl{Q}_1 \bsl{Q}_2} \psi^\dagger_{\bsl{k} , \bsl{Q}_1} \left[ h_{\text{DFT}}(\bsl{k}) \right]_{\bsl{Q}_1,\bsl{Q}_2 }\psi_{\bsl{k} , \bsl{Q}_2} \ .
}
(See \cref{general method} for the perturbation details.)
As illustrated in \cref{fig:model_contruction}(c), $H_{\text{DFT}}^{\text{low-energy}}$ reproduces the DFT band structure with very high accuracy, offering a precise framework for further many-body calculations if paired with a normal-ordered interaction term to avoid double counting of the interactions already present in the DFT. %
However, there is a limitation in directly using $h_{\text{DFT}}(\bsl{k})$---the momentum mesh we use for the many-body calculation must be the same as or a subset of the DFT momentum mesh for $h_{\text{DFT}}(\bsl{k})$.

\begin{figure*}[!t]
    \centering
    \includegraphics[width=\linewidth]{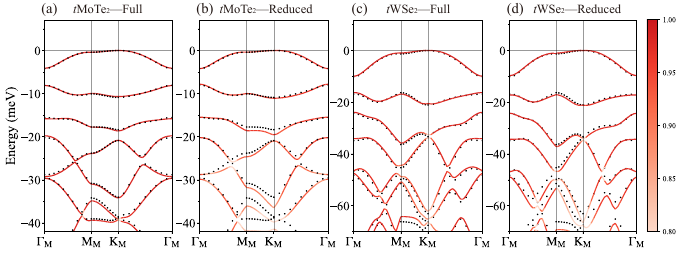}
    \caption{\textbf{Continuum Models for 2.13$^\circ$ {\tmt} and {\tws}.} 
    (a) Full continuum model of \tmt. 
    (b) Reduced continuum model of \tmt. 
    (c) Full continuum model of \tws. 
    (d) Reduced continuum model of \tws. 
    Black lines represent the K valley bands calculated using the TAPW method, while red lines depict bands computed by the continuum model. The gradient from dark to light red in the red lines illustrates the overlap probability between the continuum model wavefunctions and the TAPW wavefunctions for each corresponding band. The full lattice includes harmonics up to the 8th order, corresponding to the $\widetilde{\bsl{G}}$ vectors used in the TAPW method, whereas the reduced lattice includes harmonics up to the 3rd order.}
    \label{fig:2.13_tmt_tws_mode}
\end{figure*}

To address this issue, we derive a continuum model, which enables calculations on arbitrary meshes. %
We note that the real-space form of the $\K$-valley continuum Hamiltonian, $H_{cont,\K}$, in general reads
\eqa{
\label{main_eq:general_form_continuum_model}
H_{cont,\K} & = \sum_{M_x, M_y \in \dsN} \sum_{l,l'}\int d^2 r \left(\ii^{M_x+M_y} \partial_x^{M_x} \partial_y^{M_y} \psi^\dagger_{\bsl{r},\K_{l}} \right) \\
& t^{M_x M_y}_{ll'}(\bsl{r})\psi_{\bsl{r},\K_{l'}} \ ,
}
where $\dsN$ is the set of non-negative integers. 
In \cref{main_eq:general_form_continuum_model}, derivatives on $t$ terms are not explicitly required since differentiating a potential still yields a potential. Derivatives on $\psi_{\bsl{r},\K_l}$ terms are not explicitly included since they relate to derivatives on $\psi^\dagger_{\bsl{r},\K_l}$ and $t$ terms by integration by parts.

Under the continuum approximation, we have $H_{\text{DFT}}^{\text{low-energy}}=H_{cont,\K}$, and thus $h_{\text{DFT}}(\bsl{k})$ can be approximated by Fourier transformation of \cref{main_eq:general_form_continuum_model}, expressed as:
\eqa{
\label{main_eq:general_form_continuum_model_momentum_space}
& \left[h_{\text{DFT}}(\bsl{k})\right]_{\bsl{Q}\bsl{Q}'}  =  \sum_{M_x, M_y \in \dsN}  \sum_{l,l'}\sum_{\bsl{G}_M} r^{M_x M_y}_{ll' , \bsl{G}_M }  \left[ Y^{M_x M_y}_{ll' , \bsl{p} }(\bsl{k}) \right]_{\bsl{Q}\bsl{Q}'}\ ,
}
where $\bsl{p} = \K_l -\K_{l'}  + \bsl{G}_M$ for simplicity, $\bsl{G}_M$ is the Moir\'e reciprocal lattice vector, and
\eq{
\label{eq:Y_terms}
 \left[Y^{M_x M_y}_{ll' , \bsl{p} }(\bsl{k})\right]_{\bsl{Q}\bsl{Q}'}  =  \delta_{l_{\bsl{Q}},l} \delta_{l_{\bsl{Q}'}, l' }   (\bsl{k}-\bsl{Q})_{x}^{M_x}(\bsl{k}-\bsl{Q})_{y}^{M_y}  \delta_{ \bsl{Q} , \bsl{Q}' +   \bsl{p}  }\ .
}
Here $|\bsl{p}|$ determines the harmonics of the moir\'e potential, since $\bsl{p}$ is the momentum difference between the creation and annihilation operators.
Specifically, when $l = l'$ and $\bsl{G}_M = 0$, $\bsl{p}=0$ and the term represents the kinetic energy within each individual layer; when $l = l'$ but $\bsl{G}_M \neq 0$, $\bsl{p}=\bsl{G}_M$ and the term describes intralayer moir\'e potential; finally, when $l \neq l'$, $\bsl{p}$ is again nonzero and the term describes interlayer moir\'e coupling. (See details in \cref{general method}.) In \cref{subsec:para_relation}, we provide a relation between $r^{M_x M_y}_{ll' , \bsl{G}_M }$ and the commonly used notations in the literature.

\subsection{Determining Parameter Values}

The goal is to determine the values of $r^{M_x M_y}_{ll' , \bsl{G}_M }$. 
If the $\bsl{Q}$ lattice were infinite, we could determine the values of all $r^{M_x M_y}_{ll' , \bsl{G}_M }$ at a fixed $\bsl{k}=\bsl{k}_0$: we first apply the Gram-Schmidt process to orthogonalize the $Y$ terms in \cref{eq:Y_terms}, and then project $h_{\text{DFT}}$ onto the orthogonalized $Y$ terms by taking the trace between them to determine $r^{M_x M_y}_{ll' , \bsl{G}_M }$, as discussed in \cref{subsec:ortho_proj}. 
Although an infinite $\bsl{Q}$ lattice is not feasible in practice, we adapt this method to finite sets of $(M_x, M_y, l, l', \bsl{G}_M)$ terms, providing the corresponding $r^{M_x M_y}_{ll', \bsl{G}_M}$ values. 
Should a single $\bsl{k}$ point prove insufficient, we can include multiple $\bsl{k}$ points (in a direct sum manner) to accurately determine the parameter values (see \cref{general method} for details). %
Notably, this procedure requires no parameter fitting, as $h_{\text{DFT}}(\bsl{k})$ is derived directly from DFT calculations. 
This yields a continuum model with the same $\bsl{Q}$ lattice as the DFT Hamiltonian, termed the ``full model''. 

\begin{figure*}[!t]
    \centering
    \includegraphics[width=\linewidth]{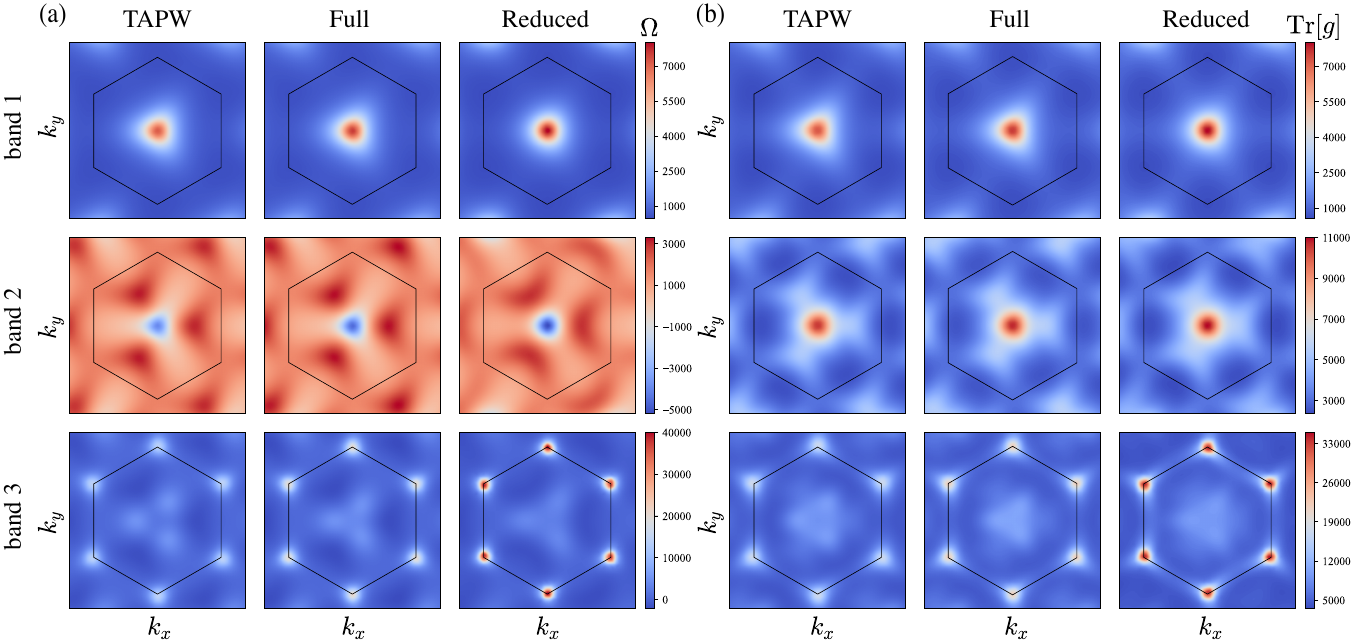}
    \caption{\textbf{Distribution of Berry curvature and quantum metric of 2.13$^{\circ}$ {\tmt}.} 
    (a):Berry curvature. 
    (b):Trace of Quantum metric. 
    Each row corresponds to different energy bands (band 1, band 2, and band 3), and each column compares different methods: TAPW, full model, and reduced model. The hexagonal boundary in each subplot represents the first moir\'e Brillouin zone.}
    \label{fig:QG_BC}
\end{figure*}

To demonstrate our method, we present the full continuum models for $2.13^{\circ}$ {\tmt} and {\tws} in \cref{fig:2.13_tmt_tws_mode}(a,c), where the results for {\tmt} and {\tws} of twisted angles from 2.45$^\circ$ to 3.89$^\circ$ are displayed in \cref{fig:model_mote2_52orb,fig:model_mote2_71orb,fig:model_wse2_47orb}. 
As shown in \cref{fig:2.13_tmt_tws_mode}(a,c), the low-energy bands and their eigen-wavefunctions calculated from the full continuum models exhibit excellent agreement with the DFT results. 
Specifically, the maximum energy difference between the full continuum model and DFT bands is less than 0.5 meV and 1.3 meV for {\tmt} and {\tws}, respectively; the minimum wavefunction overlap probability (excluding the band-touching regions) exceeds 97\% for the top four bands for both {\tmt} and {\tws}. 
The wavefunction match is also evident in the Berry curvature distributions shown in \cref{fig:QG_BC}. 
Interestingly, the top three bands of 2.13$^\circ$ {\tmt} all have Chern number 1, with integrated trace of quantum metric of 1.13, 3.25, and 5.27 (from energetically higher to lower bands) calculated directly from the DFT wave-functions. 
(See the exression of quantum metric in \cref{method:QM_BC}.)
The full continuum model yields identical Chern numbers for the top three bands and similar values for the integrated trace of quantum metric---1.12, 3.20, and 5.25 (from energetically higher to lower bands). 
This implies that the top three bands of 2.13$^\circ$ {\tmt} resemble the zeroth, first, and second Landau levels, although the distribution of Berry curvature and quantum metric is somewhat non-uniform, as shown in \cref{fig:QG_BC}. 
Similar statements were also made in previous studies~\cite{wang_Higher_2024,xu_Multiple_2024,reddy_NonAbelian_2024,ahn_NonAbelian_2024}.

In \cref{table:mote2_71orb_full_diag,table:mote2_71orb_full_intra,table:mote2_71orb_full_inter,table:wse2_47orb_full_diag,table:wse2_47orb_full_intra,table:wse2_47orb_full_inter}, we list the parameter values for the full model with twist angle ranging from 2.13$^\circ$ to $3.89^\circ$.
As we can see, the full model has a large number of parameters.
For $2.13^{\circ}$ {\tmt}, the model consists of 6 diagonal terms $m$, 72 interlayer coupling terms $V$, and 70 intralayer coupling terms $w$, resulting in 148 parameters. 
Similarly, for $2.13^{\circ}$ {\tws}, the model includes 9 diagonal terms $m$, 45 interlayer coupling terms $V$, and 33 intralayer coupling terms $w$, leading to a total of 87 parameters. 
The large number of terms pose no difficulty in numerical calculations, where the form factors of interactions are always computed numerically, but can complicate possible analytical studies. 

\subsection{Plane Wave Reduction}

To support analytical investigations, we derive a ``reduced model''. 
We will first impose a cutoff on the $\bsl{Q}$ lattice, and treat all $\psi^\dagger_{\bsl{k} , \bsl{Q}}$ with $|\bsl{Q}|$ larger than the cutoff as high-energy degrees of freedom.
Then, we then integrate out the high-energy degrees of freedom in the DFT Hamtilonian $h_{\text{DFT}}(\bsl{k})$, and obtain an effective DFT Hamiltonian $h_{\text{DFT}}^{\text{reduced}}(\bsl{k})$.
Finally, we can construct a reduced model from $h_{\text{DFT}}^{\text{reduced}}(\bsl{k})$ in the same way as constructing the full model from $h_{\text{DFT}}(\bsl{k})$. 
The reduced model includes only a subset of $\bsl{Q}$ points, significant lowering the number of parameters: for instance, the reduced model contains 23 parameters for 2.13$^\circ$ {\tmt} and 15 parameters for 2.13$^\circ$ {\tws}. 
Nevertheless, the reduced model can still well capture the essential low-energy part, as shown in \cref{fig:2.13_tmt_tws_mode}(b,d): the energy deviations under 1.2 meV and 1.8 meV for the top
four bands for both {\tmt} and {\tws}, respectively, and minimum overlap probability for the reduced model remains above 95\%.

The parameter values at different angles for the full and reduced models immediately shows a trend: we typically need more harmonics (\ie, a larger number of different values of $|\bsl{p}|$ in \cref{eq:Y_terms}) in the moir\'e potential to capture the DFT results, as the angle decreases.
Specifically, we find that for {\tmt}, the number of intralayer  (interlayer) harmonics increases from 4 (4) for $3.89^\circ$ to 5 (6) for 2.13$^\circ$ for the full model, and increases from 1 (2) for $3.89^\circ$ to 5 (4) for 2.13$^\circ$ for the reduced model.
The trend is similar for {\tws}: the number of intralayer (interlayer) Haromics increases from 2 (2) for $3.89^\circ$ to 3 (3) for 2.13$^\circ$ for the full model, though the number of harmonics does not change for the reduced model.
From the change of the number of harmonics (especially the unchanged number of harmonics of the reduced model of {\tws}), we can see a smaller angle dependence of {\tws} than {\tmt} in the same range of angles, of which the reason is left for future study.

\section{Conclusion}

In this study, we present a universal framework for constructing continuum models of moir\'e materials without the necessity of parameter fitting. 
Our method accurately captures the essential electronic properties of moir\'e systems, including a faithful reproduction of the wavefunction. 
Although we only use the method on {\tmt} and {\tws} in the current work, it is extendable to other moir\'e systems, such as multilayer graphene-hBN super lattices, which have been extensively studied both experimentally~\cite{lu_Fractional_2024,xie_Even_2024,park_Ferromagnetism_2024,choi_Electric_2024,lu_Extended_2024,zhou_Layerdependent_2023} and theoretically~\cite{park_Topological_2023,herzog-arbeitman_Moire_2024,kwan_Moire_2023,yu_Moire_2024,guo_Fractional_2024,zhou_Fractional_2024,dong_Anomalous_2024,soejima_Anomalous_2024,huang_Impurityinduced_2024,tan_Wavefunction_2024,dong_Theory_2024,huang_Selfconsistent_2024,dassarma_Thermal_2024,xie_Integer_2024,dong_Stability_2024,kudo_Quantum_2024,zhou_New_2024}.

\section{Method}

\subsection{Parameter Relations}\label{subsec:para_relation}

The parameters of the continuum Hamiltonian of {\tmt} and {\tws} in the common notation used in the literature can always be expressed in terms of $r^{M_x M_y}_{ll' , \bsl{G}_M }$ in \cref{main_eq:general_form_continuum_model_momentum_space}. For example, the second Harmonic Hamiltonian in \refcite{jia_moire_2024} of AA-stacked {\tmt} in {\K} valley reads
\eqa{
\label{eq:Kvalley_model_AA}
& h_{\K,\bsl{Q}\bsl{Q}'}^{AA}(\bsl{k})  =  \delta_{\bsl{Q}\bsl{Q}'} \left( \frac{- \hbar^2 \left( \bsl{k} -\bsl{Q} \right)^2}{ 2 m^*}  \right) \\
& \quad + V \sum_{i=1}^3 \left[  e^{- \ii \psi(-)^{\bsl{Q}}} \delta_{\bsl{Q}+\bsl{g}_i,\bsl{Q}'} +  e^{ \ii \psi(-)^{\bsl{Q}}} \delta_{\bsl{Q}-\bsl{g}_i,\bsl{Q}'}\right]  \\
& \quad + V_2 \sum_{i=1}^3 \left[   \delta_{\bsl{Q}+\bsl{g}_{2i},\bsl{Q}'} +  \delta_{\bsl{Q},\bsl{Q}'+\bsl{g}_{2i}}\right]  \\
& \quad + w \sum_{i=1}^{3} \left[  \delta_{\bsl{Q}+\bsl{q}_i,\bsl{Q}'} + \delta_{\bsl{Q},\bsl{Q}'+\bsl{q}_i} \right] \\
& \quad + w_2 \sum_{i=1}^{3} \left[  \delta_{\bsl{Q}+\bsl{q}_{2i},\bsl{Q}'} + \delta_{\bsl{Q},\bsl{Q}'+\bsl{q}_{2i}} \right] \ .
}
In the notation of \cref{main_eq:general_form_continuum_model}, \cref{eq:Kvalley_model_AA} just corresponds to $m^{11}=m^{*}$, $V_{l_{\bsl{Q}},\mathbf{b}_{M1}}^{00}=Ve^{-\ii \psi(-)^{\bsl{Q}}}$, $V_{l,\mathbf{b}_{M1}+\mathbf{b}_{M2}}^{00}=V_2$, $w^{00}_{ll'}=w$, $w_{ll',\mathbf{b}_{M1}}^{00}=w_2$, with all other parameters set to zero. Here  $r^{M_x M_y}_{ll' , \bsl{G}_M }|_{l=l',\bsl{G}_M=0}=\hbar^2/(2m^{M_xM_y})$, $r^{M_x M_y}_{ll' , \bsl{G}_M }|_{l=l',\bsl{G}_M\neq0}=V_{l,\bsl{G}_M}^{M_xM_y}$, $r^{M_x M_y}_{ll' , \bsl{G}_M }|_{l\neq l'}=w_{ll'}^{M_xM_y}$.

\subsection{Orthonormalization and Projection}\label{subsec:ortho_proj}

In this part, we outline the orthonormalization and projection process to determine the parameters $r^{M_x M_y}_{ll', \bsl{G}_M }$ in \cref{main_eq:general_form_continuum_model_momentum_space}.
We will focus on the case where the $\bsl{Q}$ lattice is infinite, since the precedure fore the finite $\bsl{Q}$ lattice is very similar, as discussed in \cref{subsec:cont_construct}.
We start by noting that the terms $Y^{M_x M_y}_{ll', \bsl{p}}(\bsl{k})$ in \cref{eq:Y_terms} are linearly independent at a fixed $\bsl{k}$, \ie, $\sum r^{M_x M_y}_{ll' , \bsl{G}_M }Y^{M_x M_y}_{ll', \bsl{p}}(\bsl{k})=0$ can only be satisfied when $r^{M_x M_y}_{ll' , \bsl{G}_M }=0$ for all $(M_x,M_y,l,l',\bsl{G}_M)$, where $\bsl{p} = \K_l -\K_{l'}  + \bsl{G}_M$ and the sum is over all values of $(M_x,M_y,l,l',\bsl{G}_M)$.
(See proof in \cref{subsec:cont_construct}.)
It measn that we could determine the values of $r^{M_x M_y}_{ll' , \bsl{G}_M }$ at a single fixed $\bsl{k}=\bsl{k}_0$ for all $M_x,M_y\in\dsN$, all moir\'e reciprocal lattice vectors $\bsl{G}_M$ and all $l,l'=b,t$.
Specifically, we apply the Gram-Schmidt process to orthonormalize the $Y^{M_x M_y}_{ll' , \bsl{p} }(\bsl{k}_0)$ terms, \ie, finding new $\widetilde{Y}^{M_x M_y}_{ll' ,  \bsl{p} }(\bsl{k}_0)$ terms that are related to $Y^{M_x M_y}_{ll', \bsl{p} }(\bsl{k}_0)$ by
\eqa{
\label{main_eq:othogonalization}
 \widetilde{Y}^{M_x M_y}_{ll' ,  \bsl{p} }(\bsl{k}_0) = \sum_{\widetilde{M}_x \widetilde{M}_y} Y^{\widetilde{M}_x \widetilde{M}_y}_{ll' ,  \bsl{p} }(\bsl{k}_0) z_{\widetilde{M}_x \widetilde{M}_y, M_x M_y}^{ll' ,  \bsl{p} }
}
such that
\eqa{
 \Tr\left\{ \left[\widetilde{Y}^{M_x M_y}_{ll' ,  \bsl{p} }(\bsl{k}_0)\right]^\dagger \widetilde{Y}^{M_x' M_y'}_{ll' ,  \bsl{p} }(\bsl{k}_0) \right\} = \delta_{M_x , M_x'}\delta_{M_y , M_y'} \ .
}
In this process, we do not need to liearly combine different values of $(l,l',\bsl{G}_M)$, since the terms $Y^{M_x M_y}_{ll', \bsl{p} }(\bsl{k})$ (with $\bsl{p} = \K_l -\K_{l'}  + \bsl{G}_M$) are naturally orthogonal for different values of $(l,l',\bsl{G}_M)$, as proved in \cref{subsec:cont_construct}.
Then, we can obtain
\eq{
\label{main_eq:coefficient_from_DFT}
\widetilde{r}^{M_x M_y}_{ll', \bsl{G}_M } = \Tr\left\{ \left[\widetilde{Y}^{M_x M_y}_{ll' ,  \bsl{p} }(\bsl{k}_0)\right]^\dagger  h_{\text{DFT}}(\bsl{k}_0)\right\} \ ,
}
resulting in 
\eq{\label{main_eq:coefficient_from_transformation}
r^{M_x M_y}_{ll', \bsl{G}_M } = \sum_{\widetilde{M}_x \widetilde{M}_y}  z_{M_x M_y, \widetilde{M}_x \widetilde{M}_y}^{ll' ,  \bsl{p} }  \widetilde{r}^{\widetilde{M}_x \widetilde{M}_y}_{ll',\bsl{G}_M  } \ ,
}
where $r^{M_x M_y}_{ll', \bsl{G}_M }$ represent the original parameters in \cref{main_eq:general_form_continuum_model_momentum_space}, and $\widetilde{r}^{M_x M_y}_{ll', \bsl{G}_M }$ is defined in \cref{main_eq:coefficient_from_DFT}.

\subsection{Quantum Metric and Berry Curvature}\label{method:QM_BC}

To characterize the quantum geometric properties of the electronic bands in our continuum models, we calculate both the Berry curvature and the quantum metric. 

The Berry curvature $ \Omega_n(\bsl{k}) $ for the $ n $-th band is expressed in terms of the projection operator $ P_n(\bsl{k}) = |u_n(\bsl{k})\rangle \langle u_n(\bsl{k})| $ as follows:
\eqa{
    \Omega_n(\bsl{k}) = \ii \, \text{Tr} \left[ P_n(\bsl{k}) \partial_{k_x} P_n(\mathbf{k}) \partial_{k_y} P_n(\bsl{k}) \right] - (x \leftrightarrow y),
    \label{eq:Berry_curvature_projection}
}
where $|u_n(\bsl{k})\rangle$ is the periodic part of the Bloch state of the $n$th band.

Similarly, the quantum metric tensor $ g_{n,ij}(\bsl{k}) $ is defined as:
\eqa{
    g_{n,ij}(\bsl{k}) = \frac{1}{2} \, \text{Tr} \left[ \partial_{k_i} P_n(\bsl{k}) \partial_{k_j} P_n(\bsl{k}) \right],
    \label{eq:FS_metric}
}
where $ i, j = x, y $. We show the distribution of $\Omega_n(\bsl{k})$ and trace of quantum metric $\text{Tr}[g_n(\bsl{k})]$ in \cref{fig:QG_BC}, and we are also interested in the integration of $\text{Tr}[g_n(\bsl{k})]$ in the first moir\'e Brillouin zone:
\eqa{
\chi = \frac{1}{2\pi} \int_{\text{BZ}} \text{d}\bsl{k} \, \text{Tr}[g_n(\bsl{k})]\ .
}

\section{Acknowledgments}
The authors thank Di Xiao and Oskar Vafek for helpful discussions.
This work was supported by the Ministry of Science and Technology of China (Grant No. 2023YFA1607400, 2022YFA1403800), the National Natural Science Foundation of China (Grant No.12274436) and the Science Center of the National Natural Science Foundation of China (Grant No. 12188101) and . H.W. acknowledge support from the Informatization Plan of the Chinese Academy of Sciences (CASWX2021SF-0102). 
B. A. B.'s work was primarily supported by the DOE Grant No. DE-SC0016239 and the Simons Investigator Grant No. 404513. N.R. also acknowledges support from the QuantERA II Programme that has received funding from the European Union’s Horizon 2020 research and innovation programme under Grant Agreement No 101017733 and from the European Research Council (ERC) under the European Union's Horizon 2020 Research and Innovation Programme (Grant Agreement No. 101020833).  The Flatiron Institute is a division of the Simons Foundation.
X. Dai is supported by a fellowship award and a CRF award from the Research Grants Council of the Hong Kong Special Administrative Region, China (Project No. SRFS2324-6S01 and No. C7037-22GF).
J. Y.'s work at Princeton University is supported by the Gordon and Betty Moore Foundation through Grant No. GBMF8685 towards the Princeton theory program.
J. Y.'s work at University of Florida is supported by startup funds at University of Florida.

\clearpage
\bibliography{main}

\newpage
\clearpage

\onecolumngrid

\tableofcontents

\appendix

\section{Crystal Structure Relaxation and DFT Calculation}
\label{appendix:DFT_details}

Accurate determination of electronic properties in moir\'e systems requires precise relaxation of the crystal structure followed by detailed electronic structure calculations. In this section, we outline our methodology for structural relaxation using Machine Learning Force Fields (MLFF) and subsequent Density Functional Theory (DFT) calculations to obtain reliable electronic band structures for {\tmt} and {\tws}.

\subsection{Crystal Structure Relaxation}

To accurately predict the band structures of twisted materials, it is essential to relax the crystal structure with high precision. 
Given the substantial number of atoms in the moir\'e supercell (e.g., 4326 atoms for a 2.13$\degree$ {\tmt}), traditional relaxation methods encounter considerable convergence issues. 
To overcome this, we adopted MLFF approach. 

The procedure is the following. 
We constructed small supercells of untwisted AA bilayer MoTe$_2$, each with different in-plane shifts between the top and bottom layers. 
Then, we ran \emph{ab initio} Molecular Dynamics (MD) simulations on these structures with VASP, collecting data to construct a robust training set.
(Specifically, the \emph{ab initio} MD simulations were done through the \emph{ab initio} MD steps in VASP MLFF module since the module is efficient to generate data~\cite{jia_moire_2024}.
Those data were \emph{not} generated using MLFF, and we will not use the MLFF in that module, since it is not precise.)
We built the test data set using the data in pure DFT relaxation of large-angle twisted structures to prevent over-fitting on untwisted structures.

Then, we trained an accurate MLFF using the Allegro package on the training set and chose the optimal model on test set. We use the Allegro package, because it is an accurate MLFF based on E(3) equivariant neural network that respects rotation, reflection, and translation symmetries in 3D space. The optical model is the MLFF that we want.

The MLFF was then used to relax {\tmt} and {\tws} structures at various angles within the Atomic Simulation Environment.
This strategy allowed us to relax even the largest moir\'e structures efficiently. 
For {\tmt} with a twist angle of 2.13$\degree$, we conducted additional DFT relaxation steps using MLFF-relaxed structures as initial inputs. 

\subsection{DFT Calculation}

To accurately determine the electronic properties of {\tmt} and {\tws}, we performed Density Functional Theory (DFT) calculations within the OpenMX software package, which employs norm-conserving pseudopotentials and pseudo-atomic localized basis functions. 
Same calculations can be done using other software such as SIESTA and ABACUS, which are based on numerical atomic orbitals.. The method here is of course not limited to {\tmt} and {\tws}. 

To ensure computational efficiency and accuracy, the convergence of the charge density in our DFT calcualtions only involves the $\Gamma$ point with a stringent convergence criterion of $10^{-7}$ Hartree in the Kohn-Sham energy.
For systems with SOC, we utilized a two-step method ~\cite{zhang_Initio_2022} available in OpenMX to efficiently account for SOC effects without compromising accuracy. 
Initially, collinear DFT calculations were conducted to obtain a converged charge density, which was then utilized as an initial guess for subsequent non-collinear DFT calculations.
This approach typically achieved convergence within a single step.
This method reduces computational cost by treating SOC perturbations in the band energy, ensuring both accuracy and efficiency.

The basis functions employed for these DFT calculations were specified as Mo7.0-$s3p2d2$, Te7.0-$s3p2d2f1$, W7.0-$s3p2d2$, and Se7.0-$s3p2d1$. 
The "7.0" indicates the cutoff radius of the pseudo-atomic orbital (PAO) functions in Bohr units. 
The notation $s3p2d2$ for Mo indicates 3 $s$-orbitals, 2 sets of $p$-orbitals, and 2 sets of $d$-orbitals, yielding a total of 19 PAO basis functions for each Mo atom $(3 + 2 \times 3 + 2 \times 5)$. 
Similarly, Te, W, and Se utilized 26, 19, and 16 basis functions, respectively.
For {\tmt}, we selected the standard basis recommended by the OpenMX manual, but the parameters for the quick-basis DFT-based continuum model are also provided in  \cref{construct tmt}.
While for {\tws}, we have so far used the quick basis selection.

\begin{figure}[!hbtp]
    \centering
    \includegraphics[width=\linewidth]{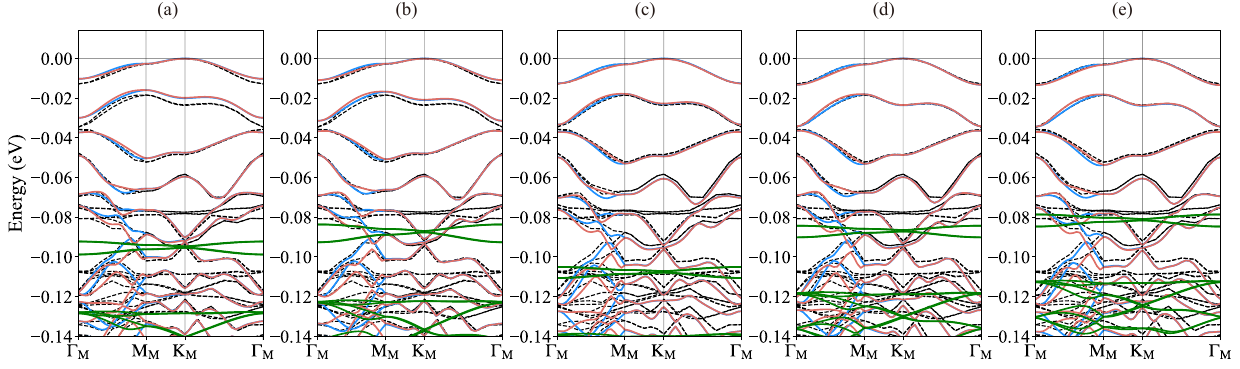}
    \caption{\textbf{Band structures of 3.89$^\circ$ {\tmt} with different PAO basis.} Black dotted lines represent the bands calculated using VASP, while blue, red, and green lines represent the $\mathrm{K}$-valley, $\mathrm{K^{\prime}}$-valley, and $\Gamma$-valley bands calculated using OpenMX, respectively. 
    (a): $\text{Mo}7.0$-$s3p2d1$ and $\text{Te}7.0$-$s3p2d2$ (quick basis). 
    (b): $\text{Mo}7.0$-$s3p3d2$ and $\text{Te}7.0$-$s3p3d2$. 
    (c): $\text{Mo}7.0$-$s3p2d2$ and $\text{Te}7.0$-$s3p2d2f1$ (standard basis). 
    (d): $\text{Mo}7.0$-$s3p2d2f1$ and $\text{Te}7.0$-$s3p3d2f1$. 
    (e): $\text{Mo}7.0$-$s3p3d2f1$ and $\text{Te}7.0$-$s3p3d2f1$ (precise basis). 
    As we are primarily interested in the $\mathrm{K}$-valley bands in the low-energy region and considering the computational resources, we ultimately selected the basis shown in subfigure (c) for {\tmt}.}
    \label{fig:benchmark of orb}
\end{figure}

To evaluate the suitability of different basis configurations, we benchmarked the electronic band structures of 3.89$^\circ$ {\tmt} as shown in Fig.~\ref{fig:benchmark of orb}.
Here, we compare the results from OpenMX using various PAO configurations against VASP-calculated bands. Notably, the PAO basis shown in subfigure (c) was selected for {\tmt}, as it provided an optimal balance between computational cost and accuracy in describing the low-energy $\mathrm{K}$-valley bands.

\section{TAPW with Non-orthogonal Basis}\label{TAPW method}

In this section, we will discuss the construction of a Truncated Atomic Plane Wave (TAPW) basis Hamiltonian derived from the DFT-calculated full Hamiltonian, which serves as the foundation for our subsequent continuum model. 
This TAPW approach retains the high precision of the full Hamiltonian in capturing low-energy physics while achieving substantial computational savings.

\subsection{Construction of the TAPW Basis}
\begin{figure}[!t]
    \centering
    \includegraphics[width=\columnwidth]{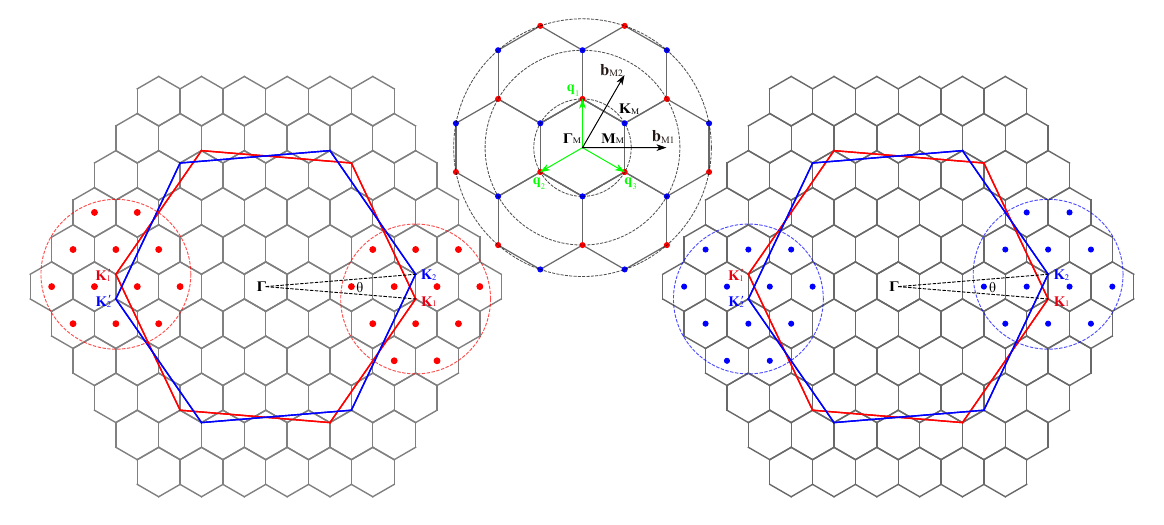}
    \caption{
        \textbf{Brillouin zone and moir\'e reciprocal lattice for 9.43$\degree$ {\tmt}}. The red and blue hexagons correspond to the monolayer Brillouin zones (BZ) of the bottom ($l=1$) and top ($l=2$) layers, respectively, each rotated by $\theta/2$ relative to the non-twisted BZ. $\mathbf{b}_{M1}$ and $\mathbf{b}_{M2}$ represent the moir\'e reciprocal lattice vectors. The $\mathbf{q}$ vectors are defined as $\mathbf{q}_1 = \mathbf{K}_2 - \mathbf{K}_1$, $\mathbf{q}_2 = C_{3z} \mathbf{q}_1$, and $\mathbf{q}_3 = C_{3z} \mathbf{q}_2$, here $\mathbf{K}_{l}$ means $\mathbf{K}$ point in the $l$-th layer's monolayer BZ. In the left panel, the red dots denote the $\widetilde{\bsl{G}}_l$ vectors chosen in the bottom layer, while the blue dots in the right panel denote those in the top layer, with their origins at the $\mathbf{\Gamma}$ point of the monolayer BZ. The middle panel shows the $\bsl{Q}$ vectors we used in the continuum model, where the red dots correspond to $\bsl{Q}$ vectors in the top layer, and the blue dots correspond to those in the bottom layer. Each layer's $\bsl{Q}_l$ vectors are defined relative to the corresponding $\mathbf{K}_l$ points, and satisfy the relation $\widetilde{\bsl{G}}_l + \bsl{Q}_l = \mathbf{K}_l$. The dashed circles centered at each layer's $\mathbf{K}$ point indicates the range from which we select the $\widetilde{\bsl{G}}$ vectors used to project the original full Hamiltonian. 
    }
    \label{fig:Gn}
\end{figure}

The low-energy physics of moir\'e materials is primarily dominated by states near specific valleys, particularly for small twist angles.  
To focus on these regions, we define the non-orthogonal truncated atomic plane wave basis centered around a specific valley for the $l$-th layer as follows:
\eqa{
|\psi_{l,\widetilde{\bsl{G}}_l,\alpha}(\bsl{k})\rangle=\frac{1}{\sqrt{N_\text{m}N_\text{a}} }\sum_{\text{I},i}\mathrm{e}^{\mathrm{i}(\bsl{k}+\widetilde{\bsl{G}}_l)(\mathbf{R}_{\text{I}}+\tau_{li{\alpha}})}\left|\phi_{\alpha}(\mathbf{R}_{\text{I}}+\tau_{li{\alpha}})\right\rangle \ ,
\label{tapw basis}
}
where $\widetilde{\bsl{G}}_{l}$ is a moir\'e reciprocal lattice vector in the $l$-th layer in the DFT convention which is chosen around the given valley (e.g. see \cref{fig:Gn} for K valley), $\bsl{k} + \widetilde{\bsl{G}}_l$ is the monolayer momentum, and $\bsl{k}$ denotes the momentum in the first moir\'e Brillouin zone. 
$N_{\text{m}}$ and $N_{\text{a}}$ denote the numbers of moir\'e unit cells and the number of non-twisted unit cells in one layer, respectively. 
$\mathbf{R}_{\text{I}}$ is the lattice vector of the moire unit cell, $\tau_{li\alpha}$ is the displacement of $\alpha$-th orbital in the $i$-th unit cell of $l$-th layer. 
$\alpha\equiv(aplm)$ represents an atom number index $a$, an organized orbital index comprising a multiplicity index $p$, an angular momentum quantum number $l$, and a magnetic quantum number $m$. 

For K valley in {\tmt} and {\tws}, the $\widetilde{\bsl{G}}_l$ vectors for each layer are selected to exhibit $C_{3z}$ symmetry with respect to the high-symmetry $\K_l$ point of the corresponding monolayer  (see Fig.~\ref{fig:Gn}) and are truncated based on the twist angle, with smaller angles requiring higher harmonics for accurate projection. 
We would like to emphasize that we cannot include many $\widetilde{\bsl{G}}_l$ indefinitely, since at some point they will connect the two valleys.  We want to avoid the unnecessary double counting from two valleys. 
This transforms the atom basis into an atomic plane wave basis and enables an efficient representation of the low-energy physics in moir\'e superlattices. 
So the Kohn-Sham Bloch state $\left|\Psi_{\mu}(\bsl{k}) \right\rangle$, which is the eigenstate of the Kohn-Sham Hamiltonian, can be written as linear combination of the atomic plane wave basis:
\eqa{
|\Psi_{\mu}(\bsl{k}) \rangle = \sum_{l\widetilde{\bsl{G}}_l\alpha}c_{l\widetilde{\bsl{G}}_l\alpha}^{\mu}(\bsl{k})|\psi_{l,\widetilde{\bsl{G}}_l\alpha}(\bsl{k})\rangle \ ,
}
where $c_{l\widetilde{\bsl{G}}_l\alpha}^{\mu}(\bsl{k})$ is an expansion coefficient. 
For the simplicity of the notation, we will use $n$ to label $\widetilde{\bsl{G}}_l$, \ie, $\widetilde{\bsl{G}}_{ln}$ is the value of the $n$-th $\widetilde{\bsl{G}}_l$, which for example, leads to
\eq{
c_{l n \alpha}^{\mu}(\bsl{k}) = c_{l\widetilde{\bsl{G}}_{ln}\alpha}^{\mu}(\bsl{k})\ .
}
Substituting it into Kohn-Sham equation, we obtain:
\eqa{
\sum_{ln\alpha}H^{\text{TAPW}}_{l' m\beta,ln\alpha}(\bsl{k}) c_{ln\alpha}^{\mu}(\bsl{k})=
\sum_{ln\alpha}\epsilon_{\mu}(\bsl{k})S^{\text{TAPW}}_{l'm\beta,ln\alpha}(\bsl{k}) c_{ln\alpha}^{\mu}(\bsl{k}) \ ,
}
where
\eqa{\begin{aligned}
    H^{\text{TAPW}}_{l'm\beta,ln\alpha}(\bsl{k}) &= \left\langle\psi_{l'm\beta}(\bsl{k})\right|\hat{H}\left|\psi_{ln\alpha}(\bsl{k})\right\rangle \ , \\ 
    S^{\text{TAPW}}_{l'm\beta,ln\alpha}(\bsl{k}) &= \left\langle\psi_{l'm\beta}(\bsl{k})\right|\left.\psi_{ln\alpha}(\bsl{k})\right\rangle \ ,
\end{aligned}}
are the Hamiltonian matrix element and the overlap matrix element, respectively. 
Utilizing \cref{tapw basis}, we can obtain
\eq{\begin{aligned}
\left\langle\psi_{ln\alpha}(\bsl{k})\right|\hat{H}\left|\psi_{l'm\beta}(\bsl{k})\right\rangle
&= \frac1{N_\text{m}{ N _\text{a}} }\sum_{\mathrm{IJ},ij}\mathrm{e}^{-\mathrm{i}(\bsl{k}+\widetilde{\bsl{G}}_{ln})(\mathbf{R}_{\mathrm{I}}+\tau_{li\alpha})}t(\mathbf{R}_{\mathrm{I}}-\mathbf{R}_{\mathrm{J}}+\tau_{li{\alpha}}-\tau_{l^{\prime}j_{\beta}})\mathrm{e}^{\mathrm{i}(\bsl{k}+\widetilde{\bsl{G}}_{l'm})(\mathbf{R}_{\mathrm{J}}+\tau_{l'j\beta})}  \\
&= \frac1{N_\mathrm{m}N_\mathrm{a}}\sum_{\mathrm{IJ},ij}\mathrm{e}^{-\mathrm{i}\widetilde{\bsl{G}}_{ln}\boldsymbol{\tau}_{li\alpha}}\mathrm{e}^{-\mathrm{i}\bsl{k}(\mathbf{R}_\mathrm{I}-\mathbf{R}_\mathrm{J}+\boldsymbol{\tau}_{li\alpha}-\boldsymbol{\tau}_{l'j\beta})}t(\mathbf{R}_\mathrm{I}-\mathbf{R}_\mathrm{J}+\boldsymbol{\tau}_{li\alpha}-\boldsymbol{\tau}_{l'j\beta})\mathrm{e}^{\mathrm{i}\widetilde{\bsl{G}}_{l'm}\boldsymbol{\tau}_{l'j\beta}} \\
&= \begin{aligned}&\frac{1}{N_{\mathrm{m}}N_\mathrm{a}}\sum_{\mathrm{I},ij}\mathrm{e}^{-\mathrm{i}\widetilde{\bsl{G}}_{ln}\boldsymbol{\tau}_{li\alpha}}\mathrm{e}^{-\mathrm{i}\bsl{k}\bar{\boldsymbol{\tau}}_{li\alpha,l'j\beta}}t(\bar{\boldsymbol{\tau}}_{li\alpha,l'j\beta})\mathrm{e}^{\mathrm{i}\widetilde{\bsl{G}}_{l'm}\boldsymbol{\tau}_{l'j\beta}}\end{aligned}  \\
&= \sum_{ij}\left(\frac{\mathrm{e}^{-\mathrm{i}\widetilde{\bsl{G}}_{ln}\boldsymbol{\tau}_{li\alpha}}}{\sqrt{N_\mathrm{a}}}\right)\cdot\left[\mathrm{e}^{-\mathrm{i}\bsl{k}\bar{\boldsymbol{\tau}}_{li\alpha,l'j\beta}}t(\bar{\boldsymbol{\tau}}_{li\alpha,l'j\beta})\right]\cdot\left(\frac{\mathrm{e}^{\mathrm{i}\widetilde{\bsl{G}}_{l'm}\boldsymbol{\tau}_{l'j\beta}}}{\sqrt{N_\mathrm{a}}}\right) \\
&= \sum_{ij}X^\dagger_{li\alpha,ln\alpha}H_{li\alpha,l'j\beta}(\bsl{k})X_{l'j\beta,l'm\beta} \ ,
\end{aligned}}
where $X_{li\alpha,ln\alpha} = \mathrm{e}^{\mathrm{i}\widetilde{\bsl{G}}_{ln}\boldsymbol{\tau}_{li\alpha}}/{\sqrt{N_\mathrm{a}}}$, $H_{li\alpha,l'j\beta}(\bsl{k})=\mathrm{e}^{-\mathrm{i}\bsl{k}\bar{\boldsymbol{\tau}}_{li\alpha,l'j\beta}}t(\bar{\boldsymbol{\tau}}_{li\alpha,l'j\beta})$, and $\bar{\boldsymbol{\tau}}_{li\alpha,l'j\beta}$ means the nearest displacement between orbital $(l,i,\alpha)$ and $(l',j,\beta)$,
$t(\bar{\boldsymbol{\tau}}_{li\alpha,l'j\beta})$ is the real space hopping between nearest orbital $(l,i,\alpha)$ and $(l',j,\beta)$ directly get from OpenMX's output Hamiltonian.
Similarly, for the overlap matrix element, we have
\eqa{ 
\left\langle\psi_{l'm\beta}(\bsl{k})\right|\left.\psi_{ln\alpha}(\bsl{k})\right\rangle = \sum_{ij}X^\dagger_{li\alpha,ln\alpha}S_{li\alpha,l'j\beta}(\bsl{k})X_{l'j\beta,l'm\beta} \ ,
}
with $S_{li\alpha,l'j\beta}(\bsl{k})=\mathrm{e}^{-\mathrm{i}\bsl{k}\bar{\boldsymbol{\tau}}_{li\alpha,l'j\beta}}s(\bar{\boldsymbol{\tau}}_{li\alpha,l'j\beta})$, 
and $s(\bar{\boldsymbol{\tau}}_{li\alpha,l'j\beta})$ is the overlap between nearest orbital $(l,i,\alpha)$ and $(l',j,\beta)$ directly get from OpenMX's output overlap matrix.
Now,we can write down the Hamiltonian and overlap matrix in a more compact form,
\eqa{\begin{aligned}
H^\mathrm{TAPW}(\bsl{k})&=X^\dagger H(\bsl{k}) X \ ,\\
S^\mathrm{TAPW}(\bsl{k})&=X^\dagger S(\bsl{k}) X \ ,
\end{aligned}}
where $H(\bsl{k})$ and $S(\bsl{k})$ are exactly the full tight-binding and overlap matrices, and $X$ is the plane wave projector. 
This projection yields a TAPW basis Hamiltonian that retains key features of the original DFT Hamiltonian while enabling substantial computational efficiency. 

To simplify the generalized eigenvalue problem, we apply Löwdin orthogonalization to the overlap matrix, this involves defining a normalized Hamiltonian by incorporating the inverse square root of the overlap matrix:
\eqa{
H_{\text{DFT}}(\bsl{k}) = (\text{S}^\mathrm{TAPW}(\bsl{k}))^{-1/2} \text{H}^\mathrm{TAPW}(\bsl{k}) (\text{S}^\mathrm{TAPW}(\bsl{k}))^{-1/2} \ .
}
This orthogonalization process yields an orthonormal set of basis functions:
\eqa{
|\widetilde{\psi}_{ln\alpha}(\bsl{k})\rangle = \sum_{l'm\beta}(S^{\text{TAPW}})^{-1/2}_{ln\alpha,l'm\beta}|\psi_{l'm\beta}(\bsl{k})\rangle
}
This orthogonal basis ensures that the overlap matrix becomes the identity matrix, $\langle\widetilde{\psi}_{ln\alpha}(\bsl{k})|\widetilde{\psi}_{l'm\beta}(\bsl{k})\rangle=\delta_{ln\alpha,l'm\beta}$, transforming the generalized eigenvalue problem into a standard one.

\subsection{Symmetrization of the Hamiltonian}

Numerical artifacts in OpenMX DFT calculations, particularly those arising from the use of conventional regular grids for Fast Fourier Transforms (FFT) and numerical integrations, can disrupt the intrinsic symmetries of the Hamiltonian\cite{openmx_Forces_2006,openmx_symmetry_2011}. 
In our case, the intrinsic $C_{3z}$ symmetry of the atomic structure in {\tmt} may not be perfectly preserved due to these numerical issues.
Increasing the cutoff energy (e.g., to 400-500 Ryd) can mitigate this issue by enhancing the resolution of the grid but requires a high computational cost.  
To balance efficiency and accuracy, we adopt an alternative strategy: performing DFT calculations with a moderate cutoff and manually restore the desired symmetries in the Hamiltonian. 

To restore the broken $C_{3z}$ symmetry, we construct a symmetrized Hamiltonian by averaging over all $C_{3z}$ symmetry operations. 
\eqa{
\tilde{H}(\bsl{k}) = \frac{1}{3}\left[ H(\bsl{k}) + D(C_{3z}) H(C_{3z}^{-1}\bsl{k}) D(C_{3z})^\dagger  + D(C_{3z}^2) H(C_{3z}^{-2}\bsl{k}) D(C_{3z}^2)^\dagger \right]\ .
}
For a certain valley like K valley, $C_{3z}$ act on basis \eqref{tapw basis} as
\eqa{
\begin{aligned}
C_{3z}\left|\psi_{ln\alpha }(\bsl{k})\right\rangle
&=\frac{1}{\sqrt{N_\text{m}N_\text{a}} }\sum_{\text{I},i}\mathrm{e}^{\mathrm{i}(\bsl{k}+\widetilde{\bsl{G}}_{ln})(\mathbf{R}_{\text{I}}+\tau_{li\alpha})}C_{3z}\left|\phi_{\alpha}(\mathbf{R}_{\text{I}}+\tau_{li\alpha})\right\rangle\\
&=\frac{1}{\sqrt{N_\text{m}N_\text{a}} }\sum_{\text{I},i}\mathrm{e}^{\mathrm{i}C_{3z}(\bsl{k}+\widetilde{\bsl{G}}_{ln}-\mathbf{K}_{l}+\mathbf{K}_{l})(\mathbf{R}_{\text{I}}+\tau_{li\alpha})}\sum_{m^{\prime}}D_{m^{\prime},m}^{j}(C_{3z})\left|\phi_{\alpha'}(\mathbf{R}_{\text{I}}+\tau_{li_{\alpha}})\right\rangle\\
&=\frac{1}{\sqrt{N_\text{m}N_\text{a}} }\sum_{\text{I},i}\mathrm{e}^{\mathrm{i}[C_{3z}(\widetilde{\bsl{G}}_{ln}-\mathbf{K}_{l})+\mathbf{K}_{l}]\tau_{li\alpha}}\mathrm{e}^{\mathrm{i}[C_{3z}(\bsl{k}+\mathbf{K}_{l})-\mathbf{K}_{l}](\mathbf{R}_{\text{I}}+\tau_{li\alpha})}\sum_{m^{\prime}}D_{m^{\prime},m}^{j}(C_{3z})\left|\phi_{\alpha'}(\mathbf{R}_{\text{I}}+\tau_{li_{\alpha}})\right\rangle\\
&=\frac{1}{\sqrt{N_\text{m}N_\text{a}} }\sum_{\text{I},i}\sum_{n^\prime}D_{n^{\prime},n}^{l}(C_{3z})\mathrm{e}^{\mathrm{i}\widetilde{\bsl{G}}_{ln'}\tau_{li\alpha}}\mathrm{e}^{\mathrm{i}[C_{3z}(\bsl{k}+\mathbf{K}_{l})-\mathbf{K}_{l}](\mathbf{R}_{\text{I}}+\tau_{li\alpha})}\sum_{m^{\prime}}D_{m^{\prime},m}^{j}(C_{3z})\left|\phi_{\alpha'}(\mathbf{R}_{\text{I}}+\tau_{li_{\alpha}})\right\rangle\\
&=\sum_{n^{\prime},m^{\prime}}D_{n^\prime,n}^{l}(C_{3z})D_{m^{\prime},m}^{j}(C_{3z})\left|\psi_{ln'\alpha'}(C_{3z}({\bsl{k}+\mathbf{K}_{l}})-\mathbf{K}_{l})\right\rangle \ ,
\end{aligned}
}
where $D_{n^{\prime},n}^{l}(C_{3z}) = \delta_{\widetilde{\bsl{G}}_{l {n^\prime}},C_{3z}(\widetilde{\bsl{G}}_{ln}-\mathbf{K}_{l})+\mathbf{K}_{l}}$, $\widetilde{\bsl{G}}_l$ represents $\widetilde{\bsl{G}}$ vectors in the $l$-th layer, $D_{m^{\prime},m}^{j}(C_{3z})$ represents Wigner D-matrix for $C_{3z}$ rotation, $j$ is the total angular momentum quantum number of orbital $\alpha$. 
The summation over $\alpha'$ indicates that the sum is taken only over the magnetic quantum number $m'$ within, while the other components $\alpha'$ remain fixed.
By enforcing this symmetrization, we ensure that the physical symmetry of the system is preserved, which is crucial for accurate band structure calculations.

\begin{figure}[!tbp]
    \centering
    \includegraphics[width=\columnwidth]{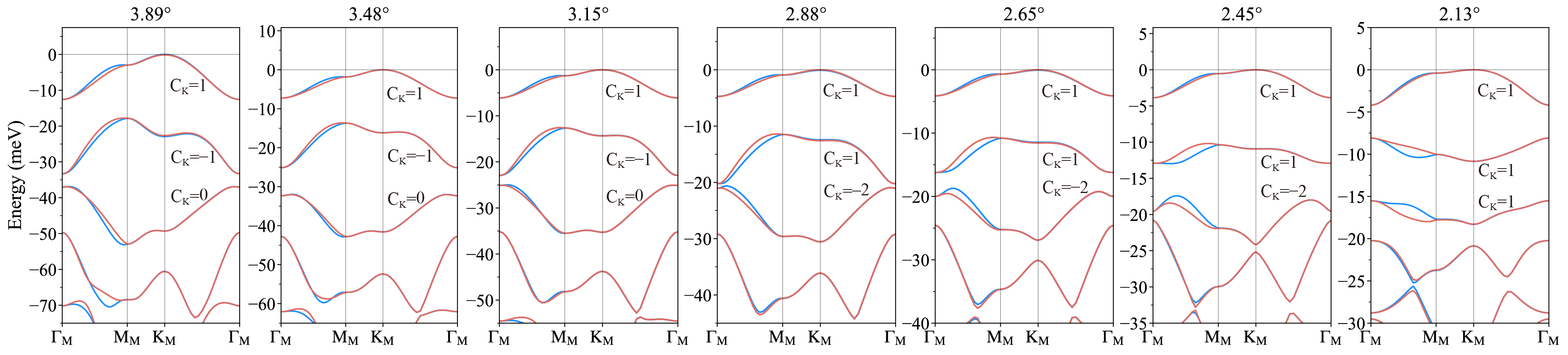}
    \includegraphics[width=\columnwidth]{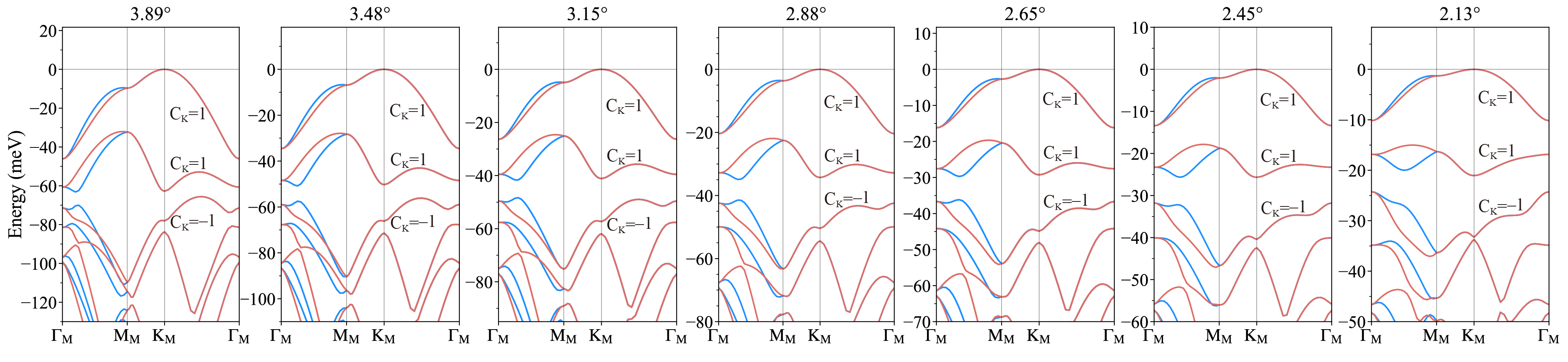}
    \caption{
    \textbf{Band structures of {\tmt} and {\tws} calculated with the TAPW method.} Results are shown for twist angles from 3.89$^\circ$ to 2.13$^\circ$ with symmetrized Hamiltonians. For {\tmt} (upper panels), $\widetilde{\bsl{G}}$ lattices up to the 7th harmonic for angles from $3.89^\circ$ to $3.15^\circ$, and up to the 8th harmonic from $2.88^\circ$ to $2.13^\circ$. For {\tws} (lower panels), $\widetilde{\bsl{G}}$ lattice up to the 7th harmonic from $3.89^\circ$ to $3.48^\circ$, and up to the 8th harmonic from $3.15^\circ$ to $2.13^\circ$. Blue lines represent $\mathrm{K}$-valley bands, while red lines represent $\mathrm{K^{\prime}}$-valley bands.
    }
    \label{fig:band_symm}
\end{figure}

As illustrated in \cref{fig:band_symm}, the symmetrized Hamiltonian accurately captures the essential features of the moir\'e band structures for both \tmt\ and \tws\ across various twist angles. 

\section{Building Analytical Models without Fitting}
\label{general method}

In this section, we present a general method to derive continuum models directly from DFT calculation \emph{without fitting}, applicable to both homostructures and heterostructures across any number of layers in any valley. 
The method generates two model types: (i) a numerically precise model suitable for further numerical calculations and (ii) an analytical model focused on low-energy physics, sacrificing precision in higher-(absolute) energy bands to facilitate analytical study. 
In the following discussion, we concentrate on a single valley.

\subsection{Continuum Model Basis Construction}
Within a given valley, the different layers of a twisted system can host distinct sub-valleys, which determine the electrons' momenta, or more precisely, the eigenvalues of the moir\'e translation operator. 
For example, in {\tmt}, the twist angle between layers shifts the monolayer K points within the K valley, effectively generating separate sub-valleys corresponding to each layer.
However, in the $\Gamma$ valley, both layers share a common sub-valley since the $\Gamma$ point remains invariant under rotations, meaning layer information must be encoded differently (than in the momentum shift)  here. 

To define the basis of the continuum model, we use $\psi^\dagger_{\bsl{r}, v, a_v}$, where $\bsl{r}$ represents the 2D continuum position, $v=\K_1, \K_2, \Gamma ...$ is the sub-valley index denoting the valley that we investigate, $a_v$ encompasses all additional indices relevant to sub-valley $v$. For example, In the K valley of {\tmt}, $a_v$ comprises only the orbital index, such as the $d_{x^2-y^2} + \ii d_{xy}$ orbital, since the sub-valleys $\K_1$ and $\K_2$ already correspond to different layer, allowing layer differentiation via $v$. In contrast, in the $\Gamma$ valley, where sub-valleys are indistinguishable, $a_v$ must include both orbital and layer indices. Thus, in the K valley, $a_v$ pertains solely to the orbital index, while in the $\Gamma$ valley, it also incorporates layer information. Additionally, in homobilayer structures, the orbital range of $a_v$ is identical across sub-valleys within any given valley, allowing simplification of the notation to $a$. 
However, for heterobilayer structures, the orbital values represented by $a_v$ can differ between sub-valleys, necessitating the retention of $a_v$ to accurately capture the distinct orbital characteristics of each sub-valley. 
Furthermore, if the spin-valley coupling is not sufficiently strong, $a_v$ can also include spin degrees of freedom.

Suppose the moir\'e lattice translation operator is $T_{\bsl{R}_M}$, and then continuum basis $\psi^\dagger_{\bsl{r},v,a_v}$ transforms as 
\eq{
\label{eq:moire_T_psi_real_space}
T_{\bsl{R}_M} \psi^\dagger_{\bsl{r},v,a_v} T_{\bsl{R}_M}^{-1} = e^{-\ii \bsl{R}_M\cdot(\text{K}_0+\bsl{q}_v)}\psi^\dagger_{\bsl{r}+\bsl{R}_M,v,a_v}\ ,
}
where $\text{K}_0$ is the monolayer momentum that corresponding to the $\Gamma_M$ in the moir\'e Brillouin zone, and $\bsl{q}_v$ labels the deviation from $\text{K}_0$ to the sub-valley $v$, with $\text{K}_0 + \bsl{q}_v = v$. 
It is important to note that we do not distinguish between two sub-valleys if they differ by moir\'e reciprocal lattice vectors, as such differences result in identical phases under moir\'e lattice translations.

With the defined $\bsl{q}_v$, we can naturally express the Fourier transformation of the continuum basis $\psi^\dagger_{\bsl{r},v,a_v}$ into momentum space, which is given by
\eq{
\label{eq:FT_continuum_basis}
\psi^\dagger_{\bsl{k}-\bsl{G}_M-\bsl{q}_v,v,a_v} = \frac{1}{\sqrt{\A}}\int d^2 r e^{\ii \bsl{r}\cdot (\bsl{k}-\bsl{G}_M-\bsl{q}_v) } \psi^\dagger_{\bsl{r},v,a_v}  \ ,
}
where $\bsl{k}$ is the momentum in the moir\'e Brillouin zone, $\bsl{G}_M$ is the moir\'e reciprocal lattice vector, $\A$ is the total area of the sample, and we do not include $\text{K}_0$ in the Fourier transformation. In this way, the basis has the following transformation property under moir\'e lattice translations:

\eq{
\label{eq:moire_T_psi_momentum}
T_{\bsl{R}_M} \psi^\dagger_{\bsl{k}-\bsl{G}_M-\bsl{q}_v,v,a_v} T_{\bsl{R}_M}^{-1} = e^{-\ii \bsl{R}_M\cdot (\bsl{k}-\bsl{G}_M+\K_0)}  \psi^\dagger_{\bsl{k}-\bsl{G}_M-\bsl{q}_v,v,a_v} \ .
}

By defining 
\eq{
\label{eq:def_Q}
 \bsl{Q} = \bsl{G}_M + \bsl{q}_v \ ,
}
we obtain $\psi^\dagger_{\bsl{k},\bsl{Q},a_{\bsl{Q}}}$, which serves as the basis of the continuum model, as $\bsl{Q}$ is one-to-one mapped to the pair $(\bsl{G}_M, \bsl{q}_v)$, where

\eq{
\label{eq:def_a_Q}
 a_{\bsl{Q}} = a_v \text{ iff } \bsl{Q}\in \left\{ \bsl{b}_{M_1} \dsZ +  \bsl{b}_{M_2} \dsZ +  \bsl{q}_v\right\} \ ,
}
where $ \bsl{b}_{M_1}$ and $ \bsl{b}_{M_2}$ are two primitive reciprocal lattice vectors, and $\dsZ$ denotes the integer set.  
We now define the main reason for us to define the concept of sub-valleys, instead of using only the layer index. 
Taking the $\Gamma$ valley in {\tmt} as an example, the two layers would have the same $\bsl{Q}$, and thus it is impossible to use $\bsl{Q}$ to cover both $\bsl{G}_M$ and layer index. 
$\bsl{Q}$ can only cover both $\bsl{G}_M$ and the sub-valleys.

\subsection{Projection of DFT Hamiltonian onto the Continuum Basis}
The next step is constructing $\psi^\dagger_{\bsl{k},\bsl{Q},a_{\bsl{Q}}}$ (see \cref{eq:FT_continuum_basis,eq:def_Q}) from the DFT calculations. 
Firstly, note that the TAPW basis can be expressed as $c^\dagger_{\bsl{k} + \widetilde{\bsl{G}}_M,\alpha_{\bsl{Q}}}$, where $\alpha_{\bsl{Q}}$ is the multiplicity index of orbital and layer, similarly as $a_{\bsl{Q}}$ in \cref{eq:def_a_Q}, only in K valley of homobilayer system index $\alpha_{\bsl{Q}}$ can be simplified as $\alpha$, which is the same notation we used in \cref{TAPW method}, as layer information is naturally composed in $q_v$. 
Second, to align $\bsl{Q} = \bsl{G}_M + \bsl{q}_v$ with the plane waves used in the TAPW calculations, we select $\bsl{G}_M = \K_0 -\widetilde{\bsl{G}}_l$, ensuring a one-to-one correspondence between $\bsl{Q}$ in the continuum model and $\widetilde{\bsl{G}}_l$ chosen in TAPW. 
We refer to this continuum model projected onto these selected $\bsl{Q}$ points as the ``full model''. 
Consequently, the DFT basis can be simplified as $c^\dagger_{\bsl{k}+\K_0 , \bsl{Q},\alpha_{\bsl{Q}}} \equiv c^\dagger_{\bsl{k}+\K_0 +\bsl{q}_{v_{\bsl{Q}}} -\bsl{Q},\alpha_{\bsl{Q}}}$, where we uesd \cref{eq:def_Q,eq:def_a_Q}, where $v_{\bsl{Q}}$ denotes the sub-valley containing $\bsl{Q}$. 
Thus, $H_\text{DFT}$ around a particular valley can be expressed as:
\eq{
\label{eq:DFT_Hamiltonian}
H_{\text{DFT}} = \sum_{\bsl{k}}\sum_{\bsl{Q}_1 \bsl{Q}_2} \sum_{\alpha_{\bsl{Q}_1}\alpha_{\bsl{Q}_2}'}c^\dagger_{\bsl{k}+\K_0 , \bsl{Q}_1,\alpha_{\bsl{Q}_1}} \left[ H_{\text{DFT}}(\bsl{k}) \right]_{\bsl{Q}_1\alpha_{\bsl{Q}_1},\bsl{Q}_2 \alpha_{\bsl{Q}_2}'}c_{\bsl{k}+\K_0 , \bsl{Q}_2,\alpha_{\bsl{Q}_2}'} \ ,
}

We want to emphasize that the key difference between $\psi^\dagger_{\bsl{k},\bsl{Q},a_{\bsl{Q}}}$ in \cref{eq:FT_continuum_basis}  and $c^\dagger_{\bsl{k}+\K_0, \bsl{Q},\alpha_{\bsl{Q}}}$ lies in the restriction on the index $a_{\bsl{Q}}$, since we normally want $a_{\bsl{Q}}$ to take much less values of $\alpha_{\bsl{Q}}$ 
For example, in the K valley of {\tmt} (for which $a_{\bsl{Q}}$ and $\alpha_{\bsl{Q}}$ can be simplified as $a$ and $\alpha$ as discussed above), $\alpha$ spans $A=71$ orbitals with spin-$\uparrow$ in one monolayer unit cell due to the Mo-$s3p2d2$ and Te-$s3p2d2f1$ basis sets, whereas $a$ is limited to only a subset of these orbitals (e.g., the $d_{z^2}$ orbital or Mo, or combinations like $d_{x^2-y^2}+\ii d_{xy}$ of Mo, among others). 
While for the $\Gamma$ valley, the inclusion of layer index doubles the range of both $a$ and $\alpha$, expanding $\alpha$ to $2 \times 71 = 142$. 
In the following, we will explain how $a_{\bsl{Q}}$ is determined.

The DFT Hamiltonian in \cref{eq:DFT_Hamiltonian} contains a lot of high-energy degrees of freedom that are irrelevant to the low-energy physics of interest.
To simplify \cref{eq:DFT_Hamiltonian}, we recall that the low-energy physics happens among the low-energy states of monolayer Hamiltonians---we always choose the low-energy states of monolayer Hamiltonians as the basis of the continuum model.
In other words, the most dominant energy scale in the continuum model is the kinetic energy directly inherited from the monolayer Hamiltonians, which do not couple different $\bsl{Q}$'s for a fixed $\bsl{k}$.
Therefore, to isolate the low-energy states, let us consider the $\bsl{Q}$ diagonal block of $H_{\text{DFT}}(\bsl{k})$, labeled as $D_{\bsl{Q}}(\bsl{k})$, which reads:

\eq{
\label{eq:block_diagonalize}
\left[ D_{\bsl{Q}}(\bsl{k}) \right]_{\alpha_{\bsl{Q}} \alpha_{\bsl{Q}}'} = \left[ H_{\text{DFT}}(\bsl{k}) \right]_{\bsl{Q}\alpha_{\bsl{Q}},\bsl{Q} \alpha_{\bsl{Q}}'}\ .
}
Diagonalizing $D_{\bsl{Q}}(\bsl{k})$ yields eigenvalues $\lambda_{n}^{\bsl{k},\bsl{Q}}$ and eigenvectors $U_{n}^{\bsl{k},\bsl{Q}}$:
\eqa{
\label{eq:Dq_eigen_eq}
D_{\bsl{Q}}(\bsl{k}) U_{n}^{\bsl{k},\bsl{Q}} =  \lambda_{n}^{\bsl{k},\bsl{Q}} U_{n}^{\bsl{k},\bsl{Q}} \ ,
}
where $n = 1, \dots, N_{\alpha_{\bsl{Q}}}$ and $N_{\alpha_{\bsl{Q}}}$ is the number of distinct $\alpha_{\bsl{Q}}$ values.
We select a subset of eigenvalues, $\lambda_{\text{low}}^{\bsl{k},\bsl{Q}}$, that meet the following criteria:
\begin{enumerate}
    \item They are the closest to the Fermi energy among all $\lambda_{n}^{\bsl{k},\bsl{Q}}$.
    \item They are separated from the remaining eigenvalues by a significant gap that is much larger than the spread of the selected $\lambda_{\text{low}}^{\bsl{k},\bsl{Q}}$ values.
\end{enumerate}

The count of $\lambda_{\text{low}}^{\bsl{k},\bsl{Q}}$ is denoted as $N_{a_{\bsl{Q}}}$. 
We focus on the wavefunctions corresponding to $a_{\bsl{Q}}$.
The corresponding eigenvectors are denoted as $U_{\text{low}}^{\bsl{k},\bsl{Q}}$, with dimensions $N_{\alpha_{\bsl{Q}}}\times N_{a_{\bsl{Q}}}$. The remaining eigenvalues and eigenvectors are labeled $\lambda_{\text{high}}^{\bsl{k},\bsl{Q}}$ and $U_{\text{high}}^{\bsl{k},\bsl{Q}}$ with dimension $N_{\alpha_{\bsl{Q}}}\times (N_{\alpha_{\bsl{Q}}}-N_{a_{\bsl{Q}}})$, capture the high-energy states.

It is important to note that $U_{\text{low}}^{\bsl{k},\bsl{Q}}$ may not exhibit the desired physical symmetry representations due to arbitrary phase factors introduced during numerical diagonalization. 
To recover the correct symmetry, we apply a unitary transformation $V^{\bsl{k},\bsl{Q}}$ to to restore symmetry:
\begin{equation}
    U_{\text{low}}^{\bsl{k},\bsl{Q}} \rightarrow V^{\bsl{k},\bsl{Q}} U_{\text{low}}^{\bsl{k},\bsl{Q}}.
\end{equation}
which is to fix the gauge. For example, we can fix the gauge by choosing a fixed orbital and then making sure its phase in the eigenvectors to be zero.
Then, the continuum basis can be constructed from the DFT basis as
\eq{
\label{eq:continuum_basis_DFT}
\psi^\dagger_{\bsl{k},\bsl{Q},a_{\bsl{Q}}} = \sum_{\alpha_{\bsl{Q}}} c^\dagger_{\bsl{k}+\K_0 , \bsl{Q},\alpha_{\bsl{Q}}} \left[ U_{\text{low}}^{\bsl{k},\bsl{Q}} \right]_{\alpha_{\bsl{Q}},a_{\bsl{Q}}} \ .
}

This construction preserves the moir\'e translation properties specified in \cref{eq:moire_T_psi_momentum}, since $c^\dagger_{\bsl{k}+\K_0 , \bsl{Q},\alpha_{\bsl{Q}}}$ carries momentum $\bsl{k}+\K_0 +q_{v_{\bsl{Q}}}-\bsl{Q}$.

With the continuum basis established, the next step is to obtain Hamiltonian by directly projecting $H_{\text{DFT}}(\bsl{k})$ onto $U_{a_{\bsl{Q}}}^{\bsl{k},\bsl{Q}}$. 
To do so, we define a matrix $U_0(\bsl{k})$ for the space spanned by $U_{a_{\bsl{Q}}}^{\bsl{k},\bsl{Q}}$, which reads
\eqa{
U_0(\bsl{k}) = \bigoplus_{\bsl{Q}} U_{\text{low}}^{\bsl{k},\bsl{Q}} \ .
}
The projected Hamiltonian is then given by:
\eq{
 h_{00}(\bsl{k}) =  U_0^\dagger(\bsl{k}) H_{\text{DFT}}(\bsl{k}) U_0(\bsl{k}) \ .
}
In general, $h_{00}(\bsl{k})$ may not fully capture low-energy band structure as it only includes the effects within the low-energy subspace and neglects the coupling to the high-energy states. Although these couplings are small, they are essential for accurately describing the physical details of the low-energy states. To account for this, we perform second-order perturbation as follows.
Define $U_1(\bsl{k})$ as the orthogonal completion of $U_0(\bsl{k})$, such that,
\eq{
U_0(\bsl{k}) U_0^\dagger (\bsl{k}) + U_1(\bsl{k}) U_1^\dagger (\bsl{k}) = 1 , \  U_1^\dagger (\bsl{k}) U_0(\bsl{k})  = 0 \ .
}
Explicitly, we can construct $U_1(\bsl{k})$ as:
\eqa{
U_1(\bsl{k}) = \bigoplus_{\bsl{Q}} U_{\text{high}}^{\bsl{k},\bsl{Q}} \ ,
}
representing the high-energy Hilbert space of $\lambda_{\text{high}}^{\bsl{k},\bsl{Q}}$ (relative to the Fermi energy). The corresponding Hamiltonian reads:
\eq{
h_{11}(\bsl{k}) = U_1^\dagger(\bsl{k}) H_{\text{DFT}}(\bsl{k}) U_1(\bsl{k})\ ,
}
and the transitions matrix between the low-$\lambda$ and high-$\lambda$ modes is:
\eq{
h_{01}(\bsl{k})= U_0^\dagger(\bsl{k}) H_{\text{DFT}}(\bsl{k}) U_1(\bsl{k}) \ ,
}
along with its Hermitian conjugate. 
Then, the corrected DFT Hamiltonian for the continumm basis reads:
\eq{
\label{eq:h_DFT}
h_{\text{DFT}}(\bsl{k}) = h_{00}(\bsl{k}) + h_{01}(\bsl{k}) \frac{1}{\overline{\lambda}_{\bsl{k}} - h_{11}(\bsl{k})} h_{01}^\dagger(\bsl{k}) \ ,
}
where $\overline{\lambda}_{\bsl{k}}$ is chosen as the mean value of the eigenvalues of $h_{00}(\bsl{k})$. 
The mean value is just one choice---one can choose other values within the range of the eigenvalues of $h_{00}(\bsl{k})$ if they can provide more accuracy.
We use the mean value because it serves as a central reference energy around which we perform the perturbative expansion.
With this setup, we derive a model with the continuum basis directly from DFT calculation 
\eq{
\label{eq:DFT_lowenergy}
H_{\text{DFT}} = \sum_{\bsl{k}}\sum_{\bsl{Q}_1 \bsl{Q}_2} \sum_{a_{\bsl{Q}_1} a_{\bsl{Q}_2}'}\psi^\dagger_{\bsl{k} , \bsl{Q}_1,a_{\bsl{Q}_1}} \left[ h_{\text{DFT}}(\bsl{k}) \right]_{\bsl{Q}_1a_{\bsl{Q}_1},\bsl{Q}_2 a_{\bsl{Q}_2}'}\psi_{\bsl{k} , \bsl{Q}_2,a_{\bsl{Q}_2}'} \ ,
}
where $\psi^\dagger_{\bsl{k} , \bsl{Q}_1,a_{\bsl{Q}_1}}$ is constructed in \cref{eq:continuum_basis_DFT} and $h_{\text{DFT}}(\bsl{k})$ is in \cref{eq:h_DFT}.

In principle, one can directly use $H_{\text{DFT}}$ together with an interaction term to perform any many-body calculation, as long as the interaction term is normal-ordered to avoid the double counting due to the interaction effect included in the DFT calculation. 
However, direct use of $H_{\text{DFT}}$ has limitations. Specifically, the DFT-calculated Hamiltonian $h_{\text{DFT}}(\bsl{k})$ is defined on a discrete momentum mesh, which restricts flexibility for continuous momentum-space applications.
To resolve this issue, we will construct a continuum model that is (nearly) numerically accurate on any momentum mesh in the next subsection.

\subsection{Continuum Model Construction}\label{subsec:cont_construct}
This section introduces a general method to construct a continuum model that enables both accurate numerical calculations and analytical understanding of the low-energy properties of moir\'e materials. The key advantage of this approach is that it does not require any fitting, instead deriving directly from DFT results.

Recall that the continuum model in the real space have the following general form
\eq{
\label{eq:general_form_continuum_model}
H_{cont} = \sum_{m_x, m_y, n_x, n_y \in \dsN} \sum_{v v'} \sum_{a_{v} a_{v'}'}\int d^2 r \left(\ii^m \partial_x^{m_x} \partial_y^{m_y} \psi^\dagger_{\bsl{r},v,a_v} \right) t^{m_x m_y n_x n_y}_{v a_v,v' a_{v'}}(\bsl{r}) (-\ii)^n\partial_x^{n_x} \partial_{y}^{n_y}\psi_{\bsl{r},v',a_{v'}'} \ ,
}
here , $m=m_x + m_y$, $n=n_x +n_y$, and we do not need to explicitly include derivative's on $t$'s since acting derivatives on a potential still gives us potentials. 
Note that \cref{eq:general_form_continuum_model} is equivalent to \cref{main_eq:general_form_continuum_model} in the main text, and the relation will be discussed in and below \cref{eq:h_DFT_continuum_expansion_linearly_independent}.
Since we require the model to be moir\'e lattice transitionally invariant, $t^{m_x m_y n_x n_y}_{v a_v,v' a_{v'}}(\bsl{r})$ has the following general form 
\eq{\label{eq:FT_continuum_coefficient}
t^{m_x m_y n_x n_y}_{v a_v,v' a_{v'}}(\bsl{r}) = \sum_{\bsl{G}_M} t^{m_x m_y n_x n_y}_{v a_v,v' a_{v'},\bsl{G}_M} e^{-\ii (\bsl{q}_v -\bsl{q}_{v'} + \bsl{G}_M)\cdot\bsl{r}} \ ,
}
where we have used \cref{eq:moire_T_psi_real_space}, this form permits efficient numerical construction of the Hamiltonian in momentum space. As a result, we can rewrite the Hamiltonian as: 
\eqa{
H_{cont} & = \sum_{m_x, m_y, n_x, n_y \in \dsN} \sum_{v v'} \sum_{a_{v} a_{v'}'}\sum_{\bsl{G}_M} t^{m_x m_y n_x n_y}_{v a_v,v' a_{v'},\bsl{G}_M}\int d^2 r \left(\ii^m \partial_x^{m_x} \partial_y^{m_y} \psi^\dagger_{\bsl{r},v,a_v}\right)  \\ 
& \qquad \times e^{-\ii (\bsl{q}_v -\bsl{q}_{v'} + \bsl{G}_M)\cdot\bsl{r}}  (-\ii)^n\partial_x^{n_x} \partial_{y}^{n_y}\psi_{\bsl{r},v',a_{v'}'} \ .
}
For numerical implementation, it is  efficient to perform the construction of the Hamiltonian in the momentum space. 
Using \cref{eq:FT_continuum_basis}, we can obtain 
\eqa{
H_{cont} & = \sum_{m_x, m_y, n_x, n_y \in \dsN}\sum_{v v'} \sum_{a_{v} a_{v'}'}\sum_{\bsl{G}_M} t^{m_x m_y n_x n_y}_{v a_v,v' a_{v'},\bsl{G}_M} \sum_{\bsl{k}} \sum_{\bsl{G}_{M,1} \bsl{G}_{M,2} } (\bsl{k}-\bsl{G}_{M,1} - \bsl{q}_v)_x^{m_x} (\bsl{k}-\bsl{G}_{M,1} - \bsl{q}_v)_{y}^{m_y} \\
& \qquad  \times  \psi_{\bsl{k}-\bsl{G}_{M,1}-\bsl{q}_{v},v,a_{v}}^\dagger \delta_{\bsl{G}_{M,1} + \bsl{q}_v , \bsl{G}_{M,2} + \bsl{q}_v' + \bsl{q}_v -\bsl{q}_{v'} + \bsl{G}_M }(\bsl{k}-\bsl{G}_{M,2} - \bsl{q}_{v'})_{x}^{n_x}(\bsl{k}-\bsl{G}_{M,2} - \bsl{q}_{v'})_{y}^{n_y}\psi_{\bsl{k}-\bsl{G}_{M,2}-\bsl{q}_{v'},v',a_{v'}'} \\
& = \sum_{m_x, m_y, n_x, n_y \in \dsN}\sum_{\bsl{Q} \bsl{Q}'} \sum_{a_{\bsl{Q}} a_{\bsl{Q}'}'}\sum_{\bsl{G}_M} t^{m_x m_y n_x n_y}_{v_{\bsl{Q}} a_{\bsl{Q}},v'_{\bsl{Q}'} a_{\bsl{Q}'}',\bsl{G}_M} \sum_{\bsl{k}}  (\bsl{k}-\bsl{Q})_{x}^{m_x}(\bsl{k}-\bsl{Q})_{y}^{m_y} \\
& \qquad  \times  \psi_{\bsl{k},\bsl{Q},a_{\bsl{Q}}}^\dagger \delta_{ \bsl{Q} , \bsl{Q}' + \bsl{q}_{v_{\bsl{Q}}} -\bsl{q}_{v_{\bsl{Q}'}'}  + \bsl{G}_M } (\bsl{k}-\bsl{Q}')_x^{n_x}(\bsl{k}-\bsl{Q}')_y^{n_y} \psi_{\bsl{k}, \bsl{Q}',a_{\bsl{Q}'}'} \\
& = \sum_{m_x, m_y, n_x, n_y \in \dsN} \sum_{v_1 v_2}  \sum_{\bsl{G}_M  } \sum_{a_{v1} a
_{v_2}'}t^{m_x m_y n_x n_y}_{v_1 a_{v_1},v_2  a'_{v_2} ,  \bsl{G}_M  } \sum_{\bsl{k}} \sum_{\bsl{Q} \bsl{Q}'}  \sum_{a_{\bsl{Q}} a_{\bsl{Q}'}'} \psi_{\bsl{k},\bsl{Q},a_{\bsl{Q}}}^\dagger \psi_{\bsl{k}, \bsl{Q}',a_{\bsl{Q}'}'}\\
& \qquad  \times \delta_{v_{\bsl{Q}}a_{\bsl{Q}} , v_1 a_{v_1}} \delta_{v_{\bsl{Q}'}' a_{\bsl{Q}'}', v_2 a_{v_2}'}   (\bsl{k}-\bsl{Q})_{x}^{m_x}(\bsl{k}-\bsl{Q})_{y}^{m_y}  \delta_{ \bsl{Q} , \bsl{Q}' + \bsl{q}_{v_1} -\bsl{q}_{v_2}  + \bsl{G}_M  } (\bsl{k}-\bsl{Q}')_x^{n_x}(\bsl{k}-\bsl{Q}')_y^{n_y}  \\
& = \sum_{m_x, m_y, n_x, n_y \in \dsN} \sum_{v_1 v_2}  \sum_{ \bsl{G}_M  } \sum_{a_{v1} a
_{v_2}'}t^{m_x m_y n_x n_y}_{v_1 a_{v_1},v_2  a'_{v_2} ,\bsl{G}_M } \sum_{\bsl{k}} \sum_{\bsl{Q} \bsl{Q}'}  \sum_{a_{\bsl{Q}} a_{\bsl{Q}'}'} \psi_{\bsl{k},\bsl{Q},a_{\bsl{Q}}}^\dagger \psi_{\bsl{k}, \bsl{Q}',a_{\bsl{Q}'}'}\\
& \qquad  \times \left[ X^{m_x m_y n_x n_y}_{v_1 a_{v_1},v_2  a'_{v_2} ,\bsl{q}_{v_1} -\bsl{q}_{v_2}  + \bsl{G}_M }(\bsl{k}) \right]_{\bsl{Q} a_{\bsl{Q}}, \bsl{Q}' a_{\bsl{Q}'}'} \ ,
}
where
\eq{
\left[ X^{m_x m_y n_x n_y}_{v_1 a_{v_1},v_2  a'_{v_2} ,\bsl{p} }(\bsl{k}) \right]_{\bsl{Q} a_{\bsl{Q}}, \bsl{Q}' a_{\bsl{Q}'}'} = \delta_{v_{\bsl{Q}}a_{\bsl{Q}} , v_1 a_{v_1}} \delta_{v_{\bsl{Q}'}' a_{\bsl{Q}'}', v_2 a_{v_2}'}   (\bsl{k}-\bsl{Q})_{x}^{m_x}(\bsl{k}-\bsl{Q})_{y}^{m_y}  \delta_{ \bsl{Q} , \bsl{Q}' + \bsl{p} } (\bsl{k}-\bsl{Q}')_x^{n_x}(\bsl{k}-\bsl{Q}')_y^{n_y} \ .
}
Then, under the belief that the continuum model should be able to precisely capture $h_{\text{DFT}}(\bsl{k})$ in \cref{eq:h_DFT}, we would have
\eq{
\label{eq:h_DFT_continuum_expansion}
h_{\text{DFT}}(\bsl{k}) = \sum_{m_x, m_y, n_x, n_y \in \dsN} \sum_{v_1 v_2}  \sum_{\bsl{G}_M} \sum_{a_{v1} a
_{v_2}'}t^{m_x m_y n_x n_y}_{v_1 a_{v_1},v_2  a'_{v_2} , \bsl{G}_M }  X^{m_x m_y n_x n_y}_{v_1 a_{v_1},v_2  a'_{v_2} ,\bsl{q}_{v_1} -\bsl{q}_{v_2} + \bsl{G}_M}(\bsl{k}) \ .
}

Our goal is to determine the \emph{linearly independent} values of $t^{m_x m_y n_x n_y}_{v_1 a_{v_1},v_2  a'_{v_2} , \bsl{G}_M } $ from $h_{\text{DFT}}(\bsl{k})$. 
Before doing so, we will first show that $X^{m_x m_y n_x n_y}_{v_1 a_{v_1},v_2  a'_{v_2} ,\bsl{p}}(\bsl{k})$ ($\bsl{q}_{v_1} -\bsl{q}_{v_2} + \bsl{G}_M$ is writen as $\bsl{p}$ for simplicity) terms are orthogonal for different $(v_1 a_{v_1},v_2  a'_{v_2} ,\bsl{G}_M)$ but not completely linearly independent for $(m_x, m_y, n_x, n_y)$. 
To see so, first note that 
\eqa{
\label{eq:X_vap_orthogonal}
& \Tr \left\{ \left[X^{\bar{m}_x \bar{m}_y \bar{n}_x \bar{n}_y}_{\overline{v}_1 \overline{a}_{\overline{v}_1},\overline{v}_2  \overline{a}_{\overline{v}_2}' , \overline{\bsl{p}} }(\bsl{k}) \right]^\dagger  X^{m_x m_y n_x n_y}_{v_1 a_{v_1},v_2  a'_{v_2} , \bsl{p} }(\bsl{k}) \right\} \\
& = \sum_{\bsl{Q} a_{\bsl{Q}}, \bsl{Q}' a_{\bsl{Q}'}'}  \left[ X^{\bar{m}_x \bar{m}_y \bar{n}_x \bar{n}_y}_{\overline{v}_1 \overline{a}_{\overline{v}_1},\overline{v}_2  \overline{a}_{\overline{v}_2}' , \overline{\bsl{p}} }(\bsl{k}) \right]_{\bsl{Q} a_{\bsl{Q}}, \bsl{Q}' a_{\bsl{Q}'}'}^* \left[ X^{m_x m_y n_x n_y}_{v_1 a_{v_1},v_2  a'_{v_2} , \bsl{p} }(\bsl{k}) \right]_{\bsl{Q} a_{\bsl{Q}}, \bsl{Q}' a_{\bsl{Q}'}'}\\
& =  \sum_{\bsl{Q} a_{\bsl{Q}}, \bsl{Q}' a_{\bsl{Q}'}'} \delta_{v_{\bsl{Q}}a_{\bsl{Q}} , v_1 a_{v_1}}\delta_{v_{\bsl{Q}}a_{\bsl{Q}} , \overline{v}_1 \overline{a}_{\overline{v}_1} }  \delta_{v_{\bsl{Q}'}' a_{\bsl{Q}'}', v_2 a_{v_2}'}  \delta_{v_{\bsl{Q}'}' a_{\bsl{Q}'}', \overline{v}_2  \overline{a}_{\overline{v}_2}'} \delta_{ \bsl{Q} , \bsl{Q}' + \bsl{p} }  \delta_{ \bsl{Q} , \bsl{Q}' + \overline{\bsl{p}} }   (\bsl{k}-\bsl{Q})_{x}^{m_x} (\bsl{k}-\bsl{Q})_{y}^{m_y} (\bsl{k}-\bsl{Q})_{x}^{\bar{m}_x} (\bsl{k}-\bsl{Q})_{y}^{\bar{m}_y}  \\
& \qquad \times (\bsl{k}-\bsl{Q}')_{x}^{n_x} (\bsl{k}-\bsl{Q}')_{y}^{n_y}  (\bsl{k}-\bsl{Q}')_{x}^{\bar{n}_x} (\bsl{k}-\bsl{Q}')_{y}^{\bar{n}_y} \\
& = \delta_{v_1 a_{v_1} , \overline{v}_1 \overline{a}_{\overline{v}_1} }  \delta_{ v_2 a_{v_2}', \overline{v}_2  \overline{a}_{\overline{v}_2}'}  \delta_{ \bsl{p} ,  \overline{\bsl{p}} } \sum_{\bsl{Q} a_{\bsl{Q}}, \bsl{Q}' a_{\bsl{Q}'}'} \delta_{v_{\bsl{Q}}a_{\bsl{Q}} , v_1 a_{v_1}}  \delta_{v_{\bsl{Q}'}' a_{\bsl{Q}'}', v_2 a_{v_2}'}  \delta_{ \bsl{Q} , \bsl{Q}' + \bsl{p} }    (\bsl{k}-\bsl{Q})_{x}^{m_x} (\bsl{k}-\bsl{Q})_{y}^{m_y} (-\bsl{Q})_{x}^{\bar{m}_x} (-\bsl{Q})_{y}^{\bar{m}_y}  \\
& \qquad \times (\bsl{k}-\bsl{Q}')_{x}^{n_x} (\bsl{k}-\bsl{Q}')_{y}^{n_y}  (\bsl{k}-\bsl{Q}')_{x}^{\bar{n}_x} (\bsl{k}-\bsl{Q}')_{y}^{\bar{n}_y} \\
& = 0 \text{ for $(v_1 a_{v_1},v_2  a'_{v_2} ,\bsl{p} ) \neq (\overline{v}_1 \overline{a}_{\overline{v}_1},\overline{v}_2  \overline{a}_{\overline{v}_2}' , \overline{\bsl{p}})$,}
}
meaning that $X^{\bar{m}_x \bar{m}_y \bar{n}_x \bar{n}_y}_{\overline{v}_1 \overline{a}_{\overline{v}_1},\overline{v}_2  \overline{a}_{\overline{v}_2}' , \overline{\bsl{p}} }(\bsl{k})$ and $X^{m_x m_y n_x n_y}_{v_1 a_{v_1},v_2  a'_{v_2} , \bsl{p} }(\bsl{k})$ are orthogonal for $(v_1 a_{v_1},v_2  a'_{v_2} ,\bsl{p} ) \neq (\overline{v}_1 \overline{a}_{\overline{v}_1},\overline{v}_2  \overline{a}_{\overline{v}_2}' , \overline{\bsl{p}})$. 
However, for the same $(v_1 a_{v_1},v_2  a'_{v_2} ,\bsl{p} )$,  we have
\eqa{
\label{eq:linear_dependence_t}
 & \sum_{m_x m_y n_x n_y\in\dsN} t_{m_x m_y n_x n_y} X^{m_x m_y n_x n_y}_{v_1 a_{v_1},v_2  a'_{v_2} , \bsl{p} }(\bsl{k}) = 0 \\
 & \Leftrightarrow \sum_{m_x m_y n_x n_y\in\dsN} t_{m_x m_y n_x n_y}  \delta_{v_{\bsl{Q}}a_{\bsl{Q}} , v_1 a_{v_1}} \delta_{v_{\bsl{Q}'}' a_{\bsl{Q}'}', v_2 a_{v_2}'}   (\bsl{k}-\bsl{Q})_{x}^{m_x} (\bsl{k}-\bsl{Q})_{y}^{m_y}  \delta_{ \bsl{Q}  , \bsl{Q}' + \bsl{p} } (\bsl{k}-\bsl{Q}')_{x}^{n_x} (\bsl{k}-\bsl{Q}')_{y}^{n_y} = 0 \\
 & \qquad \text{ for all $\bsl{k}$, $\bsl{Q}$, $a_{\bsl{Q}}$, $\bsl{Q}'$, $a_{\bsl{Q}'}'$ }\\
 & \Leftrightarrow \sum_{m_x m_y n_x n_y\in\dsN} t_{m_x m_y n_x n_y} q_{x}^{m_x} q_{y}^{m_y} (\bsl{q}-\bsl{p})_{x}^{n_x} (\bsl{q}-\bsl{p})_{y}^{n_y}  = 0 \text{ for all $\bsl{q}\in\dsR^2$ } \\
 & \Leftrightarrow \sum_{m_x m_y n_x n_y\in\dsN} t_{m_x m_y n_x n_y} Q_{x}^{m_x} Q_{y}^{m_y} \sum_{l_x=0}^{n_x}\sum_{l_y=0}^{n_y}Q_x^{n_x-l_x} p_x^{l_x} C_{n_x}^{l_x}  Q_y^{n_y-l_y} p_y^{l_y} C_{n_y}^{l_y}  = 0 \text{ for all $\bsl{q}\in\dsR^2$ } \\
 & \Leftrightarrow \sum_{m_x m_y n_x n_y l_x l_y\in\dsN} C_{n_x+l_x}^{l_x} (-p_x)^{l_x} C_{n_y+l_y}^{l_y}  (-p_y)^{l_y}   t_{m_x , m_y , n_x + l_x, n_y+ l_y} q_{x}^{m_x} q_{y}^{m_y} q_x^{n_x}   q_y^{n_y}  = 0 \text{ for all $\bsl{q}\in\dsR^2$ } \\
 & \Leftrightarrow \sum_{M_x M_y\in\dsN} \left[ \sum_{0\leq n_x\leq M_x} \sum_{0 \leq  n_y\leq M_y }  \sum_{l_x l_y\in\dsN} C_{n_x+l_x}^{l_x} (-p_x)^{l_x} C_{n_y+l_y}^{l_y}  (-p_y)^{l_y}   t_{M_x - n_x, M_y -n_y, n_x + l_x, n_y+ l_y} \right] q_{x}^{M_x} q_{y}^{M_y}   = 0 \text{ for all $\bsl{q}\in\dsR^2$ } \\
 & \Leftrightarrow \sum_{0\leq n_x\leq M_x} \sum_{0 \leq  n_y\leq M_y }   \sum_{l_x l_y\in\dsN} C_{n_x+l_x}^{l_x} (-p_x)^{l_x} C_{n_y+l_y}^{l_y}  (-p_y)^{l_y}   t_{M_x - n_x, M_y -n_y, n_x + l_x, n_y+ l_y} = 0 \text{ for all $M_x\in \dsN$ , $M_y\in \dsN$ }\ ,
}
where $C_{m}^n$ is the binomial $m$ choose $n$, and we use $q_{x}^{M_x} q_{y}^{M_y}$ are linearly independent for different powers $M_x$ and $M_y$ (they are just multipole expansions) for the last step. 
It means that as long as $t_{m_x m_y n_x n_y}$ satisfies the last equation, the coefficients are constrained such that there is only one independent parameter among all $t_{m_x m_y n_x n_y}$'s for fixed $m_x+n_x = M_x$ and  $m_y+n_y = M_y$.
In order to avoid this redundancy, we re-wrtie the continuum expression of $h_{\text{DFT}}(\bsl{k})$ in \cref{eq:h_DFT_continuum_expansion} as
\eqa{
\label{eq:h_DFT_continuum_expansion_linearly_independent}
\left[ h_{\text{DFT}}(\bsl{k}) \right]_{\bsl{Q} a_{\bsl{Q}}, \bsl{Q}' a_{\bsl{Q}'}'} & = \sum_{m_x, m_y, n_x, n_y \in \dsN} \sum_{v_1 v_2}  \sum_{\bsl{G}_M} \sum_{a_{v1} a
_{v_2}'}t^{m_x m_y n_x n_y}_{v_1 a_{v_1},v_2  a'_{v_2} , \bsl{G}_M }  \delta_{v_{\bsl{Q}}a_{\bsl{Q}} , v_1 a_{v_1}} \delta_{v_{\bsl{Q}'}' a_{\bsl{Q}'}', v_2 a_{v_2}'}   (\bsl{k}-\bsl{Q})_{x}^{m_x}(\bsl{k}-\bsl{Q})_{y}^{m_y}  \\
& \qquad \times \delta_{ \bsl{Q} , \bsl{Q}' + \bsl{q}_{v_1} -\bsl{q}_{v_2}  + \bsl{G}_M  } (\bsl{k}-\bsl{Q}')_x^{n_x}(\bsl{k}-\bsl{Q}')_y^{n_y} \\
& = \sum_{m_x, m_y, n_x, n_y \in \dsN} \sum_{v_1 v_2}  \sum_{\bsl{G}_M} \sum_{a_{v1} a
_{v_2}'}t^{m_x m_y n_x n_y}_{v_1 a_{v_1},v_2  a'_{v_2} , \bsl{G}_M }  \delta_{v_{\bsl{Q}}a_{\bsl{Q}} , v_1 a_{v_1}} \delta_{v_{\bsl{Q}'}' a_{\bsl{Q}'}', v_2 a_{v_2}'}   (\bsl{k}-\bsl{Q})_{x}^{m_x}(\bsl{k}-\bsl{Q})_{y}^{m_y}  \\
& \qquad \times \delta_{ \bsl{Q} , \bsl{Q}' + \bsl{q}_{v_1} -\bsl{q}_{v_2}  + \bsl{G}_M  } (\bsl{k}-\bsl{Q}-(\bsl{q}_{v_1} -\bsl{q}_{v_2}  + \bsl{G}_M))_x^{n_x}(\bsl{k}-\bsl{Q}-(\bsl{q}_{v_1} -\bsl{q}_{v_2}  + \bsl{G}_M))_y^{n_y} \\
& = \sum_{M_x, M_y \in \dsN} \sum_{v_1 v_2}  \sum_{\bsl{G}_M} \sum_{a_{v1} a
_{v_2}'}r^{M_x M_y}_{v_1 a_{v_1},v_2  a'_{v_2} , \bsl{G}_M }  \delta_{v_{\bsl{Q}}a_{\bsl{Q}} , v_1 a_{v_1}} \delta_{v_{\bsl{Q}'}' a_{\bsl{Q}'}', v_2 a_{v_2}'}   (\bsl{k}-\bsl{Q})_{x}^{M_x}(\bsl{k}-\bsl{Q})_{y}^{M_y}  \\
& \qquad \times \delta_{ \bsl{Q} , \bsl{Q}' + \bsl{q}_{v_1} -\bsl{q}_{v_2}  + \bsl{G}_M  } \\
& = \sum_{M_x, M_y \in \dsN} \sum_{v_1 v_2}  \sum_{\bsl{G}_M} \sum_{a_{v1} a
_{v_2}'}r^{M_x M_y}_{v_1 a_{v_1},v_2  a'_{v_2} , \bsl{G}_M }  [Y^{M_x M_y}_{v_1 a_{v_1},v_2  a'_{v_2} , \bsl{q}_{v_1} -\bsl{q}_{v_2}  + \bsl{G}_M }(\bsl{k})]_{\bsl{Q} a_{\bsl{Q}}, \bsl{Q}' a_{\bsl{Q}'}'} \ ,
}
where
\eq{
 r^{M_x M_y}_{v_1 a_{v_1},v_2  a'_{v_2} , \bsl{G}_M } = \left. \sum_{0\leq n_x\leq M_x} \sum_{0 \leq  n_y\leq M_y }   \sum_{l_x l_y\in\dsN} C_{n_x+l_x}^{l_x} (-p_x)^{l_x} C_{n_y+l_y}^{l_y}  (-p_y)^{l_y}   t^{M_x - n_x, M_y -n_y, n_x + l_x, n_y+ l_y}_{v_1 a_{v_1},v_2  a'_{v_2} , \bsl{G}_M }  \right|_{\bsl{p} = \bsl{q}_{v_1} -\bsl{q}_{v_2}  + \bsl{G}_M } \ ,
}
is derived in the same way as \cref{eq:linear_dependence_t}, and
\eq{
 [Y^{M_x M_y}_{v_1 a_{v_1},v_2  a'_{v_2} , \bsl{p} }(\bsl{k})]_{\bsl{Q} a_{\bsl{Q}}, \bsl{Q}' a_{\bsl{Q}'}'}  =  \delta_{v_{\bsl{Q}}a_{\bsl{Q}} , v_1 a_{v_1}} \delta_{v_{\bsl{Q}'}' a_{\bsl{Q}'}', v_2 a_{v_2}'}   (\bsl{k}-\bsl{Q})_{x}^{M_x}(\bsl{k}-\bsl{Q})_{y}^{M_y}  \delta_{ \bsl{Q} , \bsl{Q}' +   \bsl{p}  } \ ,
}
Our goal is therefore to determine $ r^{M_x M_y}_{v_1 a_{v_1},v_2  a'_{v_2} , \bsl{G}_M }$'s, which are the \emph{linearly independent} values of $t^{m_x m_y n_x n_y}_{v_1 a_{v_1},v_2  a'_{v_2} , \bsl{G}_M } $'s.

We now show that it is possible to determine the coefficients just from $h_{\text{DFT}}(\bsl{k}_0)$ at a fixed $\bsl{k}_0$. 
The cornerstone of this is the fact that the set of $Y^{M_x M_y}_{v_1 a_{v_1},v_2  a'_{v_2} , \bsl{p} }(\bsl{k}_0)$'s are linearly independent. 
First, $Y^{M_x M_y}_{v_1 a_{v_1},v_2  a'_{v_2} , \bsl{p} }(\bsl{k}_0)$ are orthogonal for different $(v_1 a_{v_1},v_2  a'_{v_2} , \bsl{G}_M)$ just as \cref{eq:X_vap_orthogonal}. 
Second, for the same $(v_1 a_{v_1},v_2  a'_{v_2} , \bsl{G}_M)$, we have 
\eqa{
\label{eq:linear_independence_r}
 & \sum_{M_x M_y \in\dsN} r_{M_x M_y} Y^{M_x M_y}_{v_1 a_{v_1},v_2  a'_{v_2} , \bsl{p} }(\bsl{k}_0) = 0 \\
 & \Leftrightarrow \sum_{M_x M_y \in\dsN} r_{M_x M_y} \delta_{v_{\bsl{Q}}a_{\bsl{Q}} , v_1 a_{v_1}} \delta_{v_{\bsl{Q}'}' a_{\bsl{Q}'}', v_2 a_{v_2}'}   (\bsl{k}_0-\bsl{Q})_{x}^{M_x}(\bsl{k}_0-\bsl{Q})_{y}^{M_y}  \delta_{ \bsl{Q} , \bsl{Q}' +   \bsl{p}  } =0\text{ for all $\bsl{k}$, $\bsl{Q}$, $a_{\bsl{Q}}$, $\bsl{Q}'$, $a_{\bsl{Q}'}'$ }\\
 & \Leftrightarrow \sum_{M_x M_y \in\dsN} r_{M_x M_y} (k_{0,x}-Q_{x})^{M_x}(k_{0,y}-Q_{y})^{M_y} =0 \text{ for all $\bsl{Q}\in \{ \bsl{G}_M\}  + \bsl{q}_{v_1}$ } \\
 & \Leftrightarrow r_{M_x M_y} = 0 \text{ for all $M_x M_y \in\dsN$}\ ,
}
where the last step can be proven by induction as follows.
\eq{
\label{eq:r_M_L}
\sum_{M_x=0,1,...,L_x} \sum_{ M_y = 0,1,...,L_y} r_{M_x M_y}   (k_{0,x}-Q_{x})^{M_x}(k_{0,y}-Q_{y})^{M_y}\text{ for all $\bsl{Q}\in \{ \bsl{G}_M\}  + \bsl{q}_{v_1}$ }  \Leftrightarrow r_{M_x M_y} = 0 \text{ for all $M_x M_y \in\dsN$} \ ,
}
trivially holds for $L_x=L_y = 0$. If \cref{eq:r_M_L} holdes for certain $(L_x,L_y)$, then it should also hold for $(L_x+1,L_y)$ since the extra $(k_{0,x}-Q_{x})^{L_x+1} (k_{0,y}-Q_y)^{L_y}$ can have larger absolute values than the other terms (which has less $Q_x$ powers) by as much as we want, since we can always choose $(k_{0,y}-Q_y)\neq 0$ and increase $Q_x$. 
Similarly, if \cref{eq:r_M_L} holds for certain $(L_x,L_y)$, then it should also hold for $(L_x,L_y+1)$. 
Therefore, the last step of \cref{eq:linear_independence_r} is proven. 
As a result, we have shown that the set of $Y^{M_x M_y}_{v_1 a_{v_1},v_2  a'_{v_2} , \bsl{p} }(\bsl{k}_0)$'s are linearly independent for a fixed $\bsl{k}_0$, allowing us to determine $ r^{M_x M_y}_{v_1 a_{v_1},v_2  a'_{v_2} , \bsl{G}_M }$ from $h_{\text{DFT}}(\bsl{k}_0)$ in principle. 

The above discussion assumes an infinite lattice of $\bsl{G}_M$. 
In practice, we only have finite $\bsl{G}_M$'s, and yet we only need to include a finite number of $M_x$ and $M_y$ to capture the low-energy physics. 
Therefore, we are still allowed to determine the meaningful $ r^{M_x M_y}_{v_1 a_{v_1},v_2  a'_{v_2} , \bsl{G}_M }$ from $h_{\text{DFT}}(\bsl{k}_0)$, as long as the needed $M_x$ and $M_y$ is not too large compared to the cutoff of $\bsl{G}_M$. 
Specifically, we choose a range of $(M_x, M_y, v_1 a_{v_1},v_2  a'_{v_2} , \bsl{G}_M)$ of interest, which give us a set of $Y^{M_x M_y}_{v_1 a_{v_1},v_2  a'_{v_2} , \bsl{q}_{v_1} -\bsl{q}_{v_2}  + \bsl{G}_M  }(\bsl{k}_0)$. 
We need to orthonormalize those $Y^{M_x M_y}_{v_1 a_{v_1},v_2  a'_{v_2} , \bsl{q}_{v_1} -\bsl{q}_{v_2}  + \bsl{G}_M  }(\bsl{k}_0)$'s, e.g., by Gram–Schmidt process, and obtain a set of new $\widetilde{Y}^{M_x M_y}_{v_1 a_{v_1},v_2  a'_{v_2} , \bsl{p} }(\bsl{k}_0)$ that are related to $Y^{M_x M_y}_{v_1 a_{v_1},v_2  a'_{v_2} , \bsl{q}_{v_1} -\bsl{q}_{v_2}  + \bsl{G}_M  }(\bsl{k}_0)$ by
\eq{\label{eq: othogonalization}
\widetilde{Y}^{M_x M_y}_{v_1 a_{v_1},v_2  a'_{v_2} , \bsl{q}_{v_1} -\bsl{q}_{v_2}  + \bsl{G}_M  }(\bsl{k}_0) = \sum_{\widetilde{M}_x \widetilde{M}_y} Y^{\widetilde{M}_x \widetilde{M}_y}_{v_1 a_{v_1},v_2  a'_{v_2} , \bsl{q}_{v_1} -\bsl{q}_{v_2}  + \bsl{G}_M  }(\bsl{k}_0) z_{\widetilde{M}_x \widetilde{M}_y, M_x M_y}^{v_1 a_{v_1},v_2  a'_{v_2} ,  \bsl{G}_M } \ ,
}
and that satisfy
\eq{
\Tr\left\{ \left[\widetilde{Y}^{M_x M_y}_{v_1 a_{v_1},v_2  a'_{v_2} , \bsl{q}_{v_1} -\bsl{q}_{v_2}  + \bsl{G}_M   }(\bsl{k}_0)\right]^\dagger \widetilde{Y}^{M_x' M_y'}_{v_1 a_{v_1},v_2  a'_{v_2} , \bsl{q}_{v_1} -\bsl{q}_{v_2}  + \bsl{G}_M   }(\bsl{k}_0) \right\} = \delta_{M_x M_y , M_x' M_y'} \ .
}
Here we don't need to orthonormalize $(v_1 a_{v_1},v_2  a'_{v_2} , \bsl{p} )$ indices because they are already orthogonal. Then, we can obtain
\eq{\label{eq: coefficient_from_DFT}
\widetilde{r}^{M_x M_y}_{v_1 a_{v_1},v_2  a'_{v_2} , \bsl{p} } = \Tr\{ \widetilde{Y}^{M_x M_y}_{v_1 a_{v_1},v_2  a'_{v_2} , \bsl{q}_{v_1} -\bsl{q}_{v_2}  + \bsl{G}_M  }(\bsl{k}_0)  h_{\text{DFT}}(\bsl{k}_0)\} \ ,
}
resulting in 
\eq{\label{eq: coefficient_from_transformation}
r^{M_x M_y}_{v_1 a_{v_1},v_2  a'_{v_2} , \bsl{p} } = \sum_{\widetilde{M}_x \widetilde{M}_y} z_{M_x M_y, \widetilde{M}_x \widetilde{M}_y}^{v_1 a_{v_1},v_2  a'_{v_2} ,  \bsl{G}_M } \widetilde{r}^{\widetilde{M}_x \widetilde{M}_y}_{v_1 a_{v_1},v_2  a'_{v_2} , \bsl{q}_{v_1} -\bsl{q}_{v_2}  + \bsl{G}_M  } \ .
}
With $r^{M_x M_y}_{v_1 a_{v_1},v_2  a'_{v_2} , \bsl{G}_M  }$, we have a continuum model 
\eq{
\label{eq:h_cont}
h_{cont}(\bsl{k}) = \sum_{M_x, M_y } \sum_{v_1 v_2}  \sum_{\bsl{G}_M} \sum_{a_{v1} a
_{v_2}'}r^{M_x M_y}_{v_1 a_{v_1},v_2  a'_{v_2} , \bsl{G}_M }  Y^{M_x M_y}_{v_1 a_{v_1},v_2  a'_{v_2} , \bsl{q}_{v_1} -\bsl{q}_{v_2}  + \bsl{G}_M }(\bsl{k}) \ ,
}
where  $(M_x, M_y, v_1 a_{v_1},v_2  a'_{v_2} , \bsl{G}_M)$ is summed only over a range of interested values in contrast to all values in \cref{eq:h_DFT_continuum_expansion_linearly_independent}.
We may speed up the determination of $r^{M_x M_y}_{v_1 a_{v_1},v_2  a'_{v_2} , \bsl{G}_M }$ by summarizing $Y^{M_x M_y}_{v_1 a_{v_1},v_2  a'_{v_2} , \bsl{q}_{v_1} -\bsl{q}_{v_2}  + \bsl{G}_M }(\bsl{k})$ based on hermiticity and symmetries. 
We emphasize that the construction of \eqnref{eq:h_cont} requires no fitting, and thus the more terms (labelled by $(M_x, M_y, v_1 a_{v_1},v_2  a'_{v_2} , \bsl{G}_M)$) included, the more accurate model we have since there is no overfitting issue here. 
In practice, we can include a large number of terms to keep \cref{eq:h_cont} very precise, which can be easily used in numerical calculations.
Of course, one can use more than one $\bsl{k}_0$ point in the procedure---using a direct sum of $h_{\text{DFT}}(\bsl{k})$ at more than one $\bsl{k}_0$ in the replacement of $h_{\text{DFT}}(\bsl{k}_0)$.

However, for many analytical applications, it may be desirable to retain only a limited set of $\bsl{Q}$ points that capture the essential low-energy behavior while ignoring higher-energy contributions. 
This can be achieved by projecting out the outer-shell $\bsl{Q}$ points from $h_{\text{DFT}}(\bsl{k})$ in \cref{eq:h_DFT} by perturbation theory. The resulting reduced Hamiltonian $h_{\text{DFT}}^{\text{eff}}(\bsl{k})$, lives on a much smaller $\bsl{Q}$ points. 
By applying the same construction procedure to $h_{\text{DFT}}^{\text{reduced}}(\bsl{k})$, we derive an effective continuum model Hamiltonian $h_{cont}^{\text{reduced}}(\bsl{k})$ that captures the low-energy physics 
 (it will be only precise for much fewer bands) with a much smaller number of terms. This simplified model enables analytical exploration of low-energy behaviors while retaining fidelity to the original DFT calculations.

In the following section, we apply this general method to AA-stacked {\tmt}, demonstrating the versatility and accuracy of this continuum model construction method.

\section{Constructing continuum model of AA-stacked {\tmt} and {\tws}}\label{construct tmt}

\begin{figure}[!t]
    \centering
    \includegraphics[width=\linewidth]{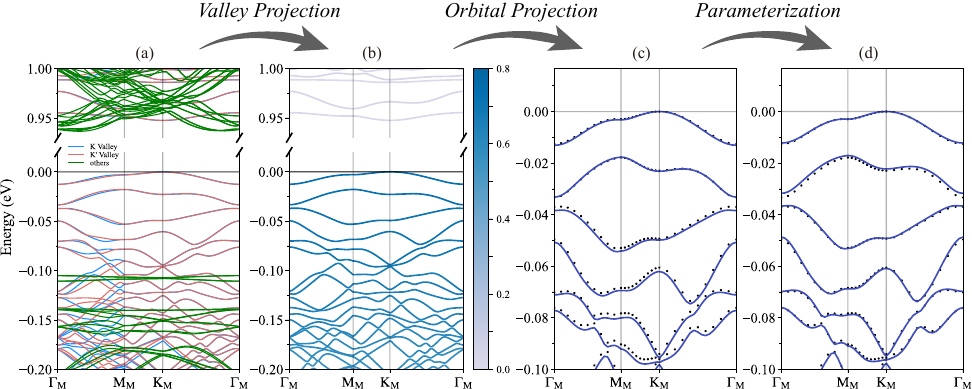}
    \caption{\textbf{DFT to Full Model.} 
    (a) Band structure of the full Hamiltonian for 3.89$\degree${\tmt} computed using OpenMX. Blue lines denote the K valley, red lines represent the K$^{\prime}$ valley bands, and green lines correspond to the bands of other valleys, including the $\Gamma$ valley. 
    (b) Valley-projected band structure obtained via the Truncated Atomic Plane Wave (TAPW) method, exclusively displaying the K valley. The color intensity indicates the contribution of Mo's $d_{xy}$ and $d_{x^2+y^2}$ orbitals to each band. 
    (c) Orbital projection focusing solely on Mo's $d_{xy}$ and $d_{x^2+y^2}$ orbitals.
    (d) Band structure derived from the full continuum model. 
    In subfigures (c) and (d), the black dotted lines correspond to the DFT results for comparison.
    }
    \label{fig:model_contruction_full}
\end{figure}

In this section, we apply the general method introduced in \cref{general method} to construct continuum models for AA-stacked \tmt\ and \tws\ in the \K\ valley. 
Due to the similarities between \tmt\ and \tws, we will primarily discuss the procedure using \tmt\ as an example while presenting results applicable to both materials. 
In AA-stacked bilayer \tmt, the low-energy moir\'e bands originate from the highest valence bands around the $K$ and $-K$ valleys in both layers. 
We focus on the continuum model of the $K$ valley in this section because the $\pm K$ valleys are related by time-reversal symmetry, and the $\Gamma$-valley bands lie well below the charge neutrality point.

We construct two types of continuum models: the first is based on the full $\bsl{Q}$ lattice used in the DFT Hamiltonian, and the second is a simplified model on a reduced $\bsl{Q}$ lattice with fewer terms.

\subsection{Procedure for Constructing the Continuum Model}

Below, we outline the steps for constructing a continuum model from $H_{\text{DFT}}$ on both the full and reduced $\bsl{Q}$ lattices.

\begin{enumerate}

\item \textbf{Obtain an Effective Hamiltonian in the Low-Energy Subspace Using Second-order Perturbation}\label{step1}

For the $K$ valley of \tmt, the sub-valley index is equivalent to the layer index, as the two layers possess distinct $\K_l$. 
Consequently, the basis of the DFT Hamiltonian, $c^\dagger_{\bsl{k}+\K_0 , \bsl{Q},\alpha_{\bsl{Q}}} \equiv c^\dagger_{\bsl{k}+\K_0 +\bsl{q}_{v_{\bsl{Q}}} -\bsl{Q},\alpha_{\bsl{Q}}}$, simplifies to $c^\dagger_{\bsl{k}+K_l- \bsl{Q},\alpha}$, where $l_{\bsl{Q}}$ is denoted as $l$ for simplicity. 
The full continuum model basis is thus represented as $\psi_{\bsl{k},\bsl{Q},a}^{\dagger}$, where $\alpha$ encompasses the 71 orbitals (Mo-$s3p2d2$ and Te-$s3p2d2f1$) with spin-$\uparrow$ in one monolayer unit cell. 
However, we are interested only in the low-energy states closest to the Fermi energy at each $\bsl{Q}$ site that form the low-energy moir\'e bands. 
We denote the number of $\bsl{Q}$ sites as $N_{\bsl{Q}}$.

To isolate these states, we diagonalize each $\bsl{Q}$-block of $H_{\text{DFT}}$ and identify the low-energy states, labeled as $U^{\bsl{k},\bsl{Q}}_{\text{low}}$, which are predominantly composed of $d_{x^2+y^2,\uparrow} + \ii d_{xy,\uparrow}$ orbitals from Mo atoms. 
As shown in \cref{fig:combined_figures}, for each $\bsl{k}$ and $\bsl{Q}$, there is usually one eigenvalue $\lambda_{n}^{\bsl{k},\bsl{Q}}$ near the Fermi level. We denote the corresponding low-energy eigenvector as $U_{\text{low}}^{\bsl{k},\bsl{Q}}$. With the remaining branches approximately about 1 eV away from the Fermi energy, we focus solely on the low-energy branch for constructing the low-energy basis: 
\eq{
\psi^\dagger_{\bsl{k},\bsl{Q}} = \sum_{\alpha_{\bsl{Q}}} c^\dagger_{\bsl{k}+\K_l - \bsl{Q},\alpha_{\bsl{Q}}} \left[ U_{\text{low}}^{\bsl{k},\bsl{Q}} \right]_{\alpha_{\bsl{Q}}} \ ,
\label{eq:dft_basis_to_low}
}

The remaining high-energy states are labeled as $U_{\text{high}}^{\bsl{k},\bsl{Q}}$. 
Given that the phases of $U^{\bsl{k},\bsl{Q}}$ vary randomly across the $\bsl{Q}$ sites, we fix the gauge by ensuring the $\text{d}_{x^2+y^2,\uparrow}$ component of $U_{\text{low}}^{\bsl{k},\bsl{Q}}$ is real and non-positive.

We then partition the Hilbert space into low-energy and high-energy subspaces with corresponding projection matrices $U_0(\bsl{k})$ and $U_1(\bsl{k})$:
\eqa{
U_0(\bsl{k}) = \bigoplus_{\bsl{Q}} U_{\text{low}}^{\bsl{k},\bsl{Q}} \ , \\
U_1(\bsl{k}) = \bigoplus_{\bsl{Q}} U_{\text{high}}^{\bsl{k},\bsl{Q}} \ .
}
 Using $U(\bsl{k})=(U_0(\bsl{k}), U_1(\bsl{k}))$, we transform the DFT Hamiltonian and write it in blocks of four submatrices,
\eqa{
U^\dagger(\bsl{k}) H_{\text{DFT}}(\bsl{k}) U(\bsl{k})&=\mat{U_0^\dagger(\bsl{k}) H_{\text{DFT}}(\bsl{k})U_0(\bsl{k})&U_0^\dagger(\bsl{k}) H_{\text{DFT}}(\bsl{k})U_1(\bsl{k})\\U_1^\dagger(\bsl{k}) H_{\text{DFT}}(\bsl{k})U_0(\bsl{k})&U_1^\dagger(\bsl{k}) H_{\text{DFT}}(\bsl{k})U_1(\bsl{k})}\\
&=\mat{h_{00}(\bsl{k})&h_{01}(\bsl{k})\\h_{10}(\bsl{k})&h_{11}(\bsl{k})} \ ,
}
and the Schr\"odinger equation becomes
\eqa{\label{eq:DFT_Hamiltonian_subspaces}
\mat{h_{00}(\bsl{k})&h_{01}(\bsl{k})\\h_{10}(\bsl{k})&h_{11}(\bsl{k})}\mat{\Psi^{L}(\bsl{k})\\\Psi^{H}(\bsl{k})}=\varepsilon\mat{\Psi^{L}(\bsl{k})\\\Psi^{H}(\bsl{k})} \ .
}
This matrix equation segregates the eigenvector $\Psi(\bsl{k})$ into low-energy $\Psi^{L}(\bsl{k})$ and high-energy $\Psi^{H}(\bsl{k})$ components. 
The corresponding matrix equation for $\Psi(\bsl{k})$ thus separates into two coupled equations for the low-energy and high-energy subspaces::
\eqa{\label{eq:low_energy_space}
h_{00}(\bsl{k})\Psi^L(\bsl{k})+h_{01}(\bsl{k})\Psi^H(\bsl{k})=\varepsilon\Psi^L(\bsl{k}) \ ,
}
\eqa{\label{eq:high_energy_space}
h_{10}(\bsl{k})\Psi^L(\bsl{k})+h_{11}(\bsl{k})\Psi^H(\bsl{k})=\varepsilon\Psi^H(\bsl{k}) \ .
}
We focus on solving $\Psi^H(\bsl{k})$ from the high-energy equation \cref{eq:high_energy_space} and substituting it into the low-energy equation \cref{eq:low_energy_space} to derive an effective Hamiltonian for the low-energy subspace: 
\eqa{
\left[h_{00}(\bsl{k})+h_{01}(\bsl{k})\frac{1}{\varepsilon-h_{11}(\bsl{k})}h_{10}(\bsl{k})\right]\Psi^L(\bsl{k})=\varepsilon\Psi^L(\bsl{k})
}
Thus, we define the effective low-energy Hamiltonian as:
\eqa{
\label{eq:low_energy_Hamiltonian_lowdin_downfolding}
h_{\text{DFT}}(\bsl{k},\varepsilon)=h_{00}(\bsl{k})+h_{01}(\bsl{k})\frac{1}{\varepsilon-h_{11}(\bsl{k})}h_{10}(\bsl{k})\ .
}
To make this Hamiltonian independent of energy, we approximate $\varepsilon$ by the mean eigenvalue $\bar{\lambda}(\bsl{k})$ of $h_{00}(\bsl{k})$. 
Using the basis transformation \cref{eq:dft_basis_to_low}, we express the low-energy effective Hamiltonian in terms of this basis as:
\eqa{
H_{\text{DFT}}^{\text{low-energy}}=\sum_{\bsl{k}}\sum_{\bsl{Q}\bsl{Q}'} 
\psi^\dagger_{\bsl{k},\bsl{Q}}h_{\text{DFT}}(\bsl{k})\psi_{\bsl{k},\bsl{Q}'} \ .
}

This Hamiltonian $H_{\text{DFT}}^{\text{low-energy}}$ now provides a description of the low-energy subspace that is computationally manageable and captures the key physical characteristics necessary for analyzing the moir\'e system.

\begin{figure}[!t]
    \centering
    \begin{minipage}[t]{0.38\linewidth}
        \centering
        \includegraphics[width=\linewidth]{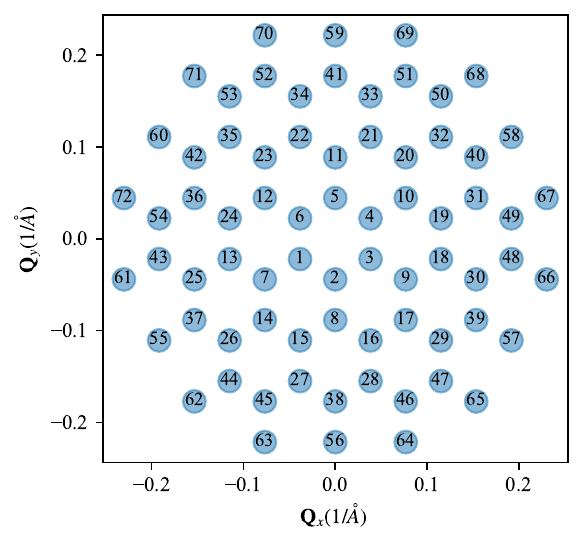}
    \end{minipage}
    \hfill
    \begin{minipage}[t]{0.55\linewidth}
        \centering
        \includegraphics[width=\linewidth]{./fig/block_Q.pdf}
    \end{minipage}
    \caption{ 
    \textbf{Visualization of the $\bsl{Q}$-lattice and its corresponding energy spectrum for the 2.13$^\circ$ {\tmt}}. 
    (a) The $\bsl{Q}$-lattice chosen for the full model at 2.13$^\circ$. The lattice is selected up to the 8th harmonic, with each number representing a specific $\bsl{Q}$-site. 
    (b) Eigenvalues of each $A\times A$ diagonal block of $H_{\text{DFT}}(\bsl{k_0})$. The horizontal axis represents the $\bsl{Q}$-label index, corresponding to the numbered indices shown in panel (a). The vertical axis shows the energy spectrum obtained by diagonalizing the $\bsl{Q}$-blocks from $D_{\bsl{Q}}(\bsl{k_0})$. Here, $\bsl{k_0}$ is chosen at the $\Gamma_M$ point in the moir\'e Brillouin zone. We focus on the low-energy states near the Fermi level.}
    \label{fig:combined_figures}
\end{figure}

\item \textbf{Construct an Effective Hamiltonian on a Reduced $\bsl{Q}$ Sublattice Using Second-order Perturbation}\label{step2}

We can already construct the continuum model for $H_{\text{DFT}}^{\text{low-energy}}$. 
To construct a smaller model, we may want to further reduce the dimension of $h_{\text{DFT}}(\bsl{k})$ and simplify the analytical model, we divide the $\bsl{Q}$ lattice into two parts: $\bsl{Q}_{inside}$, which represents the region we focus on, and $\bsl{Q}_{outside}$. 
Using Second-order Perturbation again, we project $h_{\text{DFT}}(\bsl{k})$ onto the smaller sublattice, i.e., $\bsl{Q}_{inside}$.

We represent $h_{\text{DFT}}(\bsl{k})$ in terms of four submatrices arranged in blocks
\eqa{\label{eq:DFT_Hamiltonian_sublattice}
h_{\text{DFT}}(\bsl{k})=\mat{h'_{00}(\bsl{k})&h'_{01}(\bsl{k})\\h'_{10}(\bsl{k})&h'_{11}(\bsl{k})} \ ,
}
where $h'_{00}(\bsl{k})$ and $h'_{11}(\bsl{k})$ are defined on the $\bsl{Q}_{inside}$ and $\bsl{Q}_{outside}$ lattice, respectively. 
Using a similar approach as step\ref{step1}, we derive the effective Hamiltonian for the smaller $\bsl{Q}$ sublattice:
\eqa{
\label{eq:small_lattice_Hamiltonian_lowdin_downfolding}
h_{\text{DFT}}^{\text{reduced}}(\bsl{k})=h'_{00}(\bsl{k})+h'_{01}(\bsl{k})\frac{1}{\varepsilon_{out}-h'_{11}(\bsl{k})}h'_{10}(\bsl{k}) \ ,
}
with $\varepsilon_{out}$ chosen as the maximum eigenvalue of $h'_{00}(\bsl{k})$ as we only want to focus on the top few energy bands near the Fermi level.
This setup provides a simplified yet effective Hamiltonian that governs the low-energy dynamics within the selected sublattice:
\eqa{
H_{\text{DFT}}^{\text{reduced}}=\sum_{\bsl{k}}\sum_{\bsl{Q}\bsl{Q}'\in \bsl{Q}_{inside}} 
\psi^\dagger_{\bsl{k},\bsl{Q}}h_{\text{DFT}}^{\text{reduced}}(\bsl{k})\psi_{\bsl{k},\bsl{Q}'} \ .
}

\item \textbf{Calculate the Coefficients of Linearly Independent Terms in the General Continuum Model}\label{step3}

We assume that a general continuum model can effectively describe the Hamiltonian in the low-energy subspace, whether $H_{\text{DFT}}^{\text{low-energy}}$ or $H_{\text{DFT}}^{\text{reduced}}$. 
We first express the continuum model in real space:
\eqa{
H_{cont}&=\sum_{m_x, m_y, n_x, n_y \in \dsN} \sum_{l l'} \int d^2 r \left(\ii^m \partial_x^{m_x} \partial_y^{m_y} \psi^\dagger_{\bsl{r},\K_{l}} \right) \tilde{t}^{m_x m_y n_x n_y}_{l ,l' }(\bsl{r}) (-\ii)^n\partial_x^{n_x} \partial_{y}^{n_y}\psi_{\bsl{r},\K_{l'}} \\
&=\sum_{M_x, M_y \in \dsN} \sum_{l l'} \int d^2 r \left(\ii^m \partial_x^{M_x} \partial_y^{M_y} \psi^\dagger_{\bsl{r},\K_{l}} \right) t^{M_x M_y}_{l ,l' }(\bsl{r}) \psi_{\bsl{r},\K_{l'}}\\
&=\sum_{M_z, M_{z^*} \in \dsN} \sum_{l l'} \int d^2 r \left(\ii^m \partial_{z}^{M_{z}} \partial_{z^*}^{M_{z^*}} \psi^\dagger_{\bsl{r},\K_{l}} \right) t^{M_{z} M_{z^*}}_{l ,l' }(\bsl{r}) \psi_{\bsl{r},\K_{l'}} \ ,
\label{eq:continuum_real}
}
where $M_x, M_y$ are indices of the partial derivatives in real space. 
$\psi_{\bsl{k},\K_l,l}^{\dagger}$ and $t_{\K_{l}l,\K_{l'}l'}^{M_zM_{z^*}}$ is simplified as $\psi_{\bsl{k},\K_l}^{\dagger}$ and $t_{l,l'}^{M_zM_{z^*}}$ as discussed in \cref{general method}. 
We use integration by parts and $M_{x,y}=m_{x,y}+n_{x,y}$ in the second line to make sure these terms are linearly independent as proved by \cref{eq:X_vap_orthogonal}-\cref{eq:h_DFT_continuum_expansion_linearly_independent}. 
Different form \cref{eq:general_form_continuum_model}, in the third line, we replace $\partial_{x,y}$ with $\partial_{z,z^*}$ defined by 
\eq{
\partial_z = \partial_x + \ii \partial_y \ ,\ \partial_{z^*} = \partial_x - \ii \partial_y \ .
} 
as \tmt\ has $C_{3z}$ symmetry. 
According to \cref{eq:FT_continuum_basis} and \cref{eq:FT_continuum_coefficient} required by moir\'e translation symmetry, the Fourier transformation of $\psi_{\bsl{r},\K_{l}}(\bsl{r})$ and $t^{M_{z} M_{z^*}}_{l ,l' }(\bsl{r})$ reads, 
\eqa{
\psi^\dagger_{\bsl{r},\K_{l}}(\bsl{r}) &=\sum_{\bsl{k},\bsl{G}_M}e^{-\ii (\bsl{k}-\bsl{G}_M-\bsl{q}_l) \cdot \bsl{r} }\psi^\dagger_{\bsl{k},\bsl{G}_M+\bsl{q}_l} \ ,
\\
t_{ll'}^{M_{z} M_{z^*}}(\bsl{r}) &=  \sum_{\bsl{G}_M} e^{-i\left(\K_l-\K_{l'}+\bsl{G}_M\right)\cdot\bsl{r}}t_{ll',\K_l-\K_{l'}+\bsl{G}_M}^{M_{z} M_{z^*}} \ .
}
Thus, we can obtain the $H_{cont}$ in the momentum space,
\eqa{
H_{cont}&=\sum_{\bsl{k}}\sum_{\bsl{Q}\bsl{Q}'} 
\psi^\dagger_{\bsl{k},\bsl{Q}}\psi_{\bsl{k},\bsl{Q}'} h_{cont}^{\bsl{Q}\bsl{Q}'}(\bsl{k})\\
h_{cont}^{\bsl{Q}\bsl{Q}'}(\bsl{k}) &=  \sum_{M_z, M_{z^*} \in \dsN}  \sum_{ll'}\sum_{\bsl{G}_M} t^{M_z M_{z^*}}_{ll' , \bsl{G}_M }  \left[ Y^{M_z M_{z^*}}_{ll' , \K_l -\K_{l'}  + \bsl{G}_M }(\bsl{k}) \right]_{\bsl{Q}\bsl{Q}'} \ ,
}
where
\eq{
 \left[Y^{M_z M_{z^*}}_{ll' , \bsl{p} }(\bsl{k})\right]_{\bsl{Q}\bsl{Q}'}  =  \delta_{l_{\bsl{Q}},l} \delta_{l_{\bsl{Q}'},l' }   (\bsl{k}-\bsl{Q})_{z}^{M_z}(\bsl{k}-\bsl{Q})_{z^*}^{M_{z^*}}  \delta_{ \bsl{Q} , \bsl{Q}' +   \bsl{p}  } \ ,
}
where $(\bsl{k}-\bsl{Q})_{z}^{M_z}=[(\bsl{k}-\bsl{Q})_x+\ii(\bsl{k}-\bsl{Q})_y]^{M_z}$, $(\bsl{k}-\bsl{Q})_{z^*}^{M_{z^*}}=[(\bsl{k}-\bsl{Q})_x-\ii(\bsl{k}-\bsl{Q})_y]^{M_{z^*}}$, $\bsl{p} = \K_l -\K_{l'}  + \bsl{G}_M$, and $l_{\bsl{Q}} = b $ if $\bsl{Q} \in \text{bottom layer}$; otherwise, $l_{\bsl{Q}} = t$. 
Compared with \cref{eq:h_cont}, and following the analogous derivations analogous to \cref{eq: othogonalization}-\cref{eq: coefficient_from_transformation}, we can obtain:
\eq{
r_{ll',\bsl{G}_M}^{M_xM_y} = \sum_{M_z,M_{z^*} \in \dsN}\sum_{i=0}^{M_z}\sum_{j=0}^{M_{z^*}}t_{ll',\bsl{G}_M}^{M_zM_{z^*}}\ii^{M_z+M_{z^*}-i-j}C_{M_z}^{i}C_{M_{z^*}}^{j}\frac{Y_{ll',\bsl{p}}^{i+j,M_z+M_{z^*}-i-j}(\bsl{k}_0)[Y^{M_xM_y}_{ll',\bsl{p}}(\bsl{k}_0)]^{\dagger}}{\text{Tr}\{Y^{M_xM_y}_{ll',\bsl{p}}(\bsl{k}_0)[Y^{M_xM_y}_{ll',\bsl{p}}(\bsl{k}_0)]^{\dagger}\}} \ ,
}
the coefficients $t^{M_z M_{z^*}}_{ll', \bsl{G}_M}$ are determined by orthogonalizing the terms in $h_{cont}(\bsl{k}_0)$ at a fixed reference $\bsl{k}_0$, typically chosen at the $\Gamma_M$ point in the moir\'e Brillouin zone.

To calculate these coefficients, we perform the Gram–Schmidt orthogonalization process and extract the coefficients from $h_{\text{DFT}}(\bsl{k}_0)$ or $h_{\text{DFT}}^{\text{reduced}}(\bsl{k}_0)$ using \cref{eq: othogonalization}-\cref{eq: coefficient_from_transformation}. 
Multiple $\bsl{k}$-points can be utilized to enhance accuracy.

In the general continuum model, we categorize the terms as follows:
\begin{itemize}
    \item \textbf{Kinetic Terms}: $t^{M_z M_{z^*}}_{ll', \bsl{G}_M}|_{l=l', \bsl{G}_M=0} = \hbar^2/(2m^{M_z M_{z^*}}_{l})$.
    \item \textbf{Intralayer Potential Terms}: $t^{M_z M_{z^*}}_{ll', \bsl{G}_M}|_{l=l', \bsl{G}_M \neq 0} = V^{M_z M_{z^*}}_{l, \bsl{G}_M}$.
    \item \textbf{Interlayer Coupling Terms}: $t^{M_z M_{z^*}}_{ll', \bsl{G}_M}|_{l \neq l'} = w^{M_z M_{z^*}}_{ll', \K_l - \K_{l'} + \bsl{G}_M}$ \ .
\end{itemize}

Thus, the continuum Hamiltonian becomes:
\eqa{
h_{cont}(\bsl{k})=\sum_{M_zM_{z^*}}\sum_{l}\left[\frac{\hbar^2}{2m_{l}^{M_zM_{z^*}}}Y_{ll,0}^{M_zM_{z^*}}(\bsl{k})+\sum_{\bsl{G}_M}V_{l,\bsl{G}_M}^{M_zM_{z^*}}Y_{ll,\bsl{G}_M}^{M_zM_{z^*}}(\bsl{k})+\sum_{\bsl{G}_M}w_{l\bar{l},\bsl{p}}^{M_zM_{z^*}}Y_{l\bar{l},\bsl{p}}^{M_zM_{z^*}}(\bsl{k})\right] \ ,
}
where $\bar{l}$ denotes the layer opposite to $l$. 
In principle, $H_{cont}$ is equivalent to the effective Hamiltonian with infinite $\bsl{G}_M$ and $M_z, M_{z^*}$ terms. 
However, we introduce cutoffs on these parameters to obtain an analytical model that accurately describes the low-energy moir\'e bands.

\end{enumerate}

\subsection{Symmetry Constraints}

The symmetry operations $C_{3z}$ and $C_{2y}\mathcal{T}$ enforce additional constraints on the basis functions involved in the Hamiltonian expansion, restricting their form according to the symmetries of the system. 
For the $C_{3z}$ and $C_{2y}\mathcal{T}$ symmetry, we have:
\eqa{
D[C_{3z}]_{\bsl{Q},\bsl{Q}'}=e^{\ii\frac{2}{3}\pi}\delta_{\bsl{Q},C_{3z}\bsl{Q}'},D[C_{2y}\mathcal{T}]_{\bsl{Q},\bsl{Q}'} = \delta_{ \bsl{Q}, \sigma_3 \bsl{Q}'}  \ ,
}
Under the $C_{3z}$ and $C_{2y}\mathcal{T}$ operation, the basis functions transform as
\eqa{
D[C_{3z}]Y_{ll',\bsl{p}}^{M_zM_{z^*}}(C_{3z}^{-1}\bsl{k})D[C_{3z}]^{-1} = e^{-\ii\frac{2}{3}\pi(M_z-M_{z^*})}Y_{ll',C_{3z}\bsl{p}}^{M_zM_{z^*}}(\bsl{k}) \ ,
}
\eqa{
D[C_{2y}\mathcal{T}][Y_{ll',\bsl{p}}^{M_zM_{z^*}}(\sigma_{3}\bsl{k})]^{*}D[C_{2y}\mathcal{T}]^{-1}=Y_{\bar{l}\bar{l'},\sigma_3\bsl{p}}^{M_zM_z^{*}}(\bsl{k}) \ .
}
The Hermitian conjugate of the basis functions is given by
\eqa{
[Y_{ll',\bsl{p}}^{M_zM_{z^*}}(\bsl{k})]^{\dagger} = Y_{l'l,-\bsl{p}}^{M_{z^*}M_z}(\bsl{k}) \ .
}
\begin{table}[!hbt]
\renewcommand{\arraystretch}{1.8}
    \centering
    \begin{tabular}{>{\centering\arraybackslash}p{2.2cm}>{\centering\arraybackslash}p{5cm}>{\centering\arraybackslash}p{5cm}>{\centering\arraybackslash}p{5cm}}
    \hline
    \hline
           & $C_{3z}$ & $C_{2y}\mathcal{T}$ & $\dagger$ \\
    \hline
    $m$ & $M_z-M_{z^*}=0\,\text{mod}\,3$ & $m_{l}^{M_zM_{z^*}}=m_{\bar{l}}^{M_zM_{z^*}}$ & $m_{l}^{M_zM_{z^*}}=m_{l}^{M_{z^*}M_z}$ \\
    \hline
    $V$ & $V_{l,\bsl{G}_M}^{M_zM_{z^*}}=e^{\ii\frac{2}{3}\pi(M_z-M_{z^*})}V_{l,C_{3z}\bsl{G}_M}^{M_zM_{z^*}}$ & $V_{l,\bsl{G}_M}^{M_zM_{z^*}}=V_{\bar{l},\sigma_3\bsl{G}_M}^{M_zM_{z^*}}$ & $V_{l,\bsl{G}_M}^{M_zM_{z^*}}=V_{l,-\bsl{G}_M}^{M_{z^*}M_z}$ \\
    \hline
    $w$ & $w_{l\bar{l},\bsl{p}}^{M_zM_{z^*}}=e^{\ii\frac{2}{3}\pi(M_z-M_{z^*})}w_{\bar{l}l,C_{3z}\bsl{p}}^{M_zM_{z^*}}$ & $w_{l\bar{l},\bsl{p}}^{M_zM_{z^*}}=w_{\bar{l}l,\sigma_3\bsl{p}}^{M_zM_{z^*}}$ & $w_{l\bar{l},\bsl{p}}^{M_zM_{z^*}}=w_{\bar{l}l,-\bsl{p}}^{M_{z^*}M_z}$ \\
    \hline
    \hline
    \end{tabular}
    \caption{Symmetry conditions for the model parameters under $C_{3z}$ and $C_{2y}\mathcal{T}$ symmetry operations and Hermitian conjugate.}
    \label{table:symmetry_conditions}
\end{table}

\begin{table}[hbt!]
\renewcommand{\arraystretch}{1.8}
    \centering
    \begin{tabular}{>{\centering\arraybackslash}p{2.4cm}>{\centering\arraybackslash}p{2.4cm}>{\centering\arraybackslash}p{2.4cm}>{\centering\arraybackslash}p{2.4cm}>{\centering\arraybackslash}p{2.4cm}>{\centering\arraybackslash}p{2.4cm}>{\centering\arraybackslash}p{2.4cm}}
        \hline
        \hline
        Harmonic & 1 & 2 & 3 & 4 & 5 & 6 \\
        \hline
        Intra & $\mathbf{b}_{M_1}$ & $\mathbf{b}_{M_1} + \mathbf{b}_{M_2}$ & $2\mathbf{b}_{M_1}$ & $2\mathbf{b}_{M_1} + \mathbf{b}_{M_2}$ & $3\mathbf{b}_{M_1}$ & $2\mathbf{b}_{M_1}+2\mathbf{b}_{M_2}$\\
        \hline
        Inter & $\mathbf{q}_1$ & $-2\mathbf{q}_1$ & $\mathbf{q}_1 + \mathbf{b}_{M_2}$ & $2\mathbf{b}_{M_2} + \mathbf{q}_3$ & $4\mathbf{q}_1$ & $\mathbf{q}_1 + 2\mathbf{b}_{M_2}$ \\
        \hline
        \hline
    \end{tabular}
    \caption{
    Intra- and inter-layer harmonics mapping for ${\tmt}$ and ${\tws}$ systems. Each harmonic selects a single $\bsl{p}$ vector, with other terms related by symmetry constraints (\cref{table:symmetry_conditions}). The vectors $\mathbf{b}_{M_1}$, $\mathbf{b}_{M_2}$, $\mathbf{q}_1$, $\mathbf{q}_2$, and $\mathbf{q}_3$ are illustrated in \cref{fig:Gn}.
    }
    \label{tab:harmonics_map}
\end{table}

\subsection{Results and Validation}

\begin{figure}[!t]
    \centering
    \includegraphics[width=\columnwidth]{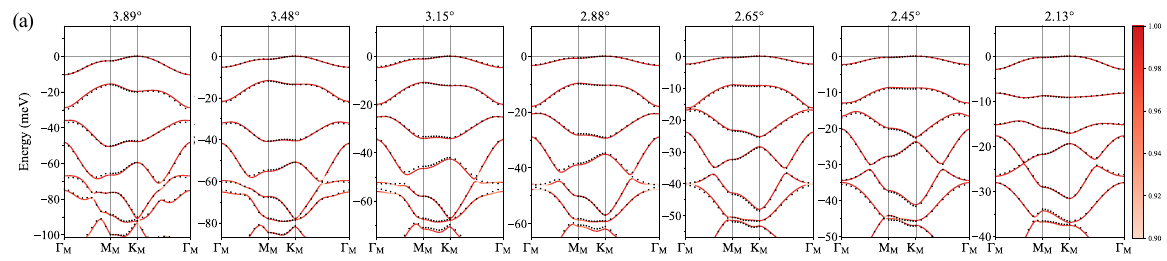}
    \includegraphics[width=\columnwidth]{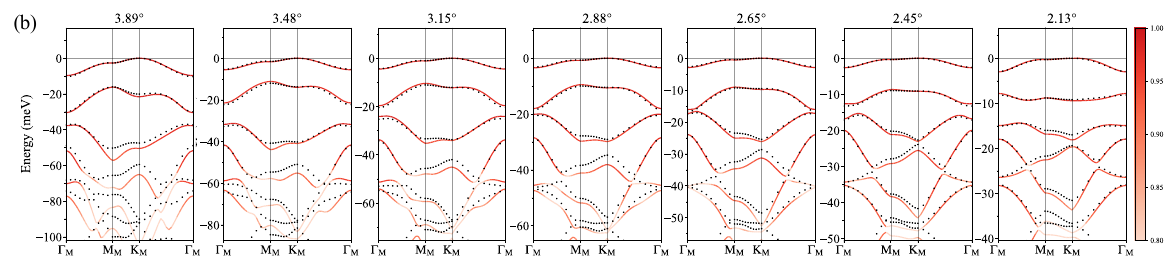}
    \caption{\textbf{Comparison of DFT with quick basis selection and continuum model results for {\tmt}.} (a): Full model. (b): Reduced model. Black dots represent DFT calculations, while red lines indicate continuum model bands. The gradient from dark to light in the red lines illustrates the overlap probability between the model and DFT wavefunctions for each corresponding band. For angles from $2.13^\circ$ to $3.15^\circ$, we choose $\bsl{Q}$ lattice up to the 8th harmonic, while for angles of  $3.48^\circ$ and $3.89^\circ$, we choose $\bsl{Q}$ lattice up to the 7th harmonic.
}
    \label{fig:model_mote2_52orb}
\end{figure}

\begin{figure}[!b]
    \centering
    \includegraphics[width=\columnwidth]{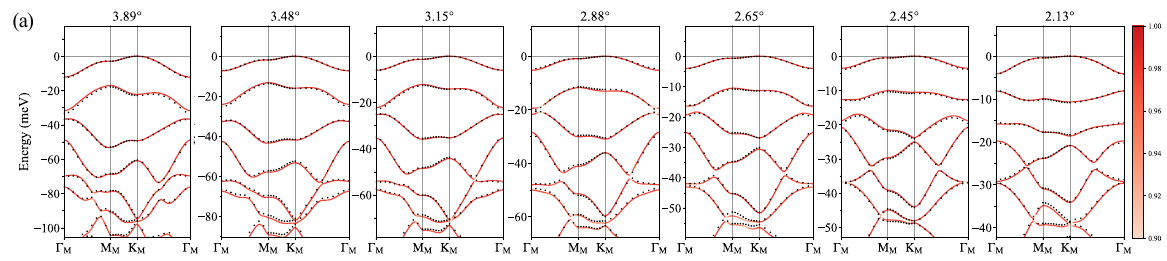}
    \includegraphics[width=\columnwidth]{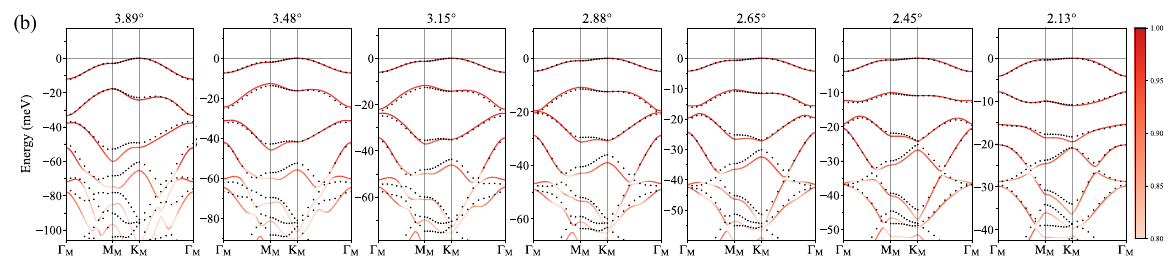}
    \caption{\textbf{Comparison of DFT with standard basis selection and continuum model results for {\tmt}.} (a): Full model. (b): Reduced model. Black dots represent DFT calculations, while red lines indicate continuum model bands. The gradient from dark to light in the red lines illustrates the overlap probability between the model and DFT wavefunctions for each corresponding band. For angles from $2.13^\circ$ to $3.15^\circ$, we choose $\bsl{Q}$ lattice up to the 8th harmonic, while for angles of  $3.48^\circ$ and $3.89^\circ$, we choose $\bsl{Q}$ lattice up to the 7th harmonic.
}
    \label{fig:model_mote2_71orb}
\end{figure}

\begin{figure}[!t]
    \centering
    \includegraphics[width=\columnwidth]{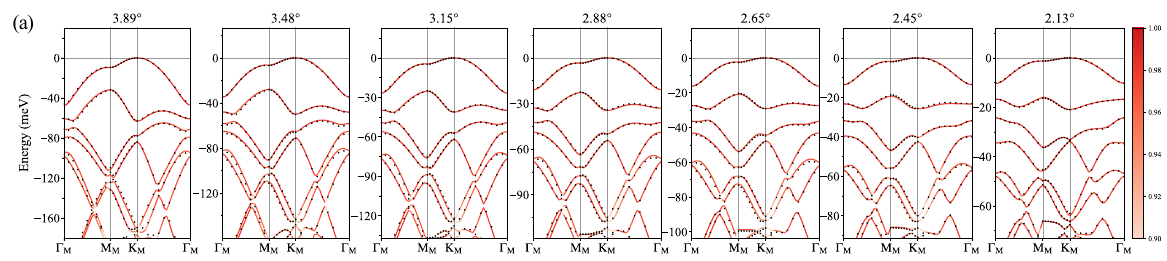}
    \includegraphics[width=\columnwidth]{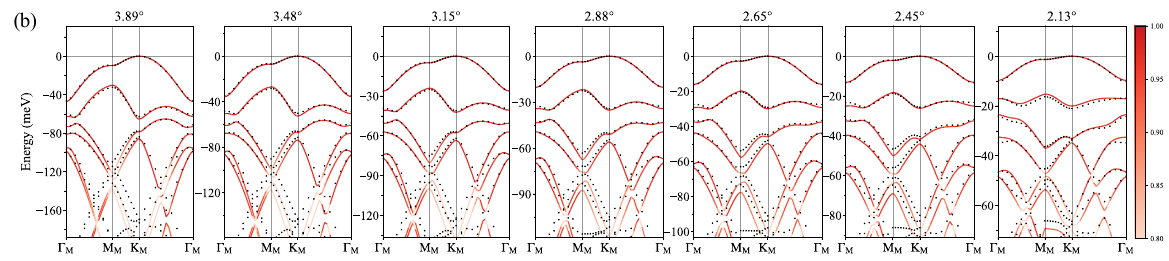}
    \caption{\textbf{Comparison of DFT with quick basis selection and continuum model results for {\tws}.} (a): Full model. (b): Reduced model. Black dots represent DFT calculations, while red lines indicate continuum model bands. The gradient from dark to light in the red lines illustrates the overlap probability between the model and DFT wavefunctions for each corresponding band. For angles from $2.13^\circ$ to $3.15^\circ$, we choose $\bsl{Q}$ lattice up to the 8th harmonic, while for angles from  $3.15^\circ$ to $3.89^\circ$, we choose $\bsl{Q}$ lattice up to the 7th harmonic.
}
    \label{fig:model_wse2_47orb}
\end{figure}

Following Steps \ref{step1} and \ref{step3}, we have developed a continuum model for the full $\bsl{Q}$ lattice used in DFT calculations. 
As illustrated in \cref{fig:model_mote2_52orb,fig:model_mote2_71orb,fig:model_wse2_47orb}(a), we compare the moir\'e bands calculated by DFT with those from our continuum model. This model accurately describes the top valence bands of \tmt\ for twist angles below $3.89^\circ$. 
Additionally, we compute the overlap $\rho_{n}(\bsl{k})$ between the wavefunctions from DFT, $\psi_{n, \bsl{k}}^{\text{DFT}}$, and those from the continuum model, $\psi_{n, \bsl{k}}^{cont}$, defined as:

\begin{equation}
    \rho_{n}(\bsl{k}) = \psi_{n, \bsl{k}}^{cont, \dagger} U_0^\dagger(\bsl{k}) \psi_{n, \bsl{k}}^{\text{DFT}},
\end{equation}
where $n$ denotes the energy band index. 
As shown in \cref{fig:model_mote2_52orb,fig:model_mote2_71orb,fig:model_wse2_47orb}(a), the minimum overlap probability for the top four bands exceeds 95\%, excluding regions where bands touch.

Furthermore, we constructed a continuum model on a reduced $\bsl{Q}$ lattice by limiting the kinetic terms to $O(\bsl{k}^2)$, following Steps \ref{step2} to \ref{step3}. 
As depicted in \cref{fig:model_mote2_52orb,fig:model_mote2_71orb,fig:model_wse2_47orb}(b), this simplification reduces the number of terms at the expense of accuracy for the third and lower valence bands. 

The final Hamiltonian can be simplified by utilizing the symmetry constraints shown in \cref{table:symmetry_conditions}, which allow us to retain only the minimal number of independent terms. 
For each harmonic, a single $\bsl{p}$ is chosen, while the coefficients for other terms of the same degree are related by the symmetry conditions outlined in the table. 
The specific $\bsl{p}$ selected for each harmonic, both intra- and inter-layer, is detailed in \cref{tab:harmonics_map}. Furthermore, for the ${\tmt}$ and ${\tws}$ systems, the parameters used in the full and reduced models are provided in \cref{table:mote2_52orb_full_diag,table:mote2_52orb_full_intra,table:mote2_52orb_full_inter,table:mote2_52orb_reduced_diag,table:mote2_52orb_reduced_intra,table:mote2_52orb_reduced_inter,table:mote2_71orb_full_diag,table:mote2_71orb_full_intra,table:mote2_71orb_full_inter,table:mote2_71orb_reduced_diag,table:mote2_71orb_reduced_intra,table:mote2_71orb_reduced_inter,table:wse2_47orb_full_diag,table:wse2_47orb_full_intra,table:wse2_47orb_full_inter,table:wse2_47orb_reduced_diag,table:wse2_47orb_reduced_intra,table:wse2_47orb_reduced_inter}.

\newpage

\renewcommand{\arraystretch}{1.5} 
\begin{longtable}{c *{7}{S[table-column-width=2cm]}}
    \caption{Kinetic term parameters ($m_\text{e}$) for the full model of {\tmt} based on DFT with quick basis selection}\label{table:mote2_52orb_full_diag} \\
    \hline\hline\hline
    Parameter & {3.89$\degree$} & {3.48$\degree$} & {3.15$\degree$} & {2.88$\degree$} & {2.65$\degree$} & {2.45$\degree$} & {2.13$\degree$} \\ 
    \hline
    \endfirsthead
    
    \caption[]{Kinetic term parameters ($m_\text{e}$) for the full model of {\tmt} based on DFT with quick basis selection (continued)} \\
    \hline\hline\hline
    Parameter & {3.89$\degree$} & {3.48$\degree$} & {3.15$\degree$} & {2.88$\degree$} & {2.65$\degree$} & {2.45$\degree$} & {2.13$\degree$} \\ 
    \hline
    \endhead
    \hline \hline\hline
    \endfoot$m^{(1, 1)}$ & -0.6593 & -0.6537 & -0.6366 & -0.6406 & -0.6646 & -0.6640 & -0.6513 \\ 
        $m^{(3, 0)}$ & 1.4008 & 1.2052 & 1.3362 & 1.3431 & 1.4354 & 1.5619 & 1.4736 \\ 
        $m^{(2, 2)}$ & 0.0739 & 0.0658 & 0.0616 & 0.0631 & 0.0787 & 0.0805 & 0.0705 \\ 
        $m^{(4, 1)}$ & -0.1355 & -0.0880 & -0.1167 & -0.1153 & -0.1525 & -0.2344 & -0.1917 \\ 
        $m^{(3, 3)}$ & -0.0116 & -0.0070 & -0.0073 & -0.0074 & -0.0141 & -0.0179 & -0.0116 \\ 
        $m^{(6, 0)}$ & -7.4620 & -0.1239 & 0.7990 & 0.5579 & 0.7328 & 1.0873 & 0.4931 \\ 
        $m^{(5, 2)}$ & 0.0280 & 0.0137 & 0.0212 & 0.0196 & 0.0456 & 0.0000 & 0.0000 \\ 
        $m^{(4, 4)}$ & 0.0031 & 0.0010 & 0.0014 & 0.0013 & 0.0054 & 0.0000 & 0.0000 \\ 
        $m^{(7, 1)}$ & 0.1055 & 0.0056 & -0.0884 & -0.0472 & -0.0871 & 0.0000 & 0.0000 \\ 
        $m^{(6, 3)}$ & -0.0149 & -0.0063 & -0.0099 & -0.0081 & 0.0000 & 0.0000 & 0.0000 \\ 
        $m^{(9, 0)}$ & -3.0218 & 0.1219 & 1.8229 & -16.8653 & 0.0000 & 0.0000 & 0.0000 \\ 
        $m^{(5, 5)}$ & -0.0018 & -0.0003 & -0.0006 & -0.0006 & 0.0000 & 0.0000 & 0.0000 \\ 
        $m^{(8, 2)}$ & -0.0217 & -0.0011 & 0.0396 & 0.0158 & 0.0000 & 0.0000 & 0.0000 \\ 
        
\end{longtable}
    
\renewcommand{\arraystretch}{2.4} 
\begin{longtable}{c *{7}{>{\centering\arraybackslash}p{2cm}}}

    \caption{Intralayer potential term parameters ($\text{meV}$) for the full model of {\tmt} based on DFT with quick basis selection}\label{table:mote2_52orb_full_intra} \\
    \hline\hline\hline
    Parameter & {3.89$\degree$} & {3.48$\degree$} & {3.15$\degree$} & {2.88$\degree$} & {2.65$\degree$} & {2.45$\degree$} & {2.13$\degree$} \\ 
    \hline
    \endfirsthead
    
    \caption[]{Intralayer potential term parameters ($\text{meV}$) for the full model of {\tmt} based on DFT with quick basis selection (continued)} \\
    \hline\hline\hline
    Parameter & {3.89$\degree$} & {3.48$\degree$} & {3.15$\degree$} & {2.88$\degree$} & {2.65$\degree$} & {2.45$\degree$} & {2.13$\degree$} \\
    \hline
    \endhead
    \hline \hline\hline
    \endfoot
        $V^{0,0}_{b,\bsl{b}_{M_1}}$ & \cellnum{-0.0038}{-3.9103}  & \cellnum{-0.0153}{+4.2608}  & \cellnum{-0.0229}{-3.7636}  & \cellnum{-0.0177}{+3.4576}  & \cellnum{-0.0156}{-3.0791}  & \cellnum{-0.0170}{-2.6988}  & \cellnum{-0.0251}{-1.8658}  \\ 
        $V^{1,0}_{b,\bsl{b}_{M_1}}$ & \cellnum{-1.9224}{-8.8822}  & \cellnum{-0.8617}{+4.8011}  & \cellnum{-1.1931}{-10.1349}  & \cellnum{0.0506}{+1.4581}  & \cellnum{-1.1722}{-10.6454}  & \cellnum{-1.1432}{-10.6991}  & \cellnum{-1.1401}{-10.5061}  \\ 
        $V^{0,1}_{b,\bsl{b}_{M_1}}$ & \cellnum{-0.8895}{-5.2475}  & \cellnum{-2.0266}{+10.4007}  & \cellnum{0.0894}{-2.4862}  & \cellnum{-1.1905}{+10.4842}  & \cellnum{0.0309}{-0.5232}  & \cellnum{0.0200}{+0.2835}  & \cellnum{-0.0857}{+1.6318}  \\ 
        $V^{2,0}_{b,\bsl{b}_{M_1}}$ & \cellnum{3.9339}{-31.2143}  & \cellnum{0.7975}{-43.9430}  & \cellnum{0.9248}{-37.3171}  & \cellnum{-1.5565}{-44.4795}  & \cellnum{1.1120}{-41.5515}  & \cellnum{1.2088}{-42.7443}  & \cellnum{2.0513}{-43.7408}  \\ 
        $V^{1,1}_{b,\bsl{b}_{M_1}}$ & \cellnum{-4.2521}{+16.1930}  & \cellnum{-4.8541}{-18.8353}  & \cellnum{0.1025}{+19.0470}  & \cellnum{-0.6967}{-20.6775}  & \cellnum{-1.5309}{+21.8434}  & \cellnum{-2.2280}{+22.8194}  & \cellnum{-3.3843}{+24.3039}  \\ 
        $V^{0,2}_{b,\bsl{b}_{M_1}}$ & \cellnum{1.2764}{+36.1156}  & \cellnum{3.8105}{+37.7526}  & \cellnum{-1.9716}{+42.3784}  & \cellnum{0.9337}{+39.8775}  & \cellnum{-1.1922}{+45.5950}  & \cellnum{-0.7637}{+46.0822}  & \cellnum{0.5324}{+45.0861}  \\ 
        $V^{3,0}_{b,\bsl{b}_{M_1}}$ & \cellnum{6.1458}{-0.0998}  & \cellnum{0.8080}{-29.8730}  & \cellnum{4.6664}{-3.0046}  & \cellnum{-3.3064}{-22.6237}  & \cellnum{3.5331}{-3.3016}  & \cellnum{2.6259}{-3.5641}  & \cellnum{-0.1372}{-4.0415}  \\
\end{longtable}
    
\renewcommand{\arraystretch}{2.4} 
\begin{longtable}{c *{7}{>{\centering\arraybackslash}p{2cm}}}

    \caption{Interlayer potential term parameters ($\text{meV}$) for the full model of {\tmt} based on DFT with quick basis selection}\label{table:mote2_52orb_full_inter} \\
    \hline\hline\hline
    Parameter & {3.89$\degree$} & {3.48$\degree$} & {3.15$\degree$} & {2.88$\degree$} & {2.65$\degree$} & {2.45$\degree$} & {2.13$\degree$} \\ 
    \hline
    \endfirsthead
    
    \caption[]{Interlayer potential term parameters ($\text{meV}$) for the full model of {\tmt} based on DFT with quick basis selection (continued)} \\
    \hline\hline\hline
    Parameter & {3.89$\degree$} & {3.48$\degree$} & {3.15$\degree$} & {2.88$\degree$} & {2.65$\degree$} & {2.45$\degree$} & {2.13$\degree$} \\
    \hline
    \endhead
    \hline \hline\hline
    \endfoot
        $w^{0,0}_{bt,\bsl{q}_{1}}$ & \cellnum{-1.4859}{+0.4139}  & \cellnum{-1.7468}{-0.0760}  & \cellnum{-1.2903}{+0.2975}  & \cellnum{-1.4904}{+0.0092}  & \cellnum{-1.1591}{+0.2210}  & \cellnum{-1.0955}{+0.1904}  & \cellnum{-0.9901}{+0.1414}  \\ 
        $w^{1,0}_{bt,\bsl{q}_{1}}$ & \cellnum{3.2918}{-2.0525}  & \cellnum{3.5455}{+8.1102}  & \cellnum{4.1767}{-1.9454}  & \cellnum{4.3257}{+6.8691}  & \cellnum{4.6972}{-2.0180}  & \cellnum{4.8638}{-1.9961}  & \cellnum{5.1894}{-1.9259}  \\ 
        $w^{2,0}_{bt,\bsl{q}_{1}}$ & \cellnum{38.9164}{-3.6037}  & \cellnum{37.5726}{+1.7483}  & \cellnum{40.6076}{-1.7984}  & \cellnum{39.1207}{+0.8536}  & \cellnum{42.3461}{-0.8450}  & \cellnum{43.0546}{-0.4813}  & \cellnum{44.1041}{+0.1205}  \\ 
        $w^{1,1}_{bt,\bsl{q}_{1}}$ & \cellnum{3.2951}{-3.6016}  & \cellnum{8.7480}{-2.3152}  & \cellnum{1.5579}{-2.0723}  & \cellnum{5.2278}{-4.4488}  & \cellnum{1.7887}{-2.2264}  & \cellnum{1.7602}{-2.2807}  & \cellnum{1.5218}{-2.3184}  \\ 
        $w^{3,0}_{bt,\bsl{q}_{1}}$ & \cellnum{-12.2950}{+9.6872}  & \cellnum{-21.4501}{+13.6503}  & \cellnum{-16.3108}{+7.9524}  & \cellnum{-22.7329}{+14.4772}  & \cellnum{-21.2502}{+5.7739}  & \cellnum{-23.2894}{+4.7097}  & \cellnum{-26.9499}{+2.6459}  \\ 
        $w^{2,1}_{bt,\bsl{q}_{1}}$ & \cellnum{-30.6028}{+12.3370}  & \cellnum{-43.6718}{-55.2743}  & \cellnum{-46.7290}{+11.7340}  & \cellnum{-57.7625}{-43.9844}  & \cellnum{-58.6792}{+16.7599}  & \cellnum{-63.5998}{+18.6761}  & \cellnum{-73.7787}{+22.1133}  \\ 
        $w^{4,0}_{bt,\bsl{q}_{1}}$ & \cellnum{27.7311}{-18.8552}  & \cellnum{36.2954}{+20.9406}  & \cellnum{42.0112}{-22.6876}  & \cellnum{48.2057}{+22.5623}  & \cellnum{53.4292}{-21.7174}  & \cellnum{57.5906}{-20.2418}  & \cellnum{63.9670}{-15.6066}  \\
\end{longtable}

\renewcommand{\arraystretch}{1.5} 
\begin{longtable}{c *{7}{S[table-column-width=2cm]}}
    \caption{Kinetic term parameters ($m_\text{e}$) for the reduced model of {\tmt} based on DFT with quick basis selection}\label{table:mote2_52orb_reduced_diag} \\
    \hline\hline\hline
    Parameter & {3.89$\degree$} & {3.48$\degree$} & {3.15$\degree$} & {2.88$\degree$} & {2.65$\degree$} & {2.45$\degree$} & {2.13$\degree$} \\ 
    \hline
    \endfirsthead
    
    \caption[]{Kinetic term parameters ($m_\text{e}$) for the reduced model of {\tmt} based on DFT with quick basis selection (continued)} \\
    \hline\hline\hline
    Parameter & {3.89$\degree$} & {3.48$\degree$} & {3.15$\degree$} & {2.88$\degree$} & {2.65$\degree$} & {2.45$\degree$} & {2.13$\degree$} \\ 
    \hline
    \endhead
    \hline \hline\hline
    \endfoot$m^{(1, 1)}$ & -0.6739 & -0.6846 & -0.6820 & -0.6660 & -0.6601 & -0.6640 & -0.6513 \\ 
        
\end{longtable}
    
\renewcommand{\arraystretch}{2.4} 
\begin{longtable}{c *{7}{>{\centering\arraybackslash}p{2cm}}}

    \caption{Intralayer potential term parameters ($\text{meV}$) for the reduced model of {\tmt} based on DFT with quick basis selection}\label{table:mote2_52orb_reduced_intra} \\
    \hline\hline\hline
    Parameter & {3.89$\degree$} & {3.48$\degree$} & {3.15$\degree$} & {2.88$\degree$} & {2.65$\degree$} & {2.45$\degree$} & {2.13$\degree$} \\ 
    \hline
    \endfirsthead
    
    \caption[]{Intralayer potential term parameters ($\text{meV}$) for the reduced model of {\tmt} based on DFT with quick basis selection (continued)} \\
    \hline\hline\hline
    Parameter & {3.89$\degree$} & {3.48$\degree$} & {3.15$\degree$} & {2.88$\degree$} & {2.65$\degree$} & {2.45$\degree$} & {2.13$\degree$} \\
    \hline
    \endhead
    \hline \hline\hline
    \endfoot
        $V^{0,0}_{b,\bsl{b}_{M_1}}$ & \cellnum{-0.1474}{-3.7990}  & \cellnum{0.1810}{+4.0256}  & \cellnum{0.2450}{-3.5442}  & \cellnum{0.2352}{+3.2592}  & \cellnum{0.2312}{-2.9032}  & \cellnum{0.2302}{-2.5452}  & \cellnum{0.2303}{-1.7489}  \\ 
        $V^{1,0}_{b,\bsl{b}_{M_1}}$ & \cellnum{-0.6450}{-9.6915}  & \cellnum{-2.5276}{-3.2638}  & \cellnum{-3.7990}{-12.6673}  & \cellnum{-1.4631}{-4.7335}  & \cellnum{-3.5411}{-12.9294}  & \cellnum{-3.3860}{-12.8630}  & \cellnum{-3.1002}{-12.3481}  \\ 
        $V^{0,1}_{b,\bsl{b}_{M_1}}$ & \cellnum{-0.8922}{-2.5289}  & \cellnum{-4.2681}{+12.7805}  & \cellnum{-1.6634}{+4.1358}  & \cellnum{-3.6963}{+12.8945}  & \cellnum{-1.2223}{+5.2018}  & \cellnum{-0.9589}{+5.5725}  & \cellnum{-0.5841}{+5.9847}  \\ 
        $V^{2,0}_{b,\bsl{b}_{M_1}}$ & \cellnum{14.2448}{-31.6143}  & \cellnum{0.0000}{+0.0000}  & \cellnum{0.0000}{+0.0000}  & \cellnum{0.0000}{+0.0000}  & \cellnum{0.0000}{+0.0000}  & \cellnum{0.0000}{+0.0000}  & \cellnum{0.0000}{+0.0000}  \\ 
        $V^{1,1}_{b,\bsl{b}_{M_1}}$ & \cellnum{9.1970}{+9.8042}  & \cellnum{0.0000}{+0.0000}  & \cellnum{0.0000}{+0.0000}  & \cellnum{0.0000}{+0.0000}  & \cellnum{0.0000}{+0.0000}  & \cellnum{0.0000}{+0.0000}  & \cellnum{0.0000}{+0.0000}  \\ 
        $V^{0,2}_{b,\bsl{b}_{M_1}}$ & \cellnum{2.5861}{+34.6600}  & \cellnum{0.0000}{+0.0000}  & \cellnum{0.0000}{+0.0000}  & \cellnum{0.0000}{+0.0000}  & \cellnum{0.0000}{+0.0000}  & \cellnum{0.0000}{+0.0000}  & \cellnum{0.0000}{+0.0000}  \\ 
        $V^{3,0}_{b,\bsl{b}_{M_1}}$ & \cellnum{-27.9727}{+8.9876}  & \cellnum{0.0000}{+0.0000}  & \cellnum{0.0000}{+0.0000}  & \cellnum{0.0000}{+0.0000}  & \cellnum{0.0000}{+0.0000}  & \cellnum{0.0000}{+0.0000}  & \cellnum{0.0000}{+0.0000}  \\ 
        $V^{2,1}_{b,\bsl{b}_{M_1}}$ & \cellnum{-78.6727}{+74.8538}  & \cellnum{0.0000}{+0.0000}  & \cellnum{0.0000}{+0.0000}  & \cellnum{0.0000}{+0.0000}  & \cellnum{0.0000}{+0.0000}  & \cellnum{0.0000}{+0.0000}  & \cellnum{0.0000}{+0.0000}  \\
\end{longtable}
    
\renewcommand{\arraystretch}{2.4} 
\begin{longtable}{c *{7}{>{\centering\arraybackslash}p{2cm}}}

    \caption{Interlayer potential term parameters ($\text{meV}$) for the reduced model of {\tmt} based on DFT with quick basis selection}\label{table:mote2_52orb_reduced_inter} \\
    \hline\hline\hline
    Parameter & {3.89$\degree$} & {3.48$\degree$} & {3.15$\degree$} & {2.88$\degree$} & {2.65$\degree$} & {2.45$\degree$} & {2.13$\degree$} \\ 
    \hline
    \endfirsthead
    
    \caption[]{Interlayer potential term parameters ($\text{meV}$) for the reduced model of {\tmt} based on DFT with quick basis selection (continued)} \\
    \hline\hline\hline
    Parameter & {3.89$\degree$} & {3.48$\degree$} & {3.15$\degree$} & {2.88$\degree$} & {2.65$\degree$} & {2.45$\degree$} & {2.13$\degree$} \\
    \hline
    \endhead
    \hline \hline\hline
    \endfoot
        $w^{0,0}_{bt,\bsl{q}_{1}}$ & \cellnum{-1.3657}{+0.4220}  & \cellnum{-1.3448}{-0.0242}  & \cellnum{-1.1629}{+0.2215}  & \cellnum{-1.2461}{+0.0360}  & \cellnum{-1.0881}{+0.1496}  & \cellnum{-1.0445}{+0.1200}  & \cellnum{-0.9656}{+0.0732}  \\ 
        $w^{1,0}_{bt,\bsl{q}_{1}}$ & \cellnum{3.8263}{-2.7138}  & \cellnum{2.8453}{+2.3748}  & \cellnum{3.3342}{-2.2203}  & \cellnum{3.6800}{+2.0142}  & \cellnum{3.9795}{-1.8407}  & \cellnum{4.2076}{-1.6913}  & \cellnum{4.6311}{-1.6070}  \\ 
        $w^{2,0}_{bt,\bsl{q}_{1}}$ & \cellnum{35.3944}{-4.3459}  & \cellnum{0.0000}{+0.0000}  & \cellnum{0.0000}{+0.0000}  & \cellnum{0.0000}{+0.0000}  & \cellnum{0.0000}{+0.0000}  & \cellnum{0.0000}{+0.0000}  & \cellnum{0.0000}{+0.0000}  \\ 
        $w^{1,1}_{bt,\bsl{q}_{1}}$ & \cellnum{-9.8408}{-4.9173}  & \cellnum{0.0000}{+0.0000}  & \cellnum{0.0000}{+0.0000}  & \cellnum{0.0000}{+0.0000}  & \cellnum{0.0000}{+0.0000}  & \cellnum{0.0000}{+0.0000}  & \cellnum{0.0000}{+0.0000}  \\ 
        $w^{3,0}_{bt,\bsl{q}_{1}}$ & \cellnum{-22.3412}{-5.4535}  & \cellnum{0.0000}{+0.0000}  & \cellnum{0.0000}{+0.0000}  & \cellnum{0.0000}{+0.0000}  & \cellnum{0.0000}{+0.0000}  & \cellnum{0.0000}{+0.0000}  & \cellnum{0.0000}{+0.0000}  \\ 
        $w^{2,1}_{bt,\bsl{q}_{1}}$ & \cellnum{-37.5948}{+78.7889}  & \cellnum{0.0000}{+0.0000}  & \cellnum{0.0000}{+0.0000}  & \cellnum{0.0000}{+0.0000}  & \cellnum{0.0000}{+0.0000}  & \cellnum{0.0000}{+0.0000}  & \cellnum{0.0000}{+0.0000}  \\ 
        \hline 
        $w^{0,0}_{bt,-2 \bsl{q}_{1}}$ & \cellnum{0.3823}{+0.2358}  & \cellnum{0.2845}{-0.3131}  & \cellnum{0.3269}{+0.2325}  & \cellnum{0.2671}{-0.3085}  & \cellnum{0.3159}{+0.2278}  & \cellnum{0.3073}{+0.2176}  & \cellnum{0.3020}{+0.1955}  \\ 
        $w^{1,0}_{bt,-2 \bsl{q}_{1}}$ & \cellnum{0.3601}{+0.4737}  & \cellnum{0.4518}{-0.5269}  & \cellnum{0.5385}{+0.5095}  & \cellnum{0.6289}{-0.4561}  & \cellnum{0.7196}{+0.4097}  & \cellnum{0.7983}{+0.3665}  & \cellnum{0.9493}{+0.3684}  \\
\end{longtable}

\renewcommand{\arraystretch}{1.5} 
\begin{longtable}{c *{7}{S[table-column-width=2cm]}}
    \caption{Kinetic term parameters ($m_\text{e}$) for the full model of {\tmt} based on DFT with standard basis selection}\label{table:mote2_71orb_full_diag} \\
    \hline\hline\hline
    Parameter & {3.89$\degree$} & {3.48$\degree$} & {3.15$\degree$} & {2.88$\degree$} & {2.65$\degree$} & {2.45$\degree$} & {2.13$\degree$} \\ 
    \hline
    \endfirsthead
    
    \caption[]{Kinetic term parameters ($m_\text{e}$) for the full model of {\tmt} based on DFT with standard basis selection (continued)} \\
    \hline\hline\hline
    Parameter & {3.89$\degree$} & {3.48$\degree$} & {3.15$\degree$} & {2.88$\degree$} & {2.65$\degree$} & {2.45$\degree$} & {2.13$\degree$} \\ 
    \hline
    \endhead
    \hline \hline\hline
    \endfoot$m^{(1, 1)}$ & -0.6290 & -0.6119 & -0.6065 & -0.6063 & -0.6050 & -0.6296 & -0.6142 \\ 
        $m^{(3, 0)}$ & 1.3947 & 1.3256 & 1.3514 & 1.3225 & 1.2931 & 1.5681 & 1.4665 \\ 
        $m^{(2, 2)}$ & 0.0701 & 0.0622 & 0.0604 & 0.0581 & 0.0567 & 0.0749 & 0.0646 \\ 
        $m^{(4, 1)}$ & -0.1323 & -0.1116 & -0.1208 & -0.1086 & -0.0982 & -0.2311 & -0.1845 \\ 
        $m^{(3, 3)}$ & -0.0111 & -0.0080 & -0.0077 & -0.0065 & -0.0058 & -0.0164 & -0.0104 \\ 
        $m^{(6, 0)}$ & 12.6650 & -2.0632 & 0.4985 & 0.4344 & 0.4296 & 0.9395 & 0.4309 \\ 
        $m^{(5, 2)}$ & 0.0267 & 0.0188 & 0.0224 & 0.0176 & 0.0141 & 0.0000 & 0.0000 \\ 
        $m^{(4, 4)}$ & 0.0030 & 0.0016 & 0.0016 & 0.0011 & 0.0009 & 0.0000 & 0.0000 \\ 
        $m^{(7, 1)}$ & 0.1481 & 0.0438 & -0.0433 & -0.0332 & -0.0323 & 0.0000 & 0.0000 \\ 
        $m^{(6, 3)}$ & -0.0140 & -0.0079 & -0.0106 & -0.0071 & -0.0048 & 0.0000 & 0.0000 \\ 
        $m^{(9, 0)}$ & -3.2103 & -2.2691 & -6.4626 & -6.7820 & -9.6389 & 0.0000 & 0.0000 \\ 
        $m^{(5, 5)}$ & -0.0018 & -0.0007 & -0.0008 & -0.0005 & -0.0003 & 0.0000 & 0.0000 \\ 
        $m^{(8, 2)}$ & -0.0273 & -0.0077 & 0.0143 & 0.0097 & 0.0099 & 0.0000 & 0.0000 \\ 
        
\end{longtable}
    
\renewcommand{\arraystretch}{2.4} 
\begin{longtable}{c *{7}{>{\centering\arraybackslash}p{2cm}}}

    \caption{Intralayer potential term parameters ($\text{meV}$) for the full model of {\tmt} based on DFT with standard basis selection}\label{table:mote2_71orb_full_intra} \\
    \hline\hline\hline
    Parameter & {3.89$\degree$} & {3.48$\degree$} & {3.15$\degree$} & {2.88$\degree$} & {2.65$\degree$} & {2.45$\degree$} & {2.13$\degree$} \\ 
    \hline
    \endfirsthead
    
    \caption[]{Intralayer potential term parameters ($\text{meV}$) for the full model of {\tmt} based on DFT with standard basis selection (continued)} \\
    \hline\hline\hline
    Parameter & {3.89$\degree$} & {3.48$\degree$} & {3.15$\degree$} & {2.88$\degree$} & {2.65$\degree$} & {2.45$\degree$} & {2.13$\degree$} \\
    \hline
    \endhead
    \hline \hline\hline
    \endfoot
        $V^{0,0}_{b,\bsl{b}_{M_1}}$ & \cellnum{-0.1200}{-3.5940}  & \cellnum{-0.1132}{+3.9385}  & \cellnum{-0.0315}{-3.4708}  & \cellnum{-0.1023}{+2.9937}  & \cellnum{-0.0271}{-2.7491}  & \cellnum{-0.0347}{-2.3510}  & \cellnum{-0.0426}{-1.5803}  \\ 
        $V^{1,0}_{b,\bsl{b}_{M_1}}$ & \cellnum{-1.3330}{-9.8741}  & \cellnum{-0.7287}{+3.9900}  & \cellnum{-1.2982}{-10.7303}  & \cellnum{0.5266}{-1.4880}  & \cellnum{-1.0359}{-11.1370}  & \cellnum{-0.8939}{-11.0962}  & \cellnum{-1.0495}{-10.6338}  \\ 
        $V^{0,1}_{b,\bsl{b}_{M_1}}$ & \cellnum{-0.7285}{-4.5535}  & \cellnum{-1.6397}{+11.6200}  & \cellnum{-0.0933}{-2.2851}  & \cellnum{-1.0684}{+10.8185}  & \cellnum{-0.2432}{-0.3291}  & \cellnum{-0.2922}{+0.4577}  & \cellnum{0.1723}{+1.8129}  \\ 
        $V^{2,0}_{b,\bsl{b}_{M_1}}$ & \cellnum{-0.1693}{-35.5222}  & \cellnum{-2.3608}{-45.9456}  & \cellnum{3.3932}{-42.1345}  & \cellnum{0.3988}{-29.8951}  & \cellnum{2.3535}{-44.6941}  & \cellnum{1.7688}{-44.9749}  & \cellnum{1.8275}{-44.6072}  \\ 
        $V^{1,1}_{b,\bsl{b}_{M_1}}$ & \cellnum{2.5042}{+11.1474}  & \cellnum{2.2974}{-14.0014}  & \cellnum{1.7774}{+15.4316}  & \cellnum{3.2775}{-8.6276}  & \cellnum{-0.5137}{+18.8286}  & \cellnum{-1.5088}{+19.8503}  & \cellnum{-4.0348}{+28.8148}  \\ 
        $V^{0,2}_{b,\bsl{b}_{M_1}}$ & \cellnum{-1.3478}{+36.7777}  & \cellnum{-1.0718}{+42.7901}  & \cellnum{1.8324}{+44.5211}  & \cellnum{0.5466}{+25.3219}  & \cellnum{1.6131}{+48.0321}  & \cellnum{1.6205}{+48.4464}  & \cellnum{0.1358}{+44.7863}  \\ 
        $V^{3,0}_{b,\bsl{b}_{M_1}}$ & \cellnum{6.4450}{-14.4462}  & \cellnum{14.8206}{-44.9263}  & \cellnum{-0.6943}{-8.4729}  & \cellnum{8.5615}{-3.8641}  & \cellnum{2.9153}{-10.0510}  & \cellnum{5.4140}{-10.5338}  & \cellnum{-1.9079}{-0.8134}  \\
\end{longtable}
    
\renewcommand{\arraystretch}{2.4} 
\begin{longtable}{c *{7}{>{\centering\arraybackslash}p{2cm}}}

    \caption{Interlayer potential term parameters ($\text{meV}$) for the full model of {\tmt} based on DFT with standard selection}\label{table:mote2_71orb_full_inter} \\
    \hline\hline\hline
    Parameter & {3.89$\degree$} & {3.48$\degree$} & {3.15$\degree$} & {2.88$\degree$} & {2.65$\degree$} & {2.45$\degree$} & {2.13$\degree$} \\ 
    \hline
    \endfirsthead
    
    \caption[]{Interlayer potential term parameters ($\text{meV}$) for the full model of {\tmt} based on DFT with standard selection (continued)} \\
    \hline\hline\hline
    Parameter & {3.89$\degree$} & {3.48$\degree$} & {3.15$\degree$} & {2.88$\degree$} & {2.65$\degree$} & {2.45$\degree$} & {2.13$\degree$} \\
    \hline
    \endhead
    \hline \hline\hline
    \endfoot
        $w^{0,0}_{bt,\bsl{q}_{1}}$ & \cellnum{-1.5294}{+0.4806}  & \cellnum{-1.7639}{-0.1428}  & \cellnum{-1.2961}{+0.3248}  & \cellnum{-1.5731}{-0.0740}  & \cellnum{-1.1882}{+0.2384}  & \cellnum{-1.1329}{+0.2051}  & \cellnum{-1.0280}{+0.1502}  \\ 
        $w^{1,0}_{bt,\bsl{q}_{1}}$ & \cellnum{3.4444}{-1.8623}  & \cellnum{3.2432}{+7.8920}  & \cellnum{3.6652}{-2.2542}  & \cellnum{3.7670}{+6.3924}  & \cellnum{4.0707}{-2.0912}  & \cellnum{4.2463}{-2.0093}  & \cellnum{4.4565}{-1.8084}  \\ 
        $w^{2,0}_{bt,\bsl{q}_{1}}$ & \cellnum{41.2787}{-4.6185}  & \cellnum{40.6596}{-1.2708}  & \cellnum{43.4659}{-1.5015}  & \cellnum{30.6718}{+1.5089}  & \cellnum{44.9943}{-0.4453}  & \cellnum{45.6129}{-0.0338}  & \cellnum{46.5930}{+1.1415}  \\ 
        $w^{1,1}_{bt,\bsl{q}_{1}}$ & \cellnum{1.0860}{-6.6119}  & \cellnum{5.0733}{-1.1683}  & \cellnum{-2.4607}{-3.9865}  & \cellnum{3.1456}{-1.7752}  & \cellnum{-1.2311}{-3.3883}  & \cellnum{-0.9171}{-3.1607}  & \cellnum{-0.2623}{-2.6906}  \\ 
        $w^{3,0}_{bt,\bsl{q}_{1}}$ & \cellnum{-34.0472}{+8.2111}  & \cellnum{-40.2516}{+16.3390}  & \cellnum{-32.7332}{+7.0471}  & \cellnum{-20.8656}{+4.7666}  & \cellnum{-34.1349}{+5.2250}  & \cellnum{-34.4262}{+4.5796}  & \cellnum{-25.3541}{+3.4457}  \\ 
        $w^{2,1}_{bt,\bsl{q}_{1}}$ & \cellnum{-51.9047}{+8.9135}  & \cellnum{-48.7946}{-38.9820}  & \cellnum{-61.5625}{+33.2170}  & \cellnum{-56.1823}{-28.9836}  & \cellnum{-66.2989}{+29.9031}  & \cellnum{-68.6334}{+28.3180}  & \cellnum{-70.0068}{+19.2686}  \\ 
        $w^{4,0}_{bt,\bsl{q}_{1}}$ & \cellnum{20.5112}{-34.5249}  & \cellnum{45.3831}{+1.8772}  & \cellnum{42.2887}{-32.4149}  & \cellnum{0.0000}{+0.0000}  & \cellnum{56.3085}{-32.6279}  & \cellnum{61.7179}{-31.5765}  & \cellnum{73.6682}{-22.0969}  \\
\end{longtable}

\renewcommand{\arraystretch}{1.5} 
\begin{longtable}{c *{7}{S[table-column-width=2cm]}}
    \caption{Kinetic term parameters ($m_\text{e}$) for the reduced model of {\tmt} based on DFT with standard basis selection}\label{table:mote2_71orb_reduced_diag} \\
    \hline\hline\hline
    Parameter & {3.89$\degree$} & {3.48$\degree$} & {3.15$\degree$} & {2.88$\degree$} & {2.65$\degree$} & {2.45$\degree$} & {2.13$\degree$} \\ 
    \hline
    \endfirsthead
    
    \caption[]{Kinetic term parameters ($m_\text{e}$) for the reduced model of {\tmt} based on DFT with standard basis selection (continued)} \\
    \hline\hline\hline
    Parameter & {3.89$\degree$} & {3.48$\degree$} & {3.15$\degree$} & {2.88$\degree$} & {2.65$\degree$} & {2.45$\degree$} & {2.13$\degree$} \\ 
    \hline
    \endhead
    \hline \hline\hline
    \endfoot$m^{(1, 1)}$ & -0.6440 & -0.6508 & -0.6473 & -0.6337 & -0.6262 & -0.6296 & -0.6133 \\ 
        
\end{longtable}
    
\renewcommand{\arraystretch}{2.4} 
\begin{longtable}{c *{7}{>{\centering\arraybackslash}p{2cm}}}

    \caption{Intralayer potential term parameters ($\text{meV}$) for the reduced model of {\tmt} based on DFT with standard basis selection}\label{table:mote2_71orb_reduced_intra} \\
    \hline\hline\hline
    Parameter & {3.89$\degree$} & {3.48$\degree$} & {3.15$\degree$} & {2.88$\degree$} & {2.65$\degree$} & {2.45$\degree$} & {2.13$\degree$} \\ 
    \hline
    \endfirsthead
    
    \caption[]{Intralayer potential term parameters ($\text{meV}$) for the reduced model of {\tmt} based on DFT with standard basis selection (continued)} \\
    \hline\hline\hline
    Parameter & {3.89$\degree$} & {3.48$\degree$} & {3.15$\degree$} & {2.88$\degree$} & {2.65$\degree$} & {2.45$\degree$} & {2.13$\degree$} \\
    \hline
    \endhead
    \hline \hline\hline
    \endfoot
        $V^{0,0}_{b,\bsl{b}_{M_1}}$ & \cellnum{-0.1690}{-3.5509}  & \cellnum{0.1774}{+3.7484}  & \cellnum{0.2322}{-3.2944}  & \cellnum{0.2123}{+2.9649}  & \cellnum{0.2137}{-2.6042}  & \cellnum{0.2034}{-2.2253}  & \cellnum{0.2019}{-1.4481}  \\ 
        $V^{1,0}_{b,\bsl{b}_{M_1}}$ & \cellnum{-0.5161}{-9.4270}  & \cellnum{-2.5538}{-3.4533}  & \cellnum{-3.7361}{-12.3446}  & \cellnum{-1.5363}{-4.8866}  & \cellnum{-3.4750}{-12.5966}  & \cellnum{-3.3111}{-12.5023}  & \cellnum{-3.0232}{-12.0633}  \\ 
        $V^{0,1}_{b,\bsl{b}_{M_1}}$ & \cellnum{-0.7979}{-2.4180}  & \cellnum{-4.1990}{+12.4002}  & \cellnum{-1.7093}{+4.3234}  & \cellnum{-3.6258}{+12.5415}  & \cellnum{-1.3206}{+5.3460}  & \cellnum{-1.0905}{+5.6707}  & \cellnum{-0.7583}{+6.0800}  \\ 
        $V^{2,0}_{b,\bsl{b}_{M_1}}$ & \cellnum{13.7913}{-31.8741}  & \cellnum{0.0000}{+0.0000}  & \cellnum{0.0000}{+0.0000}  & \cellnum{0.0000}{+0.0000}  & \cellnum{0.0000}{+0.0000}  & \cellnum{0.0000}{+0.0000}  & \cellnum{0.0000}{+0.0000}  \\ 
        $V^{1,1}_{b,\bsl{b}_{M_1}}$ & \cellnum{9.0001}{+10.1801}  & \cellnum{0.0000}{+0.0000}  & \cellnum{0.0000}{+0.0000}  & \cellnum{0.0000}{+0.0000}  & \cellnum{0.0000}{+0.0000}  & \cellnum{0.0000}{+0.0000}  & \cellnum{0.0000}{+0.0000}  \\ 
        $V^{0,2}_{b,\bsl{b}_{M_1}}$ & \cellnum{2.3027}{+34.7068}  & \cellnum{0.0000}{+0.0000}  & \cellnum{0.0000}{+0.0000}  & \cellnum{0.0000}{+0.0000}  & \cellnum{0.0000}{+0.0000}  & \cellnum{0.0000}{+0.0000}  & \cellnum{0.0000}{+0.0000}  \\ 
        $V^{3,0}_{b,\bsl{b}_{M_1}}$ & \cellnum{-28.2480}{+8.2977}  & \cellnum{0.0000}{+0.0000}  & \cellnum{0.0000}{+0.0000}  & \cellnum{0.0000}{+0.0000}  & \cellnum{0.0000}{+0.0000}  & \cellnum{0.0000}{+0.0000}  & \cellnum{0.0000}{+0.0000}  \\ 
        $V^{2,1}_{b,\bsl{b}_{M_1}}$ & \cellnum{-77.1131}{+77.1949}  & \cellnum{0.0000}{+0.0000}  & \cellnum{0.0000}{+0.0000}  & \cellnum{0.0000}{+0.0000}  & \cellnum{0.0000}{+0.0000}  & \cellnum{0.0000}{+0.0000}  & \cellnum{0.0000}{+0.0000}  \\
\end{longtable}
    
\renewcommand{\arraystretch}{2.4} 
\begin{longtable}{c *{7}{>{\centering\arraybackslash}p{2cm}}}

    \caption{Interlayer potential term parameters ($\text{meV}$) for the reduced model of {\tmt} based on DFT with standard basis selection}\label{table:mote2_71orb_reduced_inter} \\
    \hline\hline\hline
    Parameter & {3.89$\degree$} & {3.48$\degree$} & {3.15$\degree$} & {2.88$\degree$} & {2.65$\degree$} & {2.45$\degree$} & {2.13$\degree$} \\ 
    \hline
    \endfirsthead
    
    \caption[]{Interlayer potential term parameters ($\text{meV}$) for the reduced model of {\tmt} based on DFT with standard basis selection (continued)} \\
    \hline\hline\hline
    Parameter & {3.89$\degree$} & {3.48$\degree$} & {3.15$\degree$} & {2.88$\degree$} & {2.65$\degree$} & {2.45$\degree$} & {2.13$\degree$} \\
    \hline
    \endhead
    \hline \hline\hline
    \endfoot
        $w^{0,0}_{bt,\bsl{q}_{1}}$ & \cellnum{-1.4306}{+0.4334}  & \cellnum{-1.4121}{-0.0878}  & \cellnum{-1.2051}{+0.2366}  & \cellnum{-1.3069}{-0.0196}  & \cellnum{-1.1314}{+0.1660}  & \cellnum{-1.0901}{+0.1365}  & \cellnum{-1.0026}{+0.0867}  \\ 
        $w^{1,0}_{bt,\bsl{q}_{1}}$ & \cellnum{2.9821}{-2.4936}  & \cellnum{2.2236}{+2.5378}  & \cellnum{2.6149}{-2.3613}  & \cellnum{3.0370}{+2.2134}  & \cellnum{3.3213}{-2.0704}  & \cellnum{3.6018}{-1.9651}  & \cellnum{3.9768}{-1.9244}  \\ 
        $w^{2,0}_{bt,\bsl{q}_{1}}$ & \cellnum{36.1688}{-3.7629}  & \cellnum{0.0000}{+0.0000}  & \cellnum{0.0000}{+0.0000}  & \cellnum{0.0000}{+0.0000}  & \cellnum{0.0000}{+0.0000}  & \cellnum{0.0000}{+0.0000}  & \cellnum{0.0000}{+0.0000}  \\ 
        $w^{1,1}_{bt,\bsl{q}_{1}}$ & \cellnum{-10.7707}{-4.8946}  & \cellnum{0.0000}{+0.0000}  & \cellnum{0.0000}{+0.0000}  & \cellnum{0.0000}{+0.0000}  & \cellnum{0.0000}{+0.0000}  & \cellnum{0.0000}{+0.0000}  & \cellnum{0.0000}{+0.0000}  \\ 
        $w^{3,0}_{bt,\bsl{q}_{1}}$ & \cellnum{-24.7826}{-4.0028}  & \cellnum{0.0000}{+0.0000}  & \cellnum{0.0000}{+0.0000}  & \cellnum{0.0000}{+0.0000}  & \cellnum{0.0000}{+0.0000}  & \cellnum{0.0000}{+0.0000}  & \cellnum{0.0000}{+0.0000}  \\ 
        $w^{2,1}_{bt,\bsl{q}_{1}}$ & \cellnum{-31.3375}{+70.9930}  & \cellnum{0.0000}{+0.0000}  & \cellnum{0.0000}{+0.0000}  & \cellnum{0.0000}{+0.0000}  & \cellnum{0.0000}{+0.0000}  & \cellnum{0.0000}{+0.0000}  & \cellnum{0.0000}{+0.0000}  \\ 
        \hline 
        $w^{0,0}_{bt,-2 \bsl{q}_{1}}$ & \cellnum{0.3923}{+0.2455}  & \cellnum{0.2807}{-0.2837}  & \cellnum{0.3322}{+0.2350}  & \cellnum{0.2670}{-0.2831}  & \cellnum{0.3257}{+0.2312}  & \cellnum{0.3197}{+0.2220}  & \cellnum{0.3147}{+0.2018}  \\ 
        $w^{1,0}_{bt,-2 \bsl{q}_{1}}$ & \cellnum{0.0681}{+0.6804}  & \cellnum{0.2317}{-0.5987}  & \cellnum{0.2872}{+0.5743}  & \cellnum{0.3902}{-0.5340}  & \cellnum{0.4747}{+0.4984}  & \cellnum{0.5566}{+0.4749}  & \cellnum{0.6813}{+0.5111}  \\
\end{longtable}

\renewcommand{\arraystretch}{1.5} 
\begin{longtable}{c *{7}{S[table-column-width=2cm]}}
    \caption{Kinetic term parameters ($m_\text{e}$) for the full model of {\tws} based on DFT with quick basis selection}\label{table:wse2_47orb_full_diag} \\
    \hline\hline\hline
    Parameter & {3.89$\degree$} & {3.48$\degree$} & {3.15$\degree$} & {2.88$\degree$} & {2.65$\degree$} & {2.45$\degree$} & {2.13$\degree$} \\ 
    \hline
    \endfirsthead
    
    \caption[]{Kinetic term parameters ($m_\text{e}$) for the full model of {\tws} based on DFT with quick basis selection (continued)} \\
    \hline\hline\hline
    Parameter & {3.89$\degree$} & {3.48$\degree$} & {3.15$\degree$} & {2.88$\degree$} & {2.65$\degree$} & {2.45$\degree$} & {2.13$\degree$} \\ 
    \hline
    \endhead
    \hline \hline\hline
    \endfoot$m^{(1, 1)}$ & -0.3742 & -0.3788 & -0.3715 & -0.3701 & -0.3750 & -0.3713 & -0.3668 \\ 
        $m^{(3, 0)}$ & 1.4591 & 1.2957 & 1.2412 & 1.2165 & 1.1930 & 1.1717 & 1.1468 \\ 
        $m^{(2, 2)}$ & 0.0379 & 0.0395 & 0.0356 & 0.0350 & 0.0368 & 0.0346 & 0.0317 \\ 
        $m^{(4, 1)}$ & -0.2042 & -0.1319 & -0.1140 & -0.1091 & -0.0995 & -0.0912 & -0.0805 \\ 
        $m^{(3, 3)}$ & -0.0050 & -0.0052 & -0.0040 & -0.0042 & -0.0044 & -0.0037 & -0.0027 \\ 
        $m^{(6, 0)}$ & 0.3467 & 0.2152 & 0.1545 & 0.1565 & 0.1730 & 0.1439 & 0.1044 \\ 
        $m^{(5, 2)}$ & 0.2080 & 0.0358 & 0.0263 & 0.0285 & 0.0228 & 0.0186 & 0.0134 \\ 
        $m^{(4, 4)}$ & 0.0012 & 0.0011 & 0.0007 & 0.0011 & 0.0011 & 0.0008 & 0.0004 \\ 
        $m^{(7, 1)}$ & -0.0890 & -0.0674 & -0.0354 & -0.0150 & -0.0141 & -0.0100 & -0.0054 \\ 
        $m^{(6, 3)}$ & 0.0329 & -0.0362 & -0.0225 & 0.0000 & 0.0000 & 0.0000 & 0.0000 \\ 
        $m^{(9, 0)}$ & 0.8449 & 2.6952 & 1.3035 & 0.0000 & 0.0000 & 0.0000 & 0.0000 \\ 
        $m^{(5, 5)}$ & -0.0006 & -0.0005 & -0.0003 & 0.0000 & 0.0000 & 0.0000 & 0.0000 \\ 
        $m^{(8, 2)}$ & -0.0276 & -0.0059 & -0.0030 & 0.0000 & 0.0000 & 0.0000 & 0.0000 \\ 
        
\end{longtable}
    
\renewcommand{\arraystretch}{2.4} 
\begin{longtable}{c *{7}{>{\centering\arraybackslash}p{2cm}}}

    \caption{Intralayer potential term parameters ($\text{meV}$) for the full model of {\tws} based on DFT with quick basis selection}\label{table:wse2_47orb_full_intra} \\
    \hline\hline\hline
    Parameter & {3.89$\degree$} & {3.48$\degree$} & {3.15$\degree$} & {2.88$\degree$} & {2.65$\degree$} & {2.45$\degree$} & {2.13$\degree$} \\ 
    \hline
    \endfirsthead
    
    \caption[]{Intralayer potential term parameters ($\text{meV}$) for the full model of {\tws} based on DFT with quick basis selection (continued)} \\
    \hline\hline\hline
    Parameter & {3.89$\degree$} & {3.48$\degree$} & {3.15$\degree$} & {2.88$\degree$} & {2.65$\degree$} & {2.45$\degree$} & {2.13$\degree$} \\
    \hline
    \endhead
    \hline \hline\hline
    \endfoot
        $V^{0,0}_{b,\bsl{b}_{M_1}}$ & \cellnum{-0.1222}{-1.2489}  & \cellnum{-0.1193}{-1.3645}  & \cellnum{-0.1206}{-1.4292}  & \cellnum{-0.1063}{-1.4495}  & \cellnum{-0.1038}{-1.4243}  & \cellnum{-0.0977}{-1.3206}  & \cellnum{-0.0843}{-1.0448}  \\ 
        $V^{1,0}_{b,\bsl{b}_{M_1}}$ & \cellnum{-1.0320}{-6.2431}  & \cellnum{-1.1277}{-7.0645}  & \cellnum{-1.1965}{-7.8493}  & \cellnum{-1.3193}{-8.1281}  & \cellnum{-1.3149}{-8.7639}  & \cellnum{-1.2837}{-9.1403}  & \cellnum{-1.1367}{-8.9846}  \\ 
        $V^{0,1}_{b,\bsl{b}_{M_1}}$ & \cellnum{0.0611}{-4.3291}  & \cellnum{0.0344}{-3.9170}  & \cellnum{0.0434}{-3.4040}  & \cellnum{-0.2905}{-2.6829}  & \cellnum{-0.3297}{-2.0586}  & \cellnum{-0.3741}{-1.3826}  & \cellnum{-0.2097}{-0.3157}  \\ 
        $V^{2,0}_{b,\bsl{b}_{M_1}}$ & \cellnum{-1.5161}{-26.7681}  & \cellnum{-1.6718}{-31.3323}  & \cellnum{-1.9447}{-35.7410}  & \cellnum{1.4580}{-38.4323}  & \cellnum{1.3432}{-42.0794}  & \cellnum{1.2294}{-44.4275}  & \cellnum{-0.2800}{-43.8232}  \\ 
        $V^{1,1}_{b,\bsl{b}_{M_1}}$ & \cellnum{3.1686}{+1.5509}  & \cellnum{3.1322}{+3.1542}  & \cellnum{3.1147}{+4.8481}  & \cellnum{2.3772}{+5.5773}  & \cellnum{2.1016}{+7.2425}  & \cellnum{1.7004}{+8.6743}  & \cellnum{-1.1961}{+15.9635}  \\ 
        $V^{0,2}_{b,\bsl{b}_{M_1}}$ & \cellnum{-0.7984}{+25.0116}  & \cellnum{-1.1856}{+30.4357}  & \cellnum{-1.5403}{+35.7085}  & \cellnum{1.0806}{+39.1549}  & \cellnum{0.8428}{+43.3535}  & \cellnum{0.6611}{+46.2139}  & \cellnum{-2.0173}{+48.1438}  \\ 
        $V^{3,0}_{b,\bsl{b}_{M_1}}$ & \cellnum{1.6444}{-6.2855}  & \cellnum{1.8366}{-8.2211}  & \cellnum{2.1702}{-10.0162}  & \cellnum{-1.1167}{-4.1631}  & \cellnum{-0.7761}{-4.7683}  & \cellnum{-0.2075}{-5.0291}  & \cellnum{3.6886}{-1.3012}  \\
\end{longtable}
    
\renewcommand{\arraystretch}{2.4} 
\begin{longtable}{c *{7}{>{\centering\arraybackslash}p{2cm}}}

    \caption{Interlayer potential term parameters ($\text{meV}$) for the full model of {\tws} based on DFT with quick basis selection}\label{table:wse2_47orb_full_inter} \\
    \hline\hline\hline
    Parameter & {3.89$\degree$} & {3.48$\degree$} & {3.15$\degree$} & {2.88$\degree$} & {2.65$\degree$} & {2.45$\degree$} & {2.13$\degree$} \\ 
    \hline
    \endfirsthead
    
    \caption[]{Interlayer potential term parameters ($\text{meV}$) for the full model of {\tws} based on DFT with quick basis selection (continued)} \\
    \hline\hline\hline
    Parameter & {3.89$\degree$} & {3.48$\degree$} & {3.15$\degree$} & {2.88$\degree$} & {2.65$\degree$} & {2.45$\degree$} & {2.13$\degree$} \\
    \hline
    \endhead
    \hline \hline\hline
    \endfoot
        $w^{0,0}_{bt,\bsl{q}_{1}}$ & \cellnum{-1.5555}{+0.6453}  & \cellnum{-1.5626}{+0.5906}  & \cellnum{-1.5516}{+0.5446}  & \cellnum{-1.4628}{+0.5055}  & \cellnum{-1.4392}{+0.4691}  & \cellnum{-1.4107}{+0.4365}  & \cellnum{-1.3504}{+0.3686}  \\ 
        $w^{1,0}_{bt,\bsl{q}_{1}}$ & \cellnum{10.3189}{-3.0129}  & \cellnum{11.0250}{-2.9568}  & \cellnum{11.6688}{-2.9061}  & \cellnum{12.3393}{-3.2604}  & \cellnum{12.8802}{-3.2071}  & \cellnum{13.3327}{-3.1307}  & \cellnum{13.6510}{-2.6990}  \\ 
        $w^{2,0}_{bt,\bsl{q}_{1}}$ & \cellnum{36.8862}{-4.0139}  & \cellnum{37.9556}{-2.8452}  & \cellnum{39.1189}{-1.9364}  & \cellnum{41.1790}{-0.8469}  & \cellnum{42.2805}{-0.2300}  & \cellnum{43.1377}{+0.3050}  & \cellnum{43.5504}{+0.9619}  \\ 
        $w^{1,1}_{bt,\bsl{q}_{1}}$ & \cellnum{-6.5362}{+0.5277}  & \cellnum{-5.9060}{+1.0738}  & \cellnum{-5.3277}{+1.4748}  & \cellnum{-9.8047}{-0.4981}  & \cellnum{-9.1920}{-0.5169}  & \cellnum{-8.6823}{-0.5723}  & \cellnum{-6.6245}{+2.2016}  \\ 
        $w^{3,0}_{bt,\bsl{q}_{1}}$ & \cellnum{-41.6618}{+6.2763}  & \cellnum{-44.5386}{+5.6079}  & \cellnum{-47.0099}{+5.0751}  & \cellnum{-43.3325}{+4.6681}  & \cellnum{-44.8408}{+3.7666}  & \cellnum{-45.8788}{+3.0200}  & \cellnum{-30.1807}{+5.7980}  \\ 
        $w^{2,1}_{bt,\bsl{q}_{1}}$ & \cellnum{-83.0724}{+27.9134}  & \cellnum{-93.4599}{+25.1717}  & \cellnum{-102.3584}{+21.2885}  & \cellnum{-130.7495}{+51.0639}  & \cellnum{-143.0680}{+51.6052}  & \cellnum{-154.1858}{+51.4992}  & \cellnum{-116.5374}{+24.1390}  \\ 
        $w^{4,0}_{bt,\bsl{q}_{1}}$ & \cellnum{38.3277}{-42.8287}  & \cellnum{45.7377}{-45.9913}  & \cellnum{51.8141}{-49.3781}  & \cellnum{69.0756}{-46.0896}  & \cellnum{77.4548}{-47.5598}  & \cellnum{84.9780}{-48.3337}  & \cellnum{100.7315}{-22.6182}  \\
\end{longtable}

\renewcommand{\arraystretch}{1.5} 
\begin{longtable}{c *{7}{S[table-column-width=2cm]}}
    \caption{Kinetic term parameters ($m_\text{e}$) for the reduced model of {\tws} based on DFT with quick basis selection}\label{table:wse2_47orb_reduced_diag} \\
    \hline\hline\hline
    Parameter & {3.89$\degree$} & {3.48$\degree$} & {3.15$\degree$} & {2.88$\degree$} & {2.65$\degree$} & {2.45$\degree$} & {2.13$\degree$} \\ 
    \hline
    \endfirsthead
    
    \caption[]{Kinetic term parameters ($m_\text{e}$) for the reduced model of {\tws} based on DFT with quick basis selection (continued)} \\
    \hline\hline\hline
    Parameter & {3.89$\degree$} & {3.48$\degree$} & {3.15$\degree$} & {2.88$\degree$} & {2.65$\degree$} & {2.45$\degree$} & {2.13$\degree$} \\ 
    \hline
    \endhead
    \hline \hline\hline
    \endfoot$m^{(1, 1)}$ & -0.3965 & -0.3860 & -0.3860 & -0.3803 & -0.3750 & -0.3713 & -0.3853 \\ 
        
\end{longtable}
    
\renewcommand{\arraystretch}{2.4} 
\begin{longtable}{c *{7}{>{\centering\arraybackslash}p{2cm}}}

    \caption{Intralayer potential term parameters ($\text{meV}$) for the reduced model of {\tws} based on DFT with quick basis selection}\label{table:wse2_47orb_reduced_intra} \\
    \hline\hline\hline
    Parameter & {3.89$\degree$} & {3.48$\degree$} & {3.15$\degree$} & {2.88$\degree$} & {2.65$\degree$} & {2.45$\degree$} & {2.13$\degree$} \\ 
    \hline
    \endfirsthead
    
    \caption[]{Intralayer potential term parameters ($\text{meV}$) for the reduced model of {\tws} based on DFT with quick basis selection (continued)} \\
    \hline\hline\hline
    Parameter & {3.89$\degree$} & {3.48$\degree$} & {3.15$\degree$} & {2.88$\degree$} & {2.65$\degree$} & {2.45$\degree$} & {2.13$\degree$} \\
    \hline
    \endhead
    \hline \hline\hline
    \endfoot
        $V^{0,0}_{b,\bsl{b}_{M_1}}$ & \cellnum{-0.1913}{-1.2485}  & \cellnum{-0.1762}{-1.3517}  & \cellnum{-0.1669}{-1.4073}  & \cellnum{-0.1567}{-1.4314}  & \cellnum{-0.1439}{-1.3999}  & \cellnum{-0.1286}{-1.2921}  & \cellnum{-0.0598}{-1.0488}  \\ 
        $V^{1,0}_{b,\bsl{b}_{M_1}}$ & \cellnum{-0.4066}{-6.3732}  & \cellnum{-0.6328}{-7.2648}  & \cellnum{-0.8188}{-8.1045}  & \cellnum{-0.9840}{-8.9102}  & \cellnum{-1.1117}{-9.5387}  & \cellnum{-1.1983}{-9.8904}  & \cellnum{-0.9508}{-8.7636}  \\ 
        $V^{0,1}_{b,\bsl{b}_{M_1}}$ & \cellnum{-0.0652}{-3.7127}  & \cellnum{-0.0689}{-3.3704}  & \cellnum{-0.0383}{-2.9257}  & \cellnum{-0.0593}{-2.3935}  & \cellnum{-0.1007}{-1.8066}  & \cellnum{-0.1447}{-1.1752}  & \cellnum{-0.2901}{-1.3467}  \\ 
        $V^{2,0}_{b,\bsl{b}_{M_1}}$ & \cellnum{9.7572}{-32.1432}  & \cellnum{10.0103}{-37.1088}  & \cellnum{9.9871}{-41.9246}  & \cellnum{10.1418}{-46.5919}  & \cellnum{9.9096}{-50.4116}  & \cellnum{9.3107}{-52.8176}  & \cellnum{0.3733}{-38.7032}  \\ 
        $V^{1,1}_{b,\bsl{b}_{M_1}}$ & \cellnum{6.5502}{-1.5134}  & \cellnum{6.7676}{-1.7685}  & \cellnum{6.9057}{-2.0282}  & \cellnum{7.1451}{-2.2278}  & \cellnum{7.3472}{-2.3578}  & \cellnum{7.3941}{-2.5075}  & \cellnum{-1.0326}{+26.2541}  \\ 
        $V^{0,2}_{b,\bsl{b}_{M_1}}$ & \cellnum{-1.2546}{+29.5330}  & \cellnum{-1.5859}{+34.3473}  & \cellnum{-1.8994}{+38.9635}  & \cellnum{-2.0313}{+43.4155}  & \cellnum{-2.0594}{+47.0410}  & \cellnum{-1.9897}{+49.3347}  & \cellnum{2.4546}{+52.2043}  \\ 
        $V^{3,0}_{b,\bsl{b}_{M_1}}$ & \cellnum{-27.6183}{+16.0654}  & \cellnum{-30.5199}{+17.1749}  & \cellnum{-32.6387}{+18.1845}  & \cellnum{-34.3750}{+18.7236}  & \cellnum{-34.1981}{+19.2402}  & \cellnum{-32.0344}{+19.5151}  & \cellnum{0.0000}{+0.0000}  \\
\end{longtable}
    
\renewcommand{\arraystretch}{2.4} 
\begin{longtable}{c *{7}{>{\centering\arraybackslash}p{2cm}}}

    \caption{Interlayer potential term parameters ($\text{meV}$) for the reduced model of {\tws} based on DFT with quick basis selection}\label{table:wse2_47orb_reduced_inter} \\
    \hline\hline\hline
    Parameter & {3.89$\degree$} & {3.48$\degree$} & {3.15$\degree$} & {2.88$\degree$} & {2.65$\degree$} & {2.45$\degree$} & {2.13$\degree$} \\ 
    \hline
    \endfirsthead
    
    \caption[]{Interlayer potential term parameters ($\text{meV}$) for the reduced model of {\tws} based on DFT with quick basis selection (continued)} \\
    \hline\hline\hline
    Parameter & {3.89$\degree$} & {3.48$\degree$} & {3.15$\degree$} & {2.88$\degree$} & {2.65$\degree$} & {2.45$\degree$} & {2.13$\degree$} \\
    \hline
    \endhead
    \hline \hline\hline
    \endfoot
        $w^{0,0}_{bt,\bsl{q}_{1}}$ & \cellnum{-1.4353}{+0.6807}  & \cellnum{-1.4659}{+0.6316}  & \cellnum{-1.4714}{+0.5859}  & \cellnum{-1.4457}{+0.5286}  & \cellnum{-1.4218}{+0.4902}  & \cellnum{-1.3944}{+0.4548}  & \cellnum{-1.3615}{+0.3423}  \\ 
        $w^{1,0}_{bt,\bsl{q}_{1}}$ & \cellnum{11.1468}{-3.7369}  & \cellnum{11.7726}{-3.6184}  & \cellnum{12.3556}{-3.5287}  & \cellnum{12.6550}{-3.3937}  & \cellnum{13.1239}{-3.3038}  & \cellnum{13.5219}{-3.1885}  & \cellnum{12.2244}{-1.7896}  \\ 
        $w^{2,0}_{bt,\bsl{q}_{1}}$ & \cellnum{39.3527}{-5.4324}  & \cellnum{41.0136}{-3.9625}  & \cellnum{42.4871}{-2.7414}  & \cellnum{43.3656}{-1.5919}  & \cellnum{44.4125}{-0.6968}  & \cellnum{45.0791}{+0.0157}  & \cellnum{42.4213}{+0.4730}  \\ 
        $w^{1,1}_{bt,\bsl{q}_{1}}$ & \cellnum{-14.7780}{-0.8024}  & \cellnum{-14.6769}{-1.2283}  & \cellnum{-14.7217}{-1.6199}  & \cellnum{-14.6642}{-2.0556}  & \cellnum{-14.9222}{-2.4869}  & \cellnum{-15.0942}{-2.8917}  & \cellnum{-10.5945}{+3.1713}  \\ 
        $w^{3,0}_{bt,\bsl{q}_{1}}$ & \cellnum{-34.4507}{-9.0606}  & \cellnum{-36.2372}{-11.1027}  & \cellnum{-37.5416}{-12.8291}  & \cellnum{-37.1984}{-14.2029}  & \cellnum{-37.8977}{-15.1069}  & \cellnum{-38.6026}{-15.4044}  & \cellnum{0.0000}{+0.0000}  \\ 
        $w^{2,1}_{bt,\bsl{q}_{1}}$ & \cellnum{-78.9933}{+64.9710}  & \cellnum{-90.9334}{+70.9916}  & \cellnum{-101.6991}{+77.5258}  & \cellnum{-110.5890}{+84.1201}  & \cellnum{-119.2994}{+90.9022}  & \cellnum{-127.2793}{+95.7934}  & \cellnum{0.0000}{+0.0000}  \\ 
        \hline 
        $w^{0,0}_{bt,-2 \bsl{q}_{1}}$ & \cellnum{0.2976}{+0.0966}  & \cellnum{0.2961}{+0.1075}  & \cellnum{0.2952}{+0.1191}  & \cellnum{0.2907}{+0.1315}  & \cellnum{0.2878}{+0.1396}  & \cellnum{0.2855}{+0.1421}  & \cellnum{0.2852}{+0.1393}  \\
\end{longtable}

\end{document}